\documentstyle[12pt,epsf]{article}
\begin{document}

\newcommand{\f}[2]{\frac{#1}{#2}}
\newcommand{\be}{\begin{equation}}
\newcommand{\ee}{\end{equation}}
\newcommand{\bea}{\begin{eqnarray}}
\newcommand{\eea}{\end{eqnarray}}
\newcommand{\nn}{\nonumber}
\newcommand{\dd}{\displaystyle}
\newcommand{\ct}{c_\theta}
\newcommand{\st}{s_\theta}
\newcommand{\cdt}{c_{2\theta}}
\newcommand{\sdt}{s_{2\theta}}

\def\lq{\left [}
\def\rq{\right ]}
\def\LL{{\cal L}}
\def\VV{{\cal V}}
\def\AA{{\cal A}}
\def\gs{{g''}}
\def\vmu{{\bf V}_\mu}
\def\amu{{\bf A}_\mu}
\def\lmu{{\bf L}_\mu}
\def\rmu{{\bf R}_\mu}
\def\rmus{{\bf R}^\mu}
\def\MM{{\cal M}}
\def\BB{{\cal B}}
\def\de{{\partial}}
\def\dmu{\partial_\mu}
\def\dnu{\partial_\nu}
\def\dmus{\partial^\mu}
\def\dnus{\partial^\nu}
\def\gp{g'}
\def\gpt{{{\tilde g}^\prime}}
\def\gptd{\tilde g^{\prime 2}}
\def\ggs{\frac{g}{\gs}}
\def\eps{{\epsilon}}
\def\tr{{\rm {tr}}}
\def\V{{\bf{V}}}
\def\W{{\bf{W}}}
\def\Wt{\tilde{\bf {W}}}
\def\Y{{\bf{Y}}}
\def\Yt{\tilde{\bf {Y}}}
\def\tW{\tilde W}
\def\tY{\tilde Y} 
\def\tL{\tilde L}
\def\tR{\tilde R}
\def\s{s_\theta}
\def\c{c_\theta}
\def\gt{\tilde g}
\def\et{\tilde e}
\def\At{\tilde A}
\def\Zt{\tilde Z}
\def\Wpt{\tilde W^+}
\def\Wmt{\tilde W^-}

\def\tr{{\rm {tr}}}
\def\V{{\bf{V}}}
\def\W{{\bf{W}}}
\def\Wt{\tilde{\bf {W}}}
\def\Lt{{\bf {L}}}
\def\Rt{{\bf {R}}}
\def\Y{{\bf{Y}}}
\def\Yt{\tilde{\bf {Y}}}
\def\L{{\cal L}}
\def\st{\tilde s_\theta}
\def\ct{\tilde c_\theta}
\def\gt{\tilde g}
\def\et{\tilde e}
\def\A{\bf A}
\def\Z{\bf Z}
\def\Wpt{\tilde W^+}
\def\Wmt{\tilde W^-}
\def \e{{\rm e}}
\def \rs{\sqrt{s}}
\def\csi{\xi}
\def\de{\partial}

\newcommand{\bra}[1]{\left\langle #1 \right|}
\newcommand{\ket}[1]{\left| #1 \right\rangle}
\newcommand{\spur}[1]{\not\! #1 \,}

\goodbreak
\begin{center}
\bigskip\vbox
{
\vskip 3truecm
\noindent
\vskip 2truecm
\noindent
}
\end{center}
\bigbreak
\begin{center}
  \begin{large}
  \begin{bf}
{Tests for a Strong Electroweak Sector at Future
$e^+e^-$ High Energy Colliders} \\ 
\end{bf}
  \vspace{1.2cm}
Daniele Dominici\\ 
{\small Dipartimento di Fisica, Universit\`a di Firenze and Sezione di
Firenze, INFN} \\
{\small  
largo E.Fermi 2, I-50125 Firenze, Italy}
\end{large}

\vspace{1.5cm}

\begin{abstract}
The study of the  scattering at high energy of the gauge bosons
$W$ and $Z$, in  particular 
 longitudinally polarized $W$ and $Z$, 
can clarify the mechanism of spontaneous
symmetry breaking  in  the Standard Model
of the electroweak interactions. 
 Different models of strong electroweak sector,
based on the effective lagrangian approach are briefly reviewed. 
They include models with no resonance, with scalar resonance, 
 additional vector and axial-vector resonances.
 The effective Lagrangians are
derived from the chiral symmetry of the
symmetry breaking sector.
Limits on these models from  existing
measurements, mainly LEP and Tevatron, are considered. We study
also direct and indirect effects of the new interactions
at high energy  future 
$e^+e^-$ linear colliders, through $WW$
scattering and the direct production of these new vector
gauge bosons. 
\end{abstract}
\end{center}
\vspace{2.5cm}
\newpage
\tableofcontents
\newpage

\section{Introduction}
The Standard Model (SM) of the electroweak interactions is in good
agreement
with all current experimental results from LEP, SLC  and
Tevatron. However the mechanism which is responsible
for  gauge boson and fermion masses is still
lacking of experimental confirmation. It is usually assumed
that a scalar field, acquiring vacuum expectation value, induces
a spontaneous breaking of the symmetry, generating the masses.
The Higgs field is the remnant of such  a mechanism.

Direct search of the Higgs at LEPI and LEPII has already
put a lower limit on its mass,  $M_H\geq 77~GeV$ \cite{lepho}.
There are also upper limits from lattice calculations
of the order $M_H\sim 700~GeV$ \cite{lat},
 tree level unitarity \cite{lqt} $M_H\sim 1000~GeV$ 
and Landau pole $M_H\sim 200-1000~GeV$ depending
on the new physics cutoff \cite{land}.

The known mechanism for spontaneous symmetry breaking suffers
of the so-called hierarchy problem.
 Quantum corrections to the Higgs mass are naturally of
the order of the highest scale of the theory, then requiring
a fine tuning of the parameters in absence of some symmetry
protecting the Higgs mass. An appealing  solution to this 
 problem is offered by supersymmetry, because of the
cancellation among scalar and fermion loop-corrections.

An alternative approach is suggested in dynamical symmetry
breaking schemes like technicolor \cite{tech,farhi}; 
the Higgs is substituted by
a fermion condensate and the SM is thought of as an effective 
theory valid up to the new scale. In this approach the hierarchy
problem
is solved by the natural cutoff given by the new theory.
These two different philosophies require  therefore two different
 approaches; while predictions  of supersymmetric theories 
are   mainly based on renormalizable Lagrangians, the predictions
of dynamical breaking theories are based on non perturbative approaches
like effective Lagrangians or strong interaction methods.
Existing measurements already exclude the simply QCD
rescaled version of technicolor and suggest that  only extensions of
the SM which smoothly modify it are still conceivable.

The physics of $e^+e^-$ linear colliders has been the subject of
intensive work during the last years in all its aspects
 \cite{Finland,hawaii,Morioka,DESY,DESYC,DESYD,Howiesbook,LCW}.
 The properties of top quark (mass and
decay), the self couplings of $W$ bosons can be studied with great
accuracy. The Higgs particle and supersymmetric particles can be
explored and therefore we
can gain insight in beyond the SM.

In this report we are going 
 to review   the general phenomenology at future
$e^+e^-$ linear colliders (LC's) of the strong interactions associated to 
the electroweak symmetry breaking problem. These predictions are derived
from models of dynamical breaking which are based just on the symmetry
structure of the electroweak breaking or its extensions but
irrespectively
of the underlying theory which is responsible for the
breaking. Technicolor with a chiral
symmetry flavor $SU(2)\otimes SU(2)$
is the simplest example for such dynamical
breaking.
This is a limited point of view. However it allows to get
predictions which are general enough 
at least in a class of models of dynamical breaking.

Model dependent  predictions can be obtained when one deals  also
with the problem of fermion masses and theories like  extended
technicolor
\cite{ext},
walking technicolor \cite{walk}, topcolor assisted technicolor \cite{tc2}
 or noncommuting extended technicolor \cite{chi}
are considered.

The study of $WW$ scattering and in particular
of longitudinal $W$ can clarify the mechanism of spontaneous
symmetry breaking since we know from the Equivalence Theorem \cite{equ}
that this
process is equivalent at high energy to the corresponding Goldstone
boson scattering. 
We expect an energy threshold at which
   $WW$ scattering will become strong \cite{lqt,velt} and in
principle
resonances ($WW$  or technifermion bound states like
techni-$\rho$ or techni-$\omega$) can appear.
Models for $WW$ interactions have been proposed,
using in particular the chiral lagrangian technique, taking into
account the spontaneous breaking of the global symmetry
of the Higgs sector $SU(2)\otimes SU(2)$ to $SU(2)$.
Effective Lagrangians without new resonances have been built
by considering the SM in the $M_H\to\infty$ limit \cite{appel,longh}
or in the context of technicolor theories based on the same symmetry
breaking scheme or its generalizations like 
$SU(8)\otimes SU(8)$ to $SU(8)$ \cite{psrk}.
 
Models for scalar bound states in the  $W^+W^-$, $ZZ$ channels,
after the pioneering papers \cite{lqt,velt}, were first
considered in \cite{einh,cdg},
studying the $O(4)$ linear scalar model, by replacing it with the
$O(N)$ model, studied in the limit of large $N$ for all values of the
fourlinear coupling $\lambda$. This is particularly convenient, when
considering the large Higgs mass regime of the Higgs sector.

A model describing a new triplet of vector resonances in the
  $W^+W^-$, $ZZ$ and $W^\pm Z$ channels was also proposed \cite{bess}.
It was derived from the non-linear $\sigma$-model, based
on the manifold $SU(2)\otimes SU(2)/SU(2)$ and introducing
the vector resonances by means of the hidden gauge symmetry approach
\cite{bando,bala} which is equivalent to the method used by Weinberg 
to
describe the $\rho$ particle in chiral Lagrangians \cite{weinberg}.
 The model was also
generalized to include axial-vector bound states \cite{assiali}.
  
In  the following we
will review  all these different models based on the same chiral symmetry
but with a different spectrum of resonances.

Two classes of different processes will be considered at future LC's. The
annihilation
processes $e^+e^-\rightarrow f^+f^-$ and $e^+e^-\rightarrow W^+W^-$ 
will be in particular
relevant if a new vector resonance mixed with $Z$ is present.
These channels are already important 
at a LC of a center of mass energy of $500~GeV$.
In principle if LHC has already discovered such a new vector boson one can
tune the LC energy to study its properties and decays. Otherwise the
LC can give bounds on its couplings and masses. The other
process is the fusion, for example 
$e^+e^-\rightarrow \bar \nu\nu W^+W^-$, and
 can be used to study $WW$ scattering
also in absence of new resonances but, as we will see, it requires much
higher
energy (of the order $1.5~TeV$) and luminosity.

Finally let us mention that,  as already noticed,
 the symmetry group can be larger than 
$SU(2)_L\times SU(2)_R$ like in the one family technicolor model
based on the chiral symmetry $SU(8)_L\times SU(8)_R$ \cite{farhi}.
In this review we do not extend the phenomenological analysis to such 
 models. These models have a rich particle spectrum with new
pseudo-Goldstone bosons which in principle could be produced at future
LC's \cite{CRST}. 

We do not consider here the study of $WW$ scattering
at $\gamma\gamma$, $\gamma
e^-$and  $e^-e^-$ options of the machine 
(we refer to \cite{han,jikia} for recent reviews).

The contents of the paper is the following.
We briefly review the features of the electroweak symmetry breaking in
Section \ref{esb}. In Section \ref{nore}
we discuss the chiral lagrangian approach
to the electroweak symmetry breaking, considering  different models,
built in terms of the Goldstone fields, and, using the Equivalence
Theorem,
we study  also the corresponding amplitudes for the scattering of
longitudinal bosons.
In Section \ref{scre} we discuss models containing in addition to the 
Goldstone fields, a scalar bound state which could arise from
the new strong underlying interaction. In Section  \ref{vere} we review an
effective parametrization of the strong electroweak breaking with
a new triplet of vector bosons. The model is extended in 
Section \ref{veaxre} 
to describe also axial-vector bound states.
Limits on these models from LEP, SLC, Tevatron measurements are also
studied. In Section \ref{gene} models with scalar, vector and axial-vector
bound states are presented. We then study the different processes
and observables which can be used at future LC's $e^+e^-$,
starting with the channel $e^+e^-\rightarrow f^+ f^-$ in Section \ref{eech}
and  $e^+e^-\rightarrow W^+W^-$ in Section \ref{wwch}. The limits that can be
obtained
from these processes, by considering the virtual effects of new vector 
resonances are presented in Sections \ref{bess}
 and \ref{debess}. In Section \ref{fsin}
the strong effects from  electroweak symmetry breaking are included
in $e^+e^-\rightarrow W^+W^-$ through final state rescattering and
the corresponding limits on $W_LW_L$ scattering amplitudes are
derived.
In Section \ref{anom} we review the effective parametrization for trilinear
gauge
boson couplings and their bounds  at future LC's. In Section \ref{fusi}
we consider the study of  $WW$ scattering through the fusion
processes 
$e^+e^-
\rightarrow W^+ W^- e^+ e^-$
and 
$ e^+e^-
\rightarrow W^+ W^-{\overline \nu} \nu$. The results
for various models of strong sector are compared in Section \ref{fusi2}
and \ref{chpar}. In Section \ref{lclhc}
 we perform a comparison of reach of LC and LHC for different models
and in  Section \ref{conc} we make some general conclusive comments.
In the Appendix we briefly review the technique of non linear
realization of a symmetry group
$G$ which spontaneously breaks to a subgroup $H$.

\section{Electroweak symmetry breaking}
\label{esb}
Perhaps the most important subject of research which has generated
a lot of theoretical activities  and models beyond the SM is the
 electroweak symmetry breaking. 
 Gauge bosons, quarks and leptons get their masses
from this breaking.
 One of the basic feature
of the standard Higgs sector  is the global symmetry
of the Higgs potential,  which is invariant, neglecting
the gauge interactions, under the
following transformation 
\be
M=(v+H)\exp(i\pi^a\tau^a/v) \to g_L M g_R^\dagger
\ee
with $g_{L,R}\in SU(2)_{L,R}$,  $v=246~GeV$ and $\tau^a ~(a=1,2,3)$
the Pauli matrices.
 The vacuum expectation value
$<M>=v$ breaks the symmetry $SU(2)_L\otimes SU(2)_R$ to $SU(2)_{L+R}$.
A second feature of the Higgs Lagrangian is, in the standard approach 
but also in the supersymmetric one, the renormalizability 
of the theory in the conventional sense.

Most of the non supersymmetric extensions of the
SM assume that this breaking  $SU(2)_L\otimes SU(2)_R$ to $SU(2)_{L+R}$
is generated by a new confining strong interaction
(technicolor, metacolor...) and the usual
Higgs field is replaced by a bound state fermion-antifermion.
In theories like technicolor one has a $SU(2)_L\otimes SU(2)_R$ 
chiral symmetry acting
on the technifermion fields
and a technifermion condensate induces a
breaking
to $SU(2)_{L+R}$. This dynamical breaking produces three Goldstones $\pi^a$
such that
\be
<0\vert J_A^{a\mu}\vert \pi^b >=i  \delta^{ab}v p^\mu
\ee
where $J_A^{a\mu}$ denote the axial currents.
By computing the lowest order vacuum polarization for
 the gauge bosons 
one gets the usual mass matrix  in the basis $(W_1,W_2,W_3,Y)$
\be
M^2=\f 1 4 v^2\pmatrix {g^2& & & \cr
 &g^2& & \cr
 & & g^2& - g\gp\cr
& & -g\gp &\gp^2}
\ee
 This solves only the problem
of gauge boson masses and some new mechanism is necessary for fermion
masses. 

We do not discuss here the problem of fermion masses, and therefore we will
consider an effective field theory parametrization of the symmetry
breaking
 $SU(2)_L\otimes SU(2)_R\to SU(2)_{L+R}$.
In this case the effective field theory approach is the most
convenient for working out
the consequences of this breaking without knowing 
the new underlying interaction; 
the Higgs sector is now a theory valid up to the scale
characterized by the new interaction and renormalizability
is no longer a requirement for the Lagrangian. 
Effective Lagrangians can be derived using the global symmetry 
$G=SU(2)_L\times SU(2)_R$ and an expansion in the energy 
\cite{wphy,georbook,gasleu}.
Goldstones are described by a unitary field $U(x)=\exp(i\pi^a(x)\tau^a/v)$,
 transforming  under $G$
as $(2,2)$ or $U\to g_L Ug_R^\dagger$. 
The most general chiral Lagrangian is a sum of an infinite number of
terms
with an increasing number of derivatives or equivalently in terms
of increasing powers of
energy or momentum. 
These Lagrangians allow for a description of the physics below a
cutoff scale $\Lambda$, related to
the new underlying interaction. Effective field theories are not 
renormalizable
in the usual sense (finite number of primitive
divergences).  However at a finite order
in the momentum expansion only a finite number of parameters is necessary
to get a finite result; therefore at low energy only the first terms
in the expansion are retained and one has definite predictions
in terms of few parameters (for a recent proof of renormalizability
in ``modern'' sense of effective field theory see \cite{quim}).

Applications to the electroweak sector have been 
considered by different authors \cite{appel,longh,psrk,chla}.

\section{No resonance models}
\label{nore}
In this Section we review the chiral lagrangian approach which allows
to study
the symmetry breaking sector of the SM. In fact, by making use
of the Equivalence Theorem \cite{equ}, one can work taking
as degrees of freedom  the
Goldstone bosons, equivalent to the longitudinal components
 of $W$ and $Z$.

Experiments, like the measurement of
the $\rho$ parameter at LEPI,
 suggest the existence of a custodial symmetry $SU(2)$ \cite{cust}, which
guarantees, at tree level,
\be
\rho=\frac {M_W^2}{M_Z^2\cos\theta_W^2}=1
\ee

Therefore chiral Lagrangians can be built  
as   non linear $\sigma$-models assuming the spontaneous breaking 
  $SU(2)_L\otimes SU(2)_R \to
SU(2)_{L+R}$. A brief review of the non linear 
group realizations of a group $G$ which breaks spontaneously to a
subgroup
$H$  \cite{ccwz} can be found in Section \ref{appA}.

At the lowest order, the Lagrangian is given by
\be
\LL=\frac {v^2}{4} Tr \de_\mu U^\dagger \de^\mu U
=\frac 1 2 \de_\mu \pi^a \de^\mu \pi^a+
\frac {1}{6 v^2} \left [(\pi^a\de_\mu \pi^a)^2-\pi^a \pi^a
(\de_\mu \pi^a)^2\right]+\cdots
\label{2}
\ee
Here and in the following we make
the identification $v^2=1/(\sqrt 2 G_F)$, $G_F$ being the Fermi constant.

This already allows, using the Equivalence Theorem, to calculate the 
 scattering
amplitudes for $W^\pm,Z$ gauge bosons 
 at high energy, $\sqrt{s}>>M_W,M_Z$, by considering the scattering of the 
corresponding Goldstone bosons $\pi^\pm ,\pi^3$.

We have (symmetry factors for identical particles
 are not included in the amplitudes)
\bea
M(W^+_LW^-_L\rightarrow Z_LZ_L)&=&A(s,t,u)\nn\\
M(W^+_LW^-_L\rightarrow W^+_LW^-_L )&=&A(s,t,u)+A(t,s,u)\nn\\
M(Z_LZ_L\rightarrow Z_LZ_L)&=&A(s,t,u)+A(t,s,u)+A(u,t,s)
\label{amp}
\eea
with 
\be
A(s,t,u)=s/v^2
\ee
where $s=(p_1+p_2)^2$, $t=(p_1-p_3)^2=-s (1-\cos\theta)/2$
 and $u=(p_1-p_4)^2=-s (1+\cos\theta)/2$ are the
Mandelstam variables.
Using the crossing property we have also
\bea
M(W^\pm_LZ_L\rightarrow W^\pm_LZ_L)&=&A(t,s,u)\nn\\
M(W^\pm_LW^\pm_L\rightarrow W^\pm_LW^\pm_L)&=&A(t,s,u)+A(u,t,s)
\label{amp2}
\eea
These amplitudes violate unitarity. In fact, if we define the generic
 partial wave $a_l^I$,
\be
a_l^I=\frac 1 {64\pi}\int_{-1}^1P_l(\cos\theta) T^Id\cos\theta
\ee
where $T^I$ is the generic isospin amplitude and
the Legendre polynomials are normalized as
\be
\int_{-1}^1 P_l(\cos\theta)P_m(\cos\theta)d\cos\theta
= \frac 1{m+1/2}\delta_{lm}
\ee
we find that partial wave
unitarity is violated at $\sqrt{s}=1.7~TeV$ by  requiring
$\vert a_0^0\vert\leq 1$ (at $\sqrt{s}=1.2~TeV$ by requiring
$\vert a_0^0\vert\leq 1/2$) \cite{lqt}.  

The isospin amplitudes are given by
\bea
T^0&=&3A(s,t,u)+A(t,s,u)+A(u,t,s)\nn\\
T^1&=&A(t,s,u)-A(u,t,s)\nn\\
T^2&=&A(t,s,u)+A(u,t,s)
\eea

The most general
$SU(2)_L\times SU(2)_R$ invariant Lagrangian with custodial
$ SU(2)$, up to the order $p^4$, contains
two additional invariant terms and correspondingly two more parameters
$\alpha_4$ and $\alpha_5$:
\begin{eqnarray}
\LL^\prime  & = &  {v^2 \over 4}
\tr \partial_\mu U^\dagger
\partial^\mu U \nonumber \\
&  +& \alpha_4 \tr(\partial_\mu U^\dagger
\partial_\nu U) \tr(\partial^\mu U^\dagger \partial^\nu U)\nonumber \\
& +&\alpha_5
 \tr(\partial_\mu U^\dagger \partial^\mu U)
  \tr(\partial_\nu U^\dagger \partial^\nu U) 
\label{a4a5}
\end{eqnarray}
The parameters $\alpha_4$ and $\alpha_5$ are related to the analogous
parameters $L_1$ and $L_2$ (in the notation in \cite{gasleu}) by
\be
\alpha_4=\f {L_2}{16\pi^2}~~\alpha_5=\f {L_1}{16\pi^2}
\label{L1L2}
\ee
This allows to compute the amplitudes at tree level at the order $p^4$
\be
A(s,t,u)=\frac {s}{v^2}+ \alpha_4 \frac {4(t^2+u^2)}{v^4}
+\alpha_5 \frac {8 s^2} {v^4}
\ee

However terms of order $p^4$ are also generated  when
one uses the Lagrangian  at one-loop level.

The corresponding result is a sum of the tree and one-loop amplitudes  
\begin{eqnarray}
A(s,t,u) \ & = &  {s \over v^2} +
 {1 \over  v^4} \biggl( 4\alpha_4 (\mu) (t^2+u^2) 
 +8 \alpha_5(\mu) s^2
\biggr)  \nonumber \\
&\ +&\ {1 \over 16 \pi^2 v^4}\ \biggl[-{t \over 6}\,(s+2t){\log}
\biggl(-{t \over \mu^2}\biggr)\ -\ {u \over 6}\,(s+2u){\log}
\biggl(-{u \over \mu^2}\biggr)\  \nonumber \\
&\ -&\ {s^2 \over 2}\, {\log}\biggl(-
{s \over \mu^2} \biggr)\biggr]\ 
\end{eqnarray}
where
$\alpha_4(\mu)$ and $\alpha_5(\mu)$ are the renormalized coefficients.

All these amplitudes violate partial wave unitarity.
 One can however unitarize them
 by using for instance the $K$ matrix prescription \cite{gupta}
(see \cite{gaillard} for applications to $WW$ scattering)
\be
a_l^K= \frac {a_l}{1-ia_l}
\ee
or a different method like Pad\'e approximant \cite{basdevant}
(see \cite{dht} for applications to $WW$ scattering)
 or $N/D$ technique. In the Pad\'e technique the unitarized
partial wave is built using the tree level amplitude $a^{(0)}$
and the
one-loop one $a^{(1)}$, as
\be
a_l^P= \frac {a_l^{(0)}}{1-a_l^{(1)}/a_l^{(0)}}
\ee

Both the  Pad\'e method and  $N/D$  can generate a resonant behaviour.
Therefore the choice of the unitarization procedure introduces
 some arbitrariness
in the predictions in terms of  unitarized chiral amplitudes.

For a review of different techniques of unitarization see
\cite{hikasa}.

If we allow violations of the custodial symmetry, then the Lagrangian
at the lowest order becomes 
\be
\LL =\frac {v^2}{4} Tr \de_\mu U^\dagger \de^\mu U+
\frac {v^2}{2}\frac {1-\rho}{\rho}
 [Tr \frac {\tau_3}{2} U^\dagger \de^\mu U]^2
\ee
and contains $\rho$ as a new parameter.

When the gauging of $SU(2)_L\otimes U(1)_Y$
is considered, one has a  correction to the
$\rho$ parameter; the existing  measurements of
this parameter give $1- \rho =-\Delta\rho=-(
4.1\pm 1.2)\times 10^{-3}$
\cite{dati}.

The symmetry group is
now $SU(2)\otimes U(1)_Y$, breaking to $U(1)_{em}$. From this
Lagrangian one  gets  
 the following scattering amplitudes at the lowest order
\cite{cgg}
\bea
M(W^+_LW^-_L\rightarrow Z_LZ_L)&=&(4-\frac 3\rho)  \frac {s}{v^2}\nn\\
M(W^+_LW^-_L\rightarrow W^+_LW^-_L )&=&\frac 1 {\rho}
\frac u{v^2}\nn\\
M(Z_LZ_L\rightarrow Z_LZ_L)&=&0
\eea
The remaining amplitudes can be obtained via crossing.

If one is interested in the physics of the symmetry breaking sector
at energies  of the order of the mass of the gauge vector
bosons  then the complete
gauge chiral Lagrangians have to be used, since
  the Equivalence Theorem is not applicable. The Lagrangian is now built
using the Goldstone bosons and the $SU(2)_L\otimes U(1)_Y$ gauge
vector
bosons.

The corresponding  Lagrangian is equivalent to  a nonlinear gauged 
$\sigma$-model, obtained simply substituting the usual derivatives with 
 corresponding covariant ones \cite{appel,longh}
\be
D_\mu U=(\de_\mu+\f{i}{2}g { W}_\mu ^a\tau^a) U-\f{i}{2}g^\prime U
Y_\mu\tau^3
\ee
and adding all the terms consistent with the symmetry at a given order
in  the energy expansion. The model so obtained is also equivalent 
at the tree and one-loop level to the SM in the limit of
infinite $M_H$. 

Recent review of chiral Lagrangians can be found in \cite{feruglio,wudka}.

The chiral Lagrangian up to order $p^2$, invariant under $SU(2)_L\otimes
U(1)$, is given by (we follow here \cite{appwu})
\be
\LL=\f {v^2}{4} Tr(D_\mu U)^\dagger (D_\mu U) +\f 1 4 \beta_1 
g^2v^2 [Tr(TV_\mu)]^2- \f 1 4 Y_{\mu\nu} Y^{\mu\nu}
-\f 1 2 Tr W_{\mu\nu} W^{\mu\nu}
\label{lp2}
\ee
in terms of $SU(2)_L$ covariant and $U(1)_Y$ invariants
\be 
T=U\tau^3 U^\dagger~~~V_\mu =(D_\mu U) U^\dagger
\ee
and
\bea
Y_{\mu\nu}&=&\de_\mu Y_\nu -\de_\nu Y_\mu \nn\\
W_{\mu\nu}&=& \de_\mu W_\nu -\de_\nu W_\mu + i g[W_\mu, W_\nu]
\eea
The second term of eq. (\ref{lp2})
violates the custodial $SU(2)$, and  $\beta_1$ is 
related to  $\Delta\rho$: $\beta_1 g^2=\Delta\rho /2$.

Up to $p^4$ order the Lagrangian is  given by eq. (\ref{lp2})
and by the following expression
\bea
\LL_4&=&\f 1 2 \alpha_1 g \gp Y_{\mu\nu} Tr (T  W^{\mu\nu})+
\f 1 2 i\alpha_2 \gp  Y_{\mu\nu} Tr( T [V^\mu,V^\nu])\nn\\
&+&i\alpha_3 g Tr( W_{\mu\nu} [V^\mu,V^\nu])
+\alpha_4 [Tr (V_\mu V_\nu)]^2
+\alpha_5  [Tr (V_\mu V^\mu)]^2\nn\\
&+& \alpha_6 Tr (V_\mu V_\nu) Tr(TV^\mu) Tr(TV^\nu)
+\alpha_7 Tr (V_\mu V^\mu) Tr(TV_\nu) Tr(TV^\nu)\nn\\
&+&\f 1 4 \alpha_8 g^2  [Tr (T  W^{\mu\nu})]^2
+\f  1 2 i \alpha_9 g Tr (T  W_{\mu\nu})  Tr( T [V^\mu,V^\nu])\nn\\
&+&\f 1 2 \alpha_{10} [ Tr(TV_\mu)  Tr(TV_\nu) ]^2
+ \alpha_{11} g\eps^{\mu\nu\rho\sigma}  Tr(TV_\mu)
Tr(V_\nu W_{\rho\sigma})
\label{lp4}
\eea
where $\alpha_1,\cdots,\alpha_{11}$ are arbitrary parameters.
All these terms are CP invariant. There are also eight CP non
invariant
terms \cite{appwu}. 

There are already bounds on the lagrangian parameters coming from LEP
results. These can be obtained
 using the $\eps$ parameters \cite{altarelli,fri},
 which can be 
 expressed as
\bea
\eps_1 &=& \Delta\rho\nn\\
\eps_2 &=& \c^2\Delta\rho+\f{\s^2}{c_{2\theta}}\Delta r_W-2 \s^2\Delta k\nn\\
\eps_3 &=& \c^2\Delta\rho+c_{2\theta}\Delta k
\eea
where
\be
\s^2 =1-\c^2 = \f 1 2 \left [1 -\left (1- \f {4\pi \alpha_{em}}
{\sqrt{2} G_F M_Z^2}\right )^{1/2}\right ]
\label{sint}
\ee
with $\s \equiv \sin\theta$, $\c\equiv\cos\theta$
and $\alpha_{em}$ the electromagnetic fine structure constant 
at zero momentum transfer.

Now $\Delta r_W$ can be obtained by $M_W/M_Z$ which is
measured at Tevatron.
$\Delta k$ and $\Delta \rho$
 can be obtained by the   LEPI, SLC  measurements of 
 asymmetries  at the $Z$
peak, and the $Z$  widths. 

 For convenience we give also the relations
among the $\epsilon$ and  the  $S,~T,~U$ parameters of \cite{S}:
\bea
\eps_1 &=& \alpha_{em} T\nn\\
\eps_2 &=& -\frac{\alpha_{em}}{4\s^2}U \nn\\
\eps_3 &=&\frac{\alpha_{em}}{4\s^2}S
\eea
Using the chiral Lagrangian, the sum of eqs. (\ref{lp2})-(\ref{lp4}),
we get
\bea
\eps_1 &=&2g^2\beta_1\nn\\
\eps_2 &=&g^2\alpha_8 \nn\\
\eps_3 &=&-g^2\alpha_1
\eea
 Considering the experimental values \cite{dati}
\be
\eps_1=(4.1\pm 1.2)\times 10^{-3}~~
\eps_2=(-9.3 \pm 2.2)\times 10^{-3}~~
\eps_3=(4.1\pm 1.4)\times 10^{-3}
\label{epsv}
\ee
 one can derive limits on $\beta_1$, $\alpha_1$ and $\alpha_8$.
A recent evaluation of the $S,T,U$ parameters, including loop
effects can be found in \cite{alam}.

The Lagrangian given by eq. (\ref{lp2})
and eq. (\ref{lp4}) contains also triple
gauge
vertices among $W$, $Z$ and $\gamma$.
The study of $W$ pair production at LEPII, LHC and LC can allow 
a determination of these parameters.

\section{Scalar resonance models}
\label{scre}

The effect of new resonances can be introduced in
the scattering amplitudes using Pad\'e approximant \cite{dht} or $N/D$
technique \cite{lqt}. Here we choose to present models based on the effective
lagrangian approach. 

\subsection{The chirally coupled model}
\label{ccm}

The effective Lagrangian describing   a scalar resonance can be built by 
considering in addition to the field $U$, a scalar field $S$,
 which is a singlet
under $SU(2)_L\otimes SU(2)_R$. Then the effective Lagrangian at the 
lowest order is given by \cite{bargerlhc}
\bea
\cal L&=&\frac {v^2}{4} Tr \de_\mu U^\dagger \de^\mu U\nn\\
&&+\frac 1 2 \de^\mu S \de_\mu S - \frac 1 2 M_S^2 S^2\nn\\
&&+\frac 1 2 kv S Tr \de_\mu U^\dagger \de^\mu U+\cdots
\eea
where $M_S$ is the mass of the scalar resonance and
\be
\Gamma_S=\frac {3k^2M_S^3}{32\pi v^2}
\ee
is the width.
For $k=1$ one has the usual Higgs-Goldstone couplings.
Scattering amplitudes contain now the contribution of the isoscalar
resonance
\be
A(s,t,u)= \frac s{v^2}- \left (\frac {k^2s^2}{v^2}\right )
\frac 1{s-M_S^2+i M_S \Gamma_S}
\ee
In the phenomenological analysis \cite{bargerepem,bargermu}
$M_S=1~TeV$ and $\Gamma_S=0.35~TeV$ have been chosen. 

\subsection{$O(N)$ model}
\label{onmo}

 One starts with the linear Lagrangian of the symmetry breaking sector
 of the SM which is known to have the $SU(2)\otimes SU(2)$
 symmetry.
Since $SU(2)\otimes SU(2)\sim O(4)$  one can study an effective 
scalar Lagrangian based
on the symmetry group $O(N)$ in the limit of large $N$ and then consider
$N=4$ \cite{einh,cdg}. The model has the property to be solved for all
the values of the four-linear coupling $\lambda = M_H^2/(2 v^2)$, 
to leading order in $1/N$. Proceeding in this way one can take
into account the possibility of having, in addition to the Higgs,
 a bound state  from the strong interacting system.
An effective action for its interactions with Goldstone bosons
 and Higgs 
can be derived \cite{cdg}.
Scalar propagation is described by a two by two
 matrix which is degenerate. One has only a pole given by
 \be
 M_\sigma^2[1+\Delta(M_\sigma^2)]-M_H^2=0
 \ee
where
\be\Delta(s)=\frac 1{8\pi^2}\frac {M_H^2}{v^2}[1-\log (-\frac {s}{\mu^2})]
\ee
with $\mu$ a renormalization scale of the effective Lagrangian. 
The location of the pole in the complex plane $M_\sigma$ evolves
from
\be
M_\sigma^2=M_H^2[1-i \f {M_H^2}{8\pi v^2}]+{\cal O}(\f {M_H^4}{v^4})
\ee
for small $M_H/v$ to
\be
M_\sigma^2=\f {16\pi v^2} {3}[-i + \f {16\pi } {3}
\f {v^2}{M_H^2}(1- i\f 2{3\pi})+ 
{\cal O}(\f {v^4} {M_H^4})]
\ee
for large $M_H/v$. One of the  effects
 of the strong coupling is to increase the width of the Higgs 
with
respect to the width of the Higgs as calculated at the lowest order
in the SM.
The scattering amplitude derived from this Lagrangian is
\be
A(s,t,u)=-\sqrt{2}G_F M_H^2 s
[s(1+\Delta(s))-M_H^2]^{-1}
\ee
In the limit of large $M_H$ one obtains
\be
A(s,t,u)=\frac {8\pi^2 s}{8\pi^2v^2-s[1-\log(-s/\mu^2)]}
\ee
This amplitude is consistent with unitarity up to energies of the order
$\mu$.

\section{A vector  resonance model}
\label{vere}

In this Section we review a model based on a general procedure
to introduce vector resonances in chiral Lagrangians. The model contains a new 
triplet of vector bosons and  allows to derive their mixings to ordinary gauge
bosons $W$ and $Z$ and their couplings to fermions.
The model can describe as a special case a techni-$\rho$ resonance.

As we have already seen an effective description 
of the symmetry breaking mechanism in electroweak
theories can be done in 
terms of a non linear $\sigma$-model formulated on the
quotient space of the breaking of $SU(2)_L\otimes SU(2)_R \to SU(2)_{L+R}$.
This is the case when considering the limit of a 
strong interacting Higgs sector
($M_H\to\infty$).
Vector resonances can be introduced in chiral Lagrangians  following
Weinberg \cite{weinberg} or, in equivalent way, by means of the hidden
local
 symmetry
approach \cite{bando} (see also \cite{bala}). 
The model so obtained, when this technique is applied to the 
electroweak symmetry breaking sector, is called BESS (Breaking the
Electroweak Symmetry Strongly) \cite{bess}.

It is known that any nonlinear $\sigma$-model corresponding to the
coset space  $G/H$ is gauge equivalent to a linear model based
on the symmetry $G_{global}\otimes H_{local}$.
 For the construction of the BESS model Lagrangian $G_{global}=G=
SU(2)_L\otimes SU(2)_R$ and $ H_{local}=H=SU(2)_V$.
One introduces the group variables $g\in G$ with $G=SU(2)_L\otimes SU(2)_R$
\be
g=(L,R)
\ee
with $L\in SU(2)_L$ and $R\in SU(2)_R$, which transform under the
$G\otimes H$ group, 
 as follows: $L\to g_L L h$, $R\to g_R R h$,
with $g_{L,R}\in SU(2)_{L,R}$ and $h\in SU(2)_{V}$.

One further introduces the Maurer-Cartan field
\be
\omega_\mu= g^\dagger\dmu g=(L^\dagger\dmu L,R^\dagger\dmu R)
\ee
which can be decomposed into $\omega_\mu^\parallel$ lying in the Lie algebra
of $H$, and into the orthogonal complement $\omega_\mu^\perp$ 
\bea
\omega^{\parallel}_\mu &=& \f{1}{2} (L^\dagger\dmu L+R^\dagger\dmu R)\nn\\
\omega^{\perp}_\mu &=& \f{1}{2} (L^\dagger\dmu L-R^\dagger\dmu R)
\eea
Both $\omega_\mu^\parallel$ and $\omega_\mu^\perp$ are singlets of $G$
and transform under $H$ as
\bea
\omega^{\parallel}_\mu &\to & h^\dagger\omega^{\parallel}_\mu h+
h^\dagger\dmu h\nn\\
\omega^{\perp}_\mu &\to & h^\dagger\omega^{\perp}_\mu h
\eea
The non linear $\sigma$-model Lagrangian (\ref{2})
describing the electroweak symmetry
breaking sector can be easily reconstructed in terms 
of $\omega^{\perp}_\mu$
\be
{\cal L} =
-v^2\tr (\omega^{\perp}_\mu\omega^{\perp\mu})=\f{v^2}{4}\tr (\dmu
U^\dagger\dmus U)
\ee
where $U=LR^\dagger$ is a singlet under $H$.

Introducing a triplet of gauge bosons $\V_\mu$ 
transforming as 
\be
\V_\mu\rightarrow h^\dagger \V_\mu h +  h^\dagger \partial_\mu h
\ee 
under the local group $SU(2)_V$,
one can show that the most general Lagrangian, symmetric under
$G\otimes H$ and under the parity transformation $L\leftrightarrow R$,
containing at most two derivatives, 
can be constructed as an arbitrary combination of two invariant terms.
Furthermore, assuming that the gauge bosons of the hidden symmetry
become dynamical \cite{dadda,dyn}, we get

\bea
{\cal L} &=&
-v^2\Big[\tr (\omega^{\perp}_\mu )^2
 +\alpha~ \tr (\omega^{\parallel}_\mu-\V_\mu)^2 \Big]\nn\\
& &+\f{2}{\gs^2}\tr[F^{\mu\nu}(\V)F_{\mu\nu}(\V)]\eea
with $\alpha$ an arbitrary parameter,
\be
F_{\mu\nu}(\V)  = \dmu \V_\nu-\dnu \V_\mu+[\V_\mu,\V_\nu]
\label{tensV}
\ee
and $\V_\mu=\f{\dd i}{\dd 2}\f{\dd \gs}{\dd2} V_\mu ^a\tau^a$, 
with $\gs$ the new gauge
coupling constant.

The Lagrangian 
can also be written as 
\bea
{\cal L}&=&-{{v^2}\over 4} \Big [ Tr (L^\dagger D_\mu L -
R^\dagger D_\mu R)^2+ \alpha Tr (L^\dagger D_\mu L +
R^\dagger D_\mu R)^2\Big ]\nn\\
&&+\f{2}{\gs^2}\tr[F^{\mu\nu}(\V)F_{\mu\nu}(\V)]
\label{leff}
\eea
where we use the covariant derivatives
$D_\mu L=\de_\mu L -L\V_\mu$, $D_\mu R=\de_\mu R -R\V_\mu$.

Going to the unitary gauge 
\be
L=R^\dagger=\exp(\frac i{2v} {\pi^a\tau^a})
\ee
one can derive  an effective Lagrangian describing  Goldstones
and a new triplet of gauge vector bosons
whose mass is given by
\be 
M^2_V=\alpha \frac {v^2} 4  \gs^2
\label{mv2}
\ee
The scattering amplitudes of eq. (\ref{amp})
and eq. (\ref{amp2}),
using the Equivalence Theorem,  can be expressed  in terms of 
\be
A(s,t,u)= \frac s {4v^2}(4-3\alpha)+\frac {\alpha M_V^2}{ 4v^2}
\left [\frac {u-s}{t-M_V^2+i M_V\Gamma_V}+
\frac {t-s}{u-M_V^2+i M_V\Gamma_V}\right ]
\ee
where $\Gamma_V$ is the width of $V$ \cite{pierre43}
\be
\Gamma (V\to\pi\pi)=\f{\sqrt{2}G_F}{192\pi} \frac{M_V^5}{M_W^2}
\left(\f g \gs\right)^2
\label{30}
\ee

The unitarity limit of this model, from the $J=0$ partial wave,
turns out to be
\be
E\leq \frac {1.7~TeV}{\vert 1-\frac 3 4 \alpha\vert ^{1/2}}
\ee
A more stringent bound can be obtained from the $J=0$, isospin $2$
partial wave \cite{61}.

Notice that the 
 parameter $\alpha$ can be also rewritten in terms of the width as
\be
\alpha= \frac {192\pi v^2\Gamma_V}{M_V^3}
\ee

One  can also fully develop the model by including
the $W$, $Z$ and $\gamma$ gauge bosons. This is simply obtained, by
the gauging of the standard $SU(2)_L\otimes U(1)_Y$ group, 
 substituting in (\ref{leff}) the ordinary derivatives with covariant left and
right derivatives acting on the left and right group elements respectively
\bea
\dmu L &\to & (\dmu +\Wt_\mu) L\nn\\
\dmu R &\to & (\dmu +\Yt_\mu) R
\eea
where $\Wt_\mu=\f{i}{2}g {\tilde W}_\mu ^a\tau^a$ and 
$\Yt_\mu=\f{i}{2}g^\prime {\tilde Y}_\mu\tau^3$, and by 
adding the standard kinetic terms for $\Wt$ and $\Yt$ 
\be
{\cal L}^{kin}(\Wt,\Yt) =
\f{1}{2 g^2}\tr[F^{\mu\nu}(\Wt)F_{\mu\nu}(\Wt)]
   +\f{1}{2 { g}^{\prime 2}}\tr[F^{\mu\nu}(\Yt)F_{\mu\nu}(\Yt)]
\ee
with 
\bea
F_{\mu\nu}(\Wt)  &=& \dmu \Wt_\nu-\dnu \Wt_\mu+[\Wt_\mu,\Wt_\nu]\nn\\
F_{\mu\nu}(\Yt)  &=& \dmu \Yt_\nu-\dnu \Yt_\mu
\label{tensWY}
\eea


Due  to the gauge invariance of $\L$ we can choose the gauge with 
$L=R=I$ \cite{bess} and we get

\be
\L=-\f{v^2}{4}\Big[\tr(\Wt_\mu-\Yt_\mu)^2
+\alpha \tr(\Wt_\mu+\Yt_\mu-2\V_\mu)^2\Big]
+\L^{kin}(\Wt,\Yt,\V)
\label{bessp}
\ee
We have used tilded quantities to recall that, due to the effects 
of the $\V$ particles, they are not the physical  fields.

From eq. (\ref{bessp})
 one can easily derive the mass eigenstates and the mixing 
angles among the standard gauge bosons and the new resonances \cite{bess}. 
Furthermore, since in the limit $\gs\to\infty$, the Lagrangian $\L$
reproduces the SM gauge boson mass terms, 
corrections to the SM relations come in powers
of $1/\gs$.

Finally let us consider the fermions of the SM and denote them by $\psi_L$
and $\psi_R$ with
\be
\psi =\left(
\begin{array}{c}
\psi_u\\
\psi_d\end{array}
\right),\cdots
\ee
 They couple to $\V$ via the mixing with the standard $\Wt$
and $\Yt$.
In addition,  direct couplings to the new vector bosons
 are allowed by the symmetries of $\L$ \cite{bess}. 
In fact, we can define Fermi fields transforming as doublets under the
local group $SU(2)_V$ and singlets under the global one: $\chi_L=L^\dagger
\psi_L$;  an invariant term acting on $\chi_L$ by
the covariant derivative with respect to $SU(2)_V$ can be written. 
In the $L=R=I$ gauge  we get
\bea
\L_{fermion} &=& \overline{\tilde \psi}_L i \gamma^\mu\Big(\dmu+\Wt_\mu+
                      \f{i}{2}g^\prime(B-L){\tilde Y}_\mu\Big)
                      {\tilde \psi}_L\nn\\
     &+&\overline{\tilde \psi}_R i \gamma^\mu\Big(\dmu+\Yt_\mu+\f{i}{2}g^\prime
                      (B-L){\tilde Y}_\mu\Big){\tilde \psi}_R\nn\\
     &+& b \overline{\tilde \psi}_L i \gamma^\mu\Big(\dmu+\V_\mu+
                      \f{i}{2}g^\prime (B-L){\tilde Y}_\mu\Big){\tilde \psi}_L
                      \label{lfer}
\eea
where $B(L)$ is the baryon (lepton) number and
$b$ is a new parameter. Notice that
due to the introduction of the direct coupling of the $\V$ to the 
fermions, we have to rescale ${\tilde \psi}_L=
 (1+b)^{-1/2}\psi_L$ in order to get
a canonical kinetic term for the fermions \cite{bess}.

\subsection{Masses and fermion couplings of the gauge bosons}

In the charged sector the mass eigenvalues for
the gauge bosons are, in the limit of large 
$\gs$,
\be
M_W^2= \frac {v^2} 4  g^2 (1-\frac {g^2}{\gs^2})
~~~~~~~~M_{V^\pm}^2=\alpha \frac {v^2} 4  \gs^2
\ee

In the neutral sector we have
\be
M_Z^2=   \frac {v^2} 4  (g^2+\gp^2)(1-\frac
{(g^2-\gp^2)^2}{ (g^2+\gp^2)\gs^2})
~~~~~~~~M_{V}^2=\alpha \frac {v^2} 4  \gs^2
\ee

 We  also recall the couplings to the fermions which arise from
 (\ref{lfer}). The charged couplings are given by
\be
-2 e (a_W W_\mu^-+a_V V_\mu^-)J^{\mu(+)}_L+h.c.
\ee
 with
 \bea
 a_W&=&\frac 1{2\sqrt{2}\s}\frac 1 {1+b}
 \left (\frac {\cos\phi}{\cos\psi}-\frac b 2 \frac {\sin\phi}
 {\cos\psi} \frac \gs g\right )\nn\\
 a_V&=&\frac 1{2\sqrt{2}\s}\frac 1 {1+b}
 \left (\frac {\sin\phi}{\cos\psi}+\frac b 2 \frac {\cos\phi}
 {\cos\psi} \frac \gs g\right )\label{cchbess}
\eea
with $\s=\gp/\sqrt{g^2+\gp^2}$.
The mixing angles in the $M_V>>M_W$ and  large $\gs$ limit are
\be
\phi=-{g\over\gs}
\label{lphi}
\ee

\be
\psi=2\s{g\over\gs}
\label{lpsi}
\ee

In the neutral sector the couplings of the fermions to the gauge bosons 
$Z$ and $V$ are 
\be
e \left (v^f_Z+\gamma_5 a^f_Z\right )\gamma_\mu Z^\mu +
e \left (v^f_V+\gamma_5 a^f_V\right )\gamma_\mu V^\mu
\ee
where $v^f_{Z,V}$ and $a^f_{Z,V}$ are the vector and axial-vector
couplings given by
\bea
v^f_Z&=&{1\over{\sdt}} \left( A T_3^L+2BQ_{em}\right)\nn\\
a^f_Z&=&{1\over{\sdt}} AT_3^L\nn\\
v^f_V&=&{1\over{\sdt}}\left( CT_3^L+2DQ_{em}\right)\nn\\
a^f_V&=&{1\over{\sdt}}CT_3^L
\label{cneubess}
\eea
where
\bea
A&=&{{\cos \csi}\over{\cos\psi}}(1+b)^{-1}\Big[1+b\s^2\big(1-
   {{\tan\csi}\over{\tan\theta\sin\psi}}\big)\Big]\nn\\
B&=&-\s^2{{\cos \csi}\over{\cos\psi}}\Big(1-{{\tan\csi\sin\psi}\over
    {\tan\theta}}\Big)\nn\\
C&=&{{\sin \csi}\over{\cos\psi}}(1+b)^{-1}\Big[1+b\s^2\big(1+
   {{\cot\csi}\over{\tan\theta\sin\psi}}\big)\Big]\nn\\
D&=&-\s^2{{\sin \csi}\over{\cos\psi}}\Big(1+{{\cot\csi\sin\psi}\over
    {\tan\theta}}\Big)
\eea
with $e=g\s\cos\psi$, $T_3^L=\pm1/2$ and
\be
\csi=-{\cdt\over \c} {g\over\gs}
\label{lcsi}
\ee

Let us now evaluate the partial widths of the $V$ bosons   
 from  decays into 
fermion-antifermion and $WW$.
The lepton width of the charged boson is given by
\be
\Gamma(V^-\to l\nu)=\f 2 3 \alpha M_V (a_V^l)^2=
\Gamma_V^0
\ee
while the hadron width is
\be
\Gamma(V^-\to q'\bar q)=3\vert V_{qq'}\vert^2 \Gamma_V^0
\ee
where $\vert V_{qq'}\vert$ are the relevant Kobayashi-Maskawa matrix 
elements.

The $WZ$ width is 
\bea
\Gamma(V^-\to W^-Z)&=&{M_V\over 48}\alpha_{em} g^2_{VWZ}
\left[\left(1- \f {M_Z^2-M_W^2}{M_V^2}\right)^2
-4\f{M^2_W}{M^2_V}\right]^{3/2}\left(\f{M_V^4}{M_W^2 M^2_Z}
\right)\nn\\
&\times & \left[ 1 + 10 \left(\f{M_W^2 +M_Z^2}{M_V^2}
\right) + \f{M_W^4+M_Z^4+10 M_W^2 M_Z^2}{M_V^4}
\right]
\label{Gch}
\eea
where, for large $\gs$,
\be
 g_{VWZ}=-\f 1 {2s_\theta c_\theta}\f {g}{\gs}
\ee
 The total neutral width is given by
\be
\Gamma_V^{h}+3 (\Gamma_V^{l}+\Gamma_V^{\nu})+\Gamma_V^W
\ee
where $\Gamma_V^h$ is the sum of all quark-antiquark  widths, with
\be
\Gamma_V^f={ {M_V\alpha_{em}}\over 3}\left ({v^f_V}^2+{a_V^f}^2\right )
\ee
\bea
\Gamma_V^W&=&{M_V\over 48}\alpha_{em} g^2_{VWW}
\left(1-4{M_W^2\over M_V^2}\right)^{3/2}\left({M_V\over M_W}\right)^4\nn\\
&&\times\left [1+20\left({M_W\over M_V}\right)^2+12\left
({M_W\over M_V}\right)^4
\right]
\label{Gneu}
\eea
with, for large $\gs$,
\be
g_{VWW}=
-\f 1 {2s_\theta}\f {g}{\gs}
\label{gvww}
\ee

For large $M_V$ and $\gs$ from eqs. (\ref{Gch})
and (\ref{Gneu}), one recovers (\ref{30}).

For future calculations we  also give  the $g_{ZWW}$ coupling
\be
g_{ZWW}=
\f 1 { \tan\theta} \left [ 1- 3 (1-\f 1 {2 c_\theta^2})\Big (\f
{g}{\gs}\Big )^2
\right ] 
\label{gzww}
\ee

Limits on the parameter space of the model can be obtained by
computing the $\eps$ parameters \cite{altarelli,fri}.

In the limit of large $M_V$ we get \cite{epsi,anic}
\bea
\eps_1 &=& 0\nn\\
\eps_2 &=& 0\nn\\
\eps_3 &=& -\f{b}{2}+\Big(\f{g}{\gs}\Big)^2
\label{e3}
\eea
In conclusion the model has no decoupling in the limit $M_V\to\infty$.
To get decoupling,  the limits of $\gs\to\infty$ and
$b\to 0$ have to be taken.

We can derive restrictions on the BESS parameters by using the experimental
data on $\eps_3$.
The most recent value for $\eps_3$ obtained by combining the LEP, 
CDF/UA2 and SLC data \cite{dati} is
\be
\eps_3=(4.1\pm 1.4)\times 10^{-3}
\ee

In order to get these bounds one has to make some assumptions on the
 radiative corrections to the $\eps$ parameters.
By neglecting  loop effects of new gauge vectors, and
 assuming for the BESS model the same one-loop radiative corrections
as for the SM in which the Higgs mass is used as a cut-off $\Lambda$,
 adding  to $\eps_3$ given in eq. (\ref{e3}) the contribution coming
from the radiative corrections  
$(\eps_3)_{rad.corr.}= 6.4\times 10^{-3}$ \cite{dati},
(for  $M_H=\Lambda=1~TeV$ and $m_{top}=175~GeV$),
one gets
the allowed region at $90\%$ C.L. in the plane $(b,g/\gs)$
shown in Fig. 1. Therefore the model is strongly constrained.

\begin{figure} 
\epsfysize=8truecm
\centerline{\epsffile{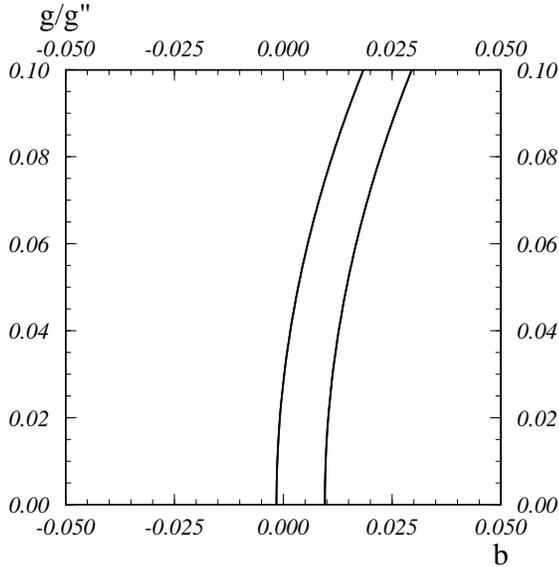}}
\noindent
\caption[bessbo]{$90\%$ C.L. contour in the plane
$(b,g/\gs)$ for large $M_V$ in the  BESS model
 from LEP/Tevatron/SLC data for $m_t=175~GeV$.
The allowed region is the internal one.}
\label{bessbo}
\end{figure}

\section{Vector axial-vector  resonance models}
\label{veaxre}
A model with vector and axial-vector resonances can be  built  by considering
in addition to the global symmetry $G=SU(2)_L\otimes SU(2)_R$, 
a local symmetry $H'=SU(2)_L\otimes SU(2)_R$ and unbroken
$H_D=SU(2)$, the diagonal  subgroup
of $G' = G\otimes H'$ \cite{assiali}.
 The nine Goldstone bosons, resulting from the spontaneous breaking 
of $G'$ to $H_D$, can be described by three independent
$2\times 2$ unitary matrices $L$, $R$ and $M$, with the following
transformations with respect  to $G$ and 
$H'$ 
\be
L'= g_L L h_L,~~~~~~R'= g_R R h_R,~~~~~~M'= h_R^\dagger M h_L
\label{0g}
\ee
with $g_{L,R}\in SU(2)_{L,R}\subset G$ and $h_{L,R}\in SU(2)_{L,R}\subset H'$. 
 We shall require the 
invariance under the discrete left-right transformation, denoted by $P$
\be
P:~~~~~~~L\leftrightarrow R,~~~~~M\leftrightarrow M^\dagger
\ee
which allows for a   low-energy theory  parity conserving.

The most general $G'\otimes P$ invariant Lagrangian, up to second order
derivatives, is given by \cite{assiali}
\be
{\cal L}_G=-\frac{v^2}{4} f(\lmu,\rmu)
\label{lg}
\ee
where
\be
f(\lmu,\rmu)=a_1 I_1 + a_2 I_2 + a_3 I_3 + a_4 I_4
\label{effe}
\ee
and by  the kinetic terms ${\cal L}_{kin}$ for
the fields.  The terms $I_i$ ($i=1,...,4$) are 
given by:
\bea
I_1&=&tr[(V_0-V_1-V_2)^2]\nn\\
I_2&=&tr[(V_0+V_2)^2]\nn\\
I_3&=&tr[(V_0-V_2)^2]\nn\\
I_4&=&tr[V_1^2]
\eea
and
\be
V_0^\mu=L^\dagger D^\mu L, ~~
V_1^\mu=M^\dagger D^\mu M,~~
V_2^\mu=M^\dagger(R^\dagger D^\mu R)M
\ee
with the following  covariant derivatives 
\bea
D_\mu L&=&\partial_\mu L -L \lmu\nn\\
D_\mu R&=&\partial_\mu R -R \rmu\nn\\
D_\mu M&=&\partial_\mu M -M \lmu+\rmu M
\eea
The kinetic term is 
\be
{\cal L}_{kin}=\frac{1}{\gs^2} tr[F_{\mu\nu}({\bf L})]^2+
         \frac{1}{\gs^2}  tr[F_{\mu\nu}({\bf R})]^2
\ee
where $\gs$ is the gauge coupling constant for the gauge fields $\lmu$ and
$\rmu$, 
\be
F_{\mu\nu}({\bf L})=\partial_\mu{\bf L}_\nu-\partial_\nu{\bf L}_\mu+
         [\lmu,{\bf L}_\nu]
\ee
and the same definition holds for $F_{\mu\nu}({\bf R})$.

The quantities $V_i^\mu~~(i=0,1,2)$ are invariant under the global
symmetry 
$G$ 
and covariant under the gauge group $H'$
\be
(V_i^\mu)'=h_L^\dagger V_i^\mu h_L
\ee
Using the $V_i^\mu$ one can build six independent quadratic invariants,
which reduce to the four $I_i$ listed above, when parity conservation is 
required.
In conclusion the Lagrangian is given by
\be
\LL= \LL_G+\LL_{kin}
\ee

The Lagrangian (\ref{lg}) can be studied in the unitary gauge
\be
L=R^\dagger=\exp(\frac i {2v} (1-z) \pi^a\tau^a)
~~~M=\exp(-\frac i {v} z \pi^a\tau^a)
\ee
where
\be
z=\frac {a_3}{a_3+a_4}
\label{zeta}
\ee

At the lowest order  in power series  of the Goldstone bosons,
 one gets the mass terms for the vector 
mesons:
\be
{\cal L}_G = -\frac{v^2}{4}[a_2~ tr(\lmu+\rmu)^2 + (a_3+a_4)~ 
         tr(\lmu-\rmu)^2]+\cdots 
\label{lge}
\ee
where the dots stand for terms at least linear in the Goldstone modes.
Using the combinations 
\be 
{\bf V}_\mu =(\lmu+\rmu)/2~~~{\bf A}_\mu =(\rmu-\lmu)/2
\label{va}
\ee
we can rewrite
\be
{\cal L}_G=-v^2[a_2~ tr(\vmu)^2 + (a_3+a_4)~ tr(\amu)^2]+\cdots
\ee
Therefore, if   $\vmu=\f{\dd i}{\dd 2}\f{\dd \gs}{\dd2} V_\mu
^a\tau^a$,
 $\amu=\f{\dd i}{\dd 2}\f{\dd \gs}{\dd2} A_\mu
^a\tau^a$,
we get
\be
M_V^2=a_2 \frac {v^2} 4 \gs^2~~~~~M_A^2=(a_3+a_4) \frac {v^2} 4 \gs^2
\ee
Again one can compute the scattering amplitudes of eq. (\ref{amp})
and eq. (\ref{amp2}) in terms of
\be
A(s,t,u)=\frac s {4v^2}(4-3\beta)+\frac {\beta M_V^2}{ 4v^2}
\left [\frac {u-s}{t-M_V^2+i M_V\Gamma_V}+
\frac {t-s}{u-M_V^2+i M_V\Gamma_V}\right ]
\label{s1}
\ee
with
\be
\beta= 4 \frac {M^2_V}{\gs^2 v^2}(1-z^2)^2= \frac{192\pi v^2\Gamma_V}{M_V^3}
\label{beta}
\ee

For generic values of the parameters $a_1,~a_2,~a_3,~a_4$, the Lagrangian 
${\cal L}$ is invariant under $G'\otimes P=G\otimes H'\otimes P$. 
There are however special choices which enlarge the symmetry group 
\cite{debess}.
One of these choices corresponds to a generalization of Georgi vector
symmetry
\cite{georgi}.

The case of interest for the electroweak sector is provided by the choice:
$a_4=0$, $a_2=a_3$. In order to discuss the symmetry properties it is 
useful to 
observe that the invariant $I_1$ could be re-written as follows
\be
I_1=-tr(\partial_\mu U^\dagger \partial^\mu U)
\ee
with
\be
U=L M^\dagger R^\dagger
\ee
Therefore  the Lagrangian can be rewritten as
\be
{\cal L}_G=\frac{v^2}{4}\{a_1~ tr(\partial_\mu U^\dagger \partial^\mu U) +
                         2~a_2~ [tr(D_\mu L^\dagger D^\mu L)+
                          tr(D_\mu R^\dagger D^\mu R)]\}
\label{lg1}
\ee
Each of the three terms in the above expressions 
is invariant under an independent $SU(2)\otimes SU(2)$
group
\be
U'=\omega_L U \omega_R^\dagger,~~~~~~L'= g_L L h_L,~~~~~~R'= g_R R h_R
\ee
 The overall symmetry is $G_{max}=[SU(2)\otimes SU(2)]^3$, with a subgroup 
$H'$ realized as a gauge symmetry.

With the particular choice $a_4=0$, $a_3=a_2$, as we see from eq. (\ref{lge}),
the mixing between $\lmu$ and $\rmu$ is vanishing, and the new states are 
degenerate in mass.
Moreover, as it follows from eq. (\ref{lg1}), the longitudinal modes
of the $\lmu$ and $\rmu$ (or $\vmu$, $\amu$) fields are entirely
provided by the would-be Goldstone bosons in $L$ and $R$. This means
that the pseudoscalar particles remaining as physical states in the
low-energy spectrum are those associated to $U$. They in turn can
provide the longitudinal components to the $W$ and $Z$ particles,
in an effective description of the electroweak breaking sector.

Since with this choice, from eq. (\ref{zeta}),
$z=1$, this model has $\beta=0$ (eq. (\ref{beta}))
and therefore reproduces at the lowest order the low energy 
amplitudes of the chiral Lagrangian (\ref{2}), 
as it can be seen from eq. (\ref{s1}).

We now consider the coupling of the model to the electroweak 
$SU(2)_L\otimes U(1)_Y$ gauge fields via the minimal substitution
\bea
D_\mu L &\to& D_\mu L+ {\Wt}_\mu L\nn\\
D_\mu R &\to& D_\mu R+ {\Yt}_\mu R\nn\\
D_\mu M &\to& D_\mu M 
\eea
where
\be
\Wt_\mu=i{\tilde W}_\mu ^a\f{\tau^a}{2},~~
\Yt_\mu=i
 {\tilde Y}_\mu\f{\tau^3}{2}
 \ee
 \be
\Lt_\mu=i L_\mu ^a\f{\tau^a}{2},~~
\Rt_\mu=i R_\mu ^a\f{\tau^a}{2}
\ee

By introducing the canonical kinetic terms for $W_\mu^a$ and $Y_\mu$ we get
the  model  \cite{debess}  called
Degenerate BESS, since in the large $\gs$ limit the new triplets $\lmu$
and $\rmu$ are degenerate in mass.

In the unitary gauge the Lagrangian 
is
\be
\L=-\f{v^2}{4}\Big[ a_1 \tr(\Wt_\mu-\Yt_\mu)^2
+2 a_2 \tr(\Wt_\mu-{\Lt}_\mu)^2
+2 a_2 \tr(\Yt_\mu-\Rt_\mu)^2\Big]
+\L^{kin}(\Wt,\Yt,\Lt,\Rt)
\label{lbd}
\ee
\bea
\L^{kin}(\Wt,\Yt,\Lt,\Rt)&=&
\f{1}{2 \gt^2}\tr[F^{\mu\nu}(\Wt)F_{\mu\nu}(\Wt)]
   +\f{1}{2 {\tilde g}^{\prime 2}}\tr[F^{\mu\nu}(\Yt)
F_{\mu\nu}(\Yt)]\nn\\
&+&\f{1}{ \gs^2}\tr[F^{\mu\nu}(\Lt)F_{\mu\nu}(\Lt)]+
\f{1}{ \gs^2}\tr[F^{\mu\nu}(\Rt)F_{\mu\nu}(\Rt)]
\eea

 Tilded quantities are used to remind that, due to the effects 
of the $\Lt$ and $\Rt$ particles, they are not the physical parameters
and fields. Notice that here we have also used tilded parameters for
a reason of convenience to be clear in the following.

The SM relations are obtained in the limit $\gs \gg {\tilde g},
{\tilde g}'$. Actually,
for a very large $\gs$, the kinetic terms for the fields $\lmu$ and $\rmu$
drop out, and ${\cal L}$ reduces to the first term in eq. (\ref{lbd}).
This term reproduces precisely, apart the  $\Wt$ and $\Yt$
 kinetic terms,  the mass term
for the ordinary gauge vector bosons, provided 
$a_1=1$.
The fermions
 couple to $\Lt$ and $\Rt$ via the mixing with the
standard $\Wt$ and $\Yt$.
For simplicity  direct couplings to the new vector bosons
  will not be considered here.

Finally in order to have canonical  kinetic terms for the gauge fields
we perform the rescaling 
$\tW\to \gt \tW$, $\tY\to \gpt \tY$, 
$\tL\to \gs\tL/\sqrt{2}$, $\tR\to\gs\tR/\sqrt{2}$.

\subsection{Masses and couplings of the new resonances }

Notice that the degenerate BESS
 model in the limit of infinite mass of the new resonances
$\lmu$ and $\rmu$ just reproduces the Higgsless SM, provided one
redefines
\be 
\frac {1}{2g^2}=\frac 1 {2\gt^2}+ \frac 1 {\gs^2}~~~~
\frac {1}{2\gp^2}=\frac 1 {2\gptd}+ \frac 1 {\gs^2}
\label{ggp}
\ee
Therefore it turns out convenient to reexpress all the physical quantities
in terms of $g$ and $g^\prime$.

 The new triplets of vector bosons are denoted by ($L^\pm,L_3$)
and ($R^\pm,R_3$).

In the charged sector the fields $R^\pm$ remain
 unmixed for any value of ${g''}$.
Their mass is given by:
\begin{equation}
M^2_{R^\pm} =\f{v^2}{4} a_2 \gs^2\equiv M^2
\label{8.2}
\end{equation} 

 In the
following we give approximate formulas in the limit $M \to \infty$ and
$g'' \to \infty$. The exact
formulas can be found in  \cite{debess}.
The charged fields $W^\pm$ and $L^\pm$ have the following masses:
\begin{eqnarray}
M^2_{{W}^\pm}&=&\frac{v^2}{4} { g}^2
,\nonumber\\
M^2_{{L}^\pm}&=&M^2 (1+2 {x^2})
\label{8.5}
\end{eqnarray}
where $x=g/g''$ and  $g$ is the $SU(2)$
gauge coupling constant defined in eq. (\ref{ggp}).

In the neutral sector we have:
\begin{eqnarray}
M^2_{Z}&=&\frac{M^2_W}{c^2_{\theta}}\nonumber\\
M^2_{L_3}&=& M^2\left(1+2 x^2\right)\nonumber\\
M^2_{R_3}&=& M^2\left(1+2 x^2 \tan^2 \theta\right)
\label{8.16}
\end{eqnarray}
where $\tan \theta = s_{\theta}/c_{\theta} = g'/g$ and $g'$ is the 
$U(1)_Y$ gauge 
coupling constant  defined in eq. (\ref{ggp}). 
As already observed for small $x$ all the new vector
resonances are degenerate in mass.

The charged part of the fermionic Lagrangian is
\begin{equation}
{\cal L}_{charged}=
-\left(a_W W_\mu^-+a_L L_\mu^-\right)J_L^{(+)\mu}+ h.c.
\end{equation}
where
\begin{eqnarray}
a_W&=&\frac {g}{\sqrt{2}}\nonumber\\
a_L&=& -g x
\label{9.2}
\end{eqnarray}
apart from higher order terms in $x$, and 
$J_L^{(+)\mu}={\bar \psi}_L \gamma^{\mu}
\tau^+ \psi_L$ with 
$\tau^+=(\tau_1 + i\tau_2)/2$. Let us notice that the $R^{\pm}$ 
fields are not coupled to the fermions.

In the neutral sector the couplings of the
 fermions to the gauge bosons are
\be
-\f 1 2 \bar \psi [(v_Z^f+\gamma_5 a_Z^f)\gamma_\mu Z^\mu+
(v_{L_3}^f+\gamma_5 a_{L_3}^f)\gamma_\mu L_3^\mu+
(v_{R_3}^f+\gamma_5 a_{R_3}^f)\gamma_\mu R_3^\mu]\psi
\ee
where $v^f$ and $a^f$ are the vector and axial-vector
couplings given by
\bea
v_Z^f &=&   A T_3^L + 2B Q_{em}\nn\\
a_Z^f&=& A T_3^L\nn\\
v_{L_3}^f &=&   C T_3^L + 2D Q_{em}\nn\\
a_{L_3}^f&=& C T_3^L\nn\\
v_{R_3}^f &=&   E T_3^L + 2F Q_{em}\nn\\
a_{R_3}^f&=& E T_3^L
\eea

Notice the different normalization with respect to eq. (\ref{cchbess}) and
eq. (\ref{cneubess}).
In the limit $M\to\infty$, $x\to 0$,
\begin{eqnarray}
&A&= \frac{g}{c_{\theta}}
~~~~B= -\frac{g s^2_\theta}{c_\theta} \nonumber \\
&C&=-\sqrt{2} g x ~~~~D= 0\nonumber \\
&E&= \sqrt{2} g \frac{x}{c_\theta} \tan^2 \theta ~~~~F= -E
\end{eqnarray}

In general, the total width of a vector boson $V$ corresponding to 
the decay into fermion - antifermion is
\be 
\Gamma^{fermion}_V= \Gamma_V^h+3(\Gamma_V^l+\Gamma_V^\nu)
\ee
where $\Gamma_V^h$ includes the contribution of all the 
allowed quark-antiquark decays.
The partial widths are given by
\be
\Gamma_V^f=\f  {M_V}{48\pi}
F(m_f^2/M^2_V)
\ee
with
\be
F(r_f)= (1-r_f)^{1/2} ((v_V^f)^2(1+2 r_f) +(a_V^f)^2(1-4 r_f))
\ee
and $m_f$ the mass of the fermion.

Concerning the new charged resonances, only $L^\pm$ decay into fermions.
The leptonic width is, neglecting fermion masses,
\be
\Gamma(L^-\to l\bar \nu_l)=\f 1 {24\pi} a_L^2 M_L\equiv\Gamma_L^0
\ee
The decay widths into quark pairs is 
\be
\Gamma(L^-\to q'\bar q)=3\vert V_{qq'}\vert^2 \Gamma_L^0
\ee

In the case of the $\bar t b$ 
decay, we have, 
neglecting the bottom mass:
\be
\Gamma(L^-\to b\bar t)= 3 |V_{tb}|^2 (1-\f 3 2 r_t + \f 1 2 r_t^3)\Gamma_L^0
\ee
where $r_t=m_t^2/M_L^2$.

For the neutral resonances,
in the usual limit, at the order $(g/\gs)^2$, and neglecting the 
fermion mass corrections we get 
\begin{eqnarray}
\Gamma^{fermion}_{L_3}&=&\frac{2\sqrt{2}G_FM_W^2}{\pi}M_{L_3}
\left(\frac g {g''}\right)^2\nonumber\\
\Gamma^{fermion}_{R_3}&=&\frac{10\sqrt{2}G_FM_W^2}{3\pi}\frac {{s_\theta}^4}
{{c_\theta}^4}M_{R_3}\left(\frac g {g''}\right)^2\nonumber\\
\Gamma^{fermion}_{L^\pm}&=&\frac{2\sqrt{2}G_FM_W^2}{\pi}M_{L^\pm}
\left(\frac g {g''}\right)^2
\end{eqnarray}
The other possible decay channel for a neutral vector boson
 $V$ is the one corresponding to 
the  a $WW$ pair. The  partial width is
\bea
\Gamma_V^W &=&\f{M_V}{192 \pi} g^2_{VW^+W^-} 
\left(1-4 \f{M^2_W}{M^2_V}\right)^{3/2}\left(\f{M_V}{M_W}
\right)^4\nn\\
&\times & \left[ 1 + 20 \left(\f{M_W}{M_V}
\right)^2 +12 \left(\f{M_W}{M_V}
\right)^4\right]
\eea
The other possible decay channel for  $L^{\pm}$ 
is the one corresponding to 
 a $WZ$ pair. The  partial width is
\bea
\Gamma_L^{WZ} &=&\f{M_{L}}{192 \pi} g^2_{ZW^+L^-} 
\left[\left(1- \f {M_Z^2-M_W^2}{M_L^2}\right)^2
-4\f{M^2_W}{M^2_L}\right]^{3/2}\left(\f{M_L^4}{M_W^2 M^2_Z}
\right)\nn\\
&\times & \left[ 1 + 10 \left(\f{M_W^2 +M_Z^2}{M_L^2}
\right) + \f{M_W^4+M_Z^4+10 M_W^2 M_Z^2}{M_L^4}
\right]
\eea
The relevant trilinear couplings are, always in the limit of large
$\gs$,
\bea g_{L_3 W^+W^-}&=&\sqrt{2} g \f g \gs \f {M^2_W}{M^2}\nn\\
 g_{R_3 W^+W^-}&=&\sqrt{2} g \f {s_\theta^2}
{c_\theta^2}\f g \gs \f {M^2_W}{M^2}\nn\\
 g_{Z W^+L^-}&=&\sqrt{2}  \f {g}
{c_\theta}\f g \gs \f {M^2_W}{M^2}
\eea
Using  these trilinear gauge couplings we get the following widths:
\begin{eqnarray}
\Gamma^{WW}_{L_3}&=&\frac{\sqrt{2}G_FM_W^2}{24\pi}M_{L_3}
\left(\frac g {g''}\right)^2\nonumber\\
\Gamma^{WW}_{R_3}&=&\frac{\sqrt{2}G_FM_W^2}{24\pi}
\frac{{s_\theta}^4}{{c_\theta}^4}
M_{R_3}\left(\frac g {g''}\right)^2\nonumber\\
\Gamma^{WZ}_{L^\pm}&=&\frac{\sqrt{2}G_FM_W^2}{24\pi}M_{L^\pm}
\left(\frac g {g''}\right)^2
\end{eqnarray}
By  comparing  the widths of the new gauge bosons into
vector boson pairs with those into fermions:
\begin{eqnarray}
\Gamma_{L_3}^{fermion}&=&48~\Gamma_{L_3}^{WW} \nonumber\\
\Gamma_{R_3}^{fermion}&=&80~\Gamma_{R_3}^{WW} \nonumber\\
\Gamma_{L^\pm}^{fermion}&=&48~\Gamma_{L^\pm}^{WZ}
\end{eqnarray}
we see that the  fermionic channel is dominant due to the multiplicity.

As already observed, the absence of couplings among 
$U$ and the states $\lmu $ and $\rmu$
results in a suppression of the decay rate of these
states into $W$ and $Z$. Consider, for instance, the decay of
the new neutral gauge bosons into  $W$ pairs. In a model with
only vector resonances this decay channel is largely the dominant one.
The corresponding width is indeed given by \cite{pierre43}
\be
\Gamma (V\to WW)=\f{\sqrt{2}G_F}{192\pi} \frac{M^5}{M_W^2}
\left(\f g \gs\right)^2\label{97}
\ee
and it is enhanced with respect to the partial width into a fermion
pair, by a factor $(M/M_W)^4$ 
\be
\Gamma(V\to {\bar f} f)\approx G_F M_W^2 \left(\frac{g}{\gs}\right)^2 M
\ee
This fact is closely related to the existence of a coupling of order $\gs$
among $V$ and the unphysical scalars absorbed by the $W$ boson.
Indeed the fictitious width of $V$ into these scalars provides,
via the Equivalence Theorem \cite{equ}, a good approximation to
the width of $V$ into a pair of longitudinal $W$ and it is
precisely given by eq. (\ref{97}).

On the contrary, if there is no direct coupling among the new gauge bosons and
the would-be Goldstone bosons which provide the longitudinal
degrees of freedom to the $W$, then their partial width 
into longitudinal $W$'s will be suppressed compared to the 
leading behaviour in eq. (\ref{97}), and the width
into a $W$ pair could be similar to the fermionic width.
The same argument also holds for the charged case. 

In usual strong interacting
models an enhancement of $W_L W_L$ scattering is expected. Due to the previous 
considerations, this  case is quite different. If we study $W_L W_L$ scattering
the lowest order result violates unitarity at energies above 1.7 TeV, 
as in the 
SM in the formal limit $M_H \to \infty$. So we expect this
model
to be valid  up to energies of this order.

\subsection{Low energy limits}

Again one can get limits
on the parameter space of the vector
axial-vector model by considering the $\eps$ parameters.
For the general model of Section \ref{veaxre},
 at the leading order in the limit of large
$\bf V$ and $\bf A$ masses, one gets \cite{epsi}
\be
\epsilon_1=\epsilon_2 =0,~~~~~~~~~~ \epsilon_3=(1-z^2)(g/g'')^2
\ee
The axial-vector resonances contribute with  opposite sign with
respect to the vector particles. This is easily understood by noticing that
$\epsilon_3$ can be expressed in terms of the combination $(\Pi_{VV}-\Pi_{AA})$
of the correlators of the vector and of the axial-vector currents \cite{S}.

The
parameter $z$ is free. Notice that in the degenerate limit $z=1$
all the $\eps$ vanish (at the leading order). One can compute the next
to leading order in the expansion $p^2/M^2$ and the result
is \cite{debess}
\bea
\epsilon_1&=&-\f{\c^4+\s^4}{\c^2}~ X\nn\\
\epsilon_2&=&-\c^2~ X\nn\\
\epsilon_3&=&-X
\label{e1e2e3}
\eea
where \be
X= 2 \left (\frac g\gs\right )^2 \frac {M_Z^2}{M^2}
\ee

\begin{figure}
\epsfysize=8truecm
\centerline{\epsffile{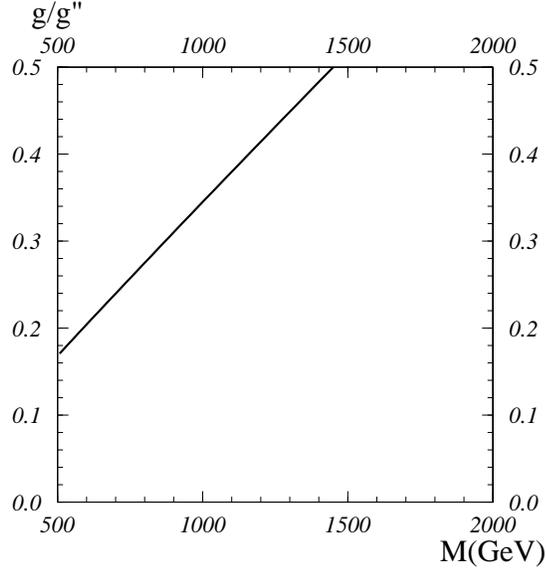}}
\noindent
\caption[degebo2]{ $ 90\% $ C.L. upper bounds on $g/g''$ vs. $M$
in the  Degenerate BESS model from 
the $\eps$ parameters of eq. (\ref{e1e2e3}) and 
LEP/Tevatron/SLC data.}
\label{degebo2}
\end{figure}

All these deviations are of order $X$ and  therefore contain a double
suppression factor $M_Z^2/M^2$ and $(g/\gs)^2$. 
The fact that in Degenerate BESS in the limit $M\to\infty$ all
the $\eps$ vanish
follows from the $(SU(2)\otimes SU(2))^3$ symmetry. Again, 
assuming the same radiative corrections to the $\eps$
parameters as in the SM with $M_H=1~TeV$, that is \cite{dati}
\be
\eps_1=3.65\times 10^{-3}~~
\eps_2=-7.10\times 10^{-3}~~
\eps_3=6.38\times 10^{-3}
\label{epsrad}
\ee
and considering their experimental values \cite{dati},
 one can derive a
90\% C.L. limit in the plane $(M,g/\gs)$, given in Fig. \ref{degebo2}.

\begin{figure}
\epsfysize=8truecm
\centerline{\epsffile{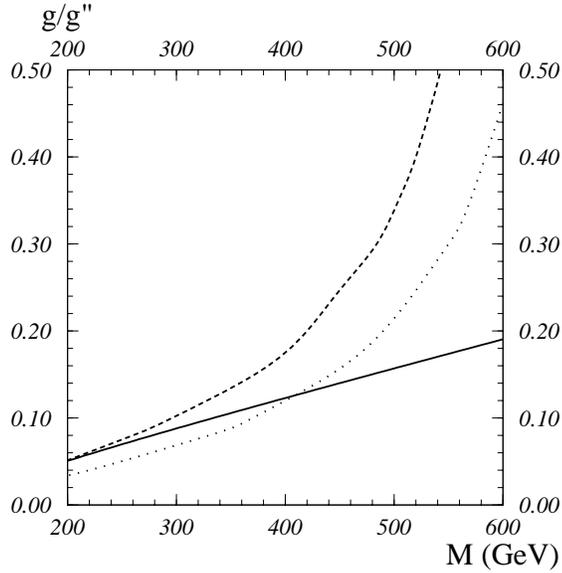}}
\noindent
\caption[degebo] 
{ $ 95 \% $ C.L. upper bounds on $g/g''$ vs. $M$ 
in the  Degenerate BESS model
from 
LEP/Tevatron/SLC data (solid line) and CDF with $L=19.7~pb^{-1}$ (dashed line).
The dotted line shows the extrapolation of the CDF bounds to $L=100~
pb^{-1}$.}
\label{degebo}
\end{figure}

For low value of $M$ we have also considered the bounds from
Tevatron. In Fig. \ref{degebo} we show 
 $ 95 \% $ C.L. upper bounds on $g/g''$ vs. $M$ from 
LEP/Tevatron/SLC data (solid line) and CDF with $L=19.7~pb^{-1}$ (dashed line).
The dotted line corresponds to 
the extrapolation of the CDF bounds to $L=100~
pb^{-1}$.

\subsection{Comparison with technicolor theories}

 The model  of Section \ref{vere} (BESS) describes in a rather
general way vector resonances;  therefore it can be
 specialized to describe
the vector resonances of a theory like technicolor.
The idea of hidden gauge symmetries \cite{bando}, 
in terms of which BESS can be formulated \cite{bess},
was successfully used to describe
the ordinary strong interacting vector resonances, like $\rho$ 
etc. \cite{bando}.

Technicolor was one of the earliest suggestions \cite{tech,farhi}
 to provide for an
alternative to the spontaneous symmetry breaking mechanism of the SM, based
on elementary scalars, which is considered as theoretically unsatisfactory.
The one-doublet model has a single techniquark doublet, of charge
$+1/2$ and $-1/2$ for anomaly cancellation. The one-family model 
has four doublets (3 colors + 1 lepton), anomaly cancellation going as
in quark-lepton families, with flavor symmetry $SU(8)\otimes SU(8)$ and
$SU(N_{TC})$ of technicolor. Technifermion condensates break flavor into
diagonal $SU(8)$, giving 63 Goldstone bosons, three of which provide the
$W,Z$ longitudinal degrees of freedom. Generation of masses for ordinary
fermions has led to extended technicolor and is generally associated with
difficulties because of the limitations on flavor-changing neutral-currents.
To solve these problems, subsequent proposals have been advanced
like walking technicolor \cite{walk},
 topcolor assisted technicolor \cite{tc2}
and non commuting extended technicolor \cite{chi}.
Pseudo-Goldstones are in general a very sensitive and potentially dangerous
feature, expecially in the original technicolor models. They are absent
in the one-doublet model. However in the one-family model some of the 63
pseudo-Goldstones would be low in mass and are expected to be produced as
charged scalar pairs from $e^+e^-$. Also, virtual pseudo-Goldstones would
contribute to radiative corrections. The overall technicolor
contribution to electroweak radiative corrections will however result from
a number of effects, among which one of the most
ambiguous comes from technipions.  
Therefore, if one realistically believes in the full
$SU(8)\otimes SU(8)$ one-family model, or in a more complicated model,
one will have to qualify all the quantitative statements for technicolor
because of the ambiguities related at least to technipion masses and
calculation of their virtual contributions. A detailed analysis under various
assumptions can be found in \cite{bessu8}.

The dynamics in technicolor theories is usually believed to be roughly
readable from QCD by simply scaling the fundamental scale $\Lambda_{QCD}$ to
$\Lambda_{TC}$. In QCD, vector dominance has revealed itself to be a
useful concept leading to results comparing very well to the experimental
data. Therefore it is natural to assume that vector dominance works as
well in technicolor theories as in QCD. In this spirit one can specialize
the BESS model to technicolor, as taken in a vector dominance 
approximation.  Strictly speaking, BESS would correspond to a technicolor 
model involving a single technidoublet.
However we will take here the simplest assumption  of
neglecting dynamical contributions from technipions. By doing so one 
can specialize
BESS also to technicolor models involving more than one technidoublet. 
To translate BESS parameters into such technicolor specialization
\cite{comp},
 we 
recall that for $SU(N_{TC})$ of technicolor one scales directly from
QCD the techni-$\rho$ mass
\be
M_{{\rho_{TC}}}=M_0\left({3\over N_{TC}}\right)^{1/2}
\left({4\over N_{d}}\right)^{1/2}
\label{trho}
\ee
where $N_d$ is the number of technidoublets and $M_0$ is a scale parameter
roughly of order $1~TeV$. As repeatedly said, technicolor dynamics is
obtained by scaling  QCD, so the Kawarabayashi, Suzuki, Fayazzuddin, 
Ryazzuddin (KSFR) relations are supposed to be valid also in technicolor.
The first KSRF relation for technicolor would read
\be
g_{\gamma{\rho_{TC}}}=2N_df^2_{TC}g_{{\rho_{TC}}{\pi_{TC}}{\pi_{TC}}}
\label{102}
\ee
relating the $\gamma-\rho_{TC}$ coupling to the technipion decay-coupling
constant and to the $\rho_{TC}-\pi_{TC}-\pi_{TC}$ trilinear coupling.
This relation is automatically  satisfied in BESS, where 
\be
g_{\gamma V}=-{1\over 2}\alpha g''v^2
\label{103}
\ee
\be
g_{V\pi\pi}=-{1\over 4}\alpha g''
\label{104}
\ee
with $v=\sqrt{N_d} f_{TC}$,
  the decay coupling constant of the Goldstone $\pi$. So this
relation does not impose any condition on the BESS parameters $\alpha$
and $g''$. On the other hand the second KSRF relation,
\be
M_{{\rho_{TC}}}^2=2N_df^2_{TC}g^2_{{\rho_{TC}}{\pi_{TC}}{\pi_{TC}}}
\ee
requires $\alpha=2$ in the BESS model by using eqs. (\ref{mv2})  and 
(\ref{102})-(\ref{104}).
No restrictions are imposed on $g''$ and by comparing the techni-$\rho$
mass expression (\ref{trho}) to the BESS expression for $M_V$,
one can establish the following relation between $g''$ and the product
$N_{TC} N_d$
\be
g''={M_0\over v}\sqrt{{24\over N_{TC}N_d}}
\ee
To compare the experimental bounds on BESS with those for technicolor 
the following relation is useful
\be
{g\over g''}={M_W\over M_0}\sqrt{{N_{TC}N_d\over 6}}=\sqrt{2}
\f {M_W}{M_{\rho_{TC}}}
\ee

Therefore the upper bound for $g/\gs$ can be translated into a lower
bound
for $M_{\rho_{ TC}}$.

In technicolor theories one also expects axial-vector
resonances. To translate this part in BESS we recall
that in the hidden gauge symmetry approach
to QCD including axial-vector resonances, one should take the
parameter $z$ equal to 1/2. This
follows from vector dominance and Weinberg sum rules.
Furthermore one has the Weinberg mass relation $M_A^2=2M_V^2$.  
 In this case
the contribution to the $\eps_3$ parameter comes out
  $\eps_3=3/4 (g/\gs)^2$. 
Therefore using eqs. (\ref{epsv}) and (\ref{epsrad}) 
we get that QCD rescaled technicolor is strongly
excluded.

To complete the technicolor translator for this specialized
BESS we consider the role 
of the $b$ parameter of BESS, characterizing a direct coupling
of $V$ to the fermions as described in eq. (\ref{lfer}). The four-fermion
interaction of extended technicolor, of the form
\be
{1\over\Lambda^2_{ETC}}\left(\bar\psi_{TC}^L\gamma_\mu{{\vec\tau}\over 2}
\psi_{TC}^L\right)\left(\bar\psi_{f}^L\gamma_\mu{{\vec\tau}\over 2}
\psi_{f}^L\right)
\ee
would induce a similar direct coupling of $\rho_{TC}$  to fermions,
through technifermion loops, given by an effective interaction
\be
{1\over \Lambda_{ETC}^2}{M^2_{{\rho_{TC}}}\over
g_{{\rho_{TC}}{\pi_{TC}}
{\pi_{TC}}}}
{\vec \rho}_{TC}\cdot {\vec J}_L
\ee
Using eqs.  (\ref{mv2}) and  (\ref{104})
for the BESS model ($\alpha=2$), we find (for
small $b$)
\be
b\approx -2\left({v\over\Lambda_{ETC}}\right)^2
\ee
When translating the specialized BESS to technicolor, 
one sees that $b$ can be interpreted as
a parameter of extended technicolor, its magnitude being expected to be
inversely related to the square of the extended technicolor scale.

\section{Generalizations}
\label{gene}

A model describing vector, axial-vector and a scalar resonance has been
recently proposed \cite{cdd}.
 This is obtained by adding to the Lagrangian $\LL$ of eq. (\ref{lg})
the term
\be
\LL_S= \frac 1 2 \de^\mu S \de_\mu S - \frac 1 2 M_S^2 S^2
-\frac {vk}{2} S f(\lmu,\rmu)
\ee
where $S$ is a  scalar field singlet under the symmetry group
$G\otimes H^\prime\otimes P$, $k$ is a dimensionless parameter
and $f$ is given in eq. (\ref{effe}).

The scattering amplitude (\ref{s1}) receives an additional contribution
\bea
A(s,t,u)&=&\frac s {4v^2}(4-3\beta)+\frac {\beta M_V^2}{ 4v^2}
\left [\frac {u-s}{t-M_V^2+i M_V\Gamma_V}+
\frac {t-s}{u-M_V^2+i M_V\Gamma_V}\right ]\nn\\
&&-\frac {k^2s^2}{v^2}\frac 1{s-M_S^2+i\Gamma_S M_S}
\eea
It has been shown \cite{cdd} that in the limit 
of low energy with respect to the
masses of the new resonances the Lagrangian becomes equivalent 
to the chiral Lagrangian (\ref{a4a5}) with
the following identification 
\be
\alpha_4=\frac {(1-z^2)^2}{4\gs^2},~~~\alpha_4+\alpha_5=\frac
{kv^2}{8M_S^2}
\label{a4}
\ee
Notice that $\alpha_4$ can be related to the vector resonance width by
\be
\alpha_4= \frac {12 \pi v^4 \Gamma_V}{M_V^5}
\ee

A model based on Adler-Weisberger sum-rules was recently
proposed \cite{H3}.
It is known from strong interactions that all Adler-Weisberger
sum-rules 
can be satisfied by the four meson states $\pi$, $\rho$, $\sigma$ and
$a_1$ \cite{gilman,weinberg2}. Weinberg has also shown \cite{men}
that this is a result of the algebric structure of the broken symmetry
and that the zero helicity states  of $\pi$, $\rho$, $\sigma$ and
$a_1$ should form a complete representation of the
$SU(2)\otimes SU(2)$ symmetry.
Such a behavior for the electroweak sector
inspired the model \cite{H3}. Eight Adler-Weisberger sum-rules are satisfied 
by a scalar resonance $S$, a triplet of vectors $V$, a triplet 
of axial-vectors $A$
and an isoscalar vector $\omega$. The sum-rules are satisfied by
the following relations
\be
M_V^2=M^2_\omega =M_S^2 \tan^2\psi+M_Z^2 (1-\tan^2\psi)
\ee
\be
M_A^2 =\f {M_V^2}{\sin^2\psi}-M_Z^2\cot^2\psi
\ee
\be g_{SWW}^2=\f 4 {v^2} \sin^2\psi~~~~
 g_{\omega VW}^2 =\f 4 {v^2}~~~
~g_{VWW}^2=\f {M_V^2} {v^2} \cos^2\psi
\ee
\be
g_{ASW}^2= \f {M_A^2} {v^2} \cos^2\psi
~~~~
g_{AVW}^2= \f {16} {v^2} \f {M_A^2M_V^2}{(M_A^2-M_V^2)^2}\sin^2\psi
\ee
where $\psi$ is a mixing angle.
An effective Lagrangian was built in terms of these fields; however
many but not  all of the relations coming from the sum-rules can be satisfied.

\section{$e^+e^-\rightarrow f^+f^-$ channel}
\label{eech}

Before entering into the details of the analysis we briefly review 
in Table 1 the
parameters of the proposed $e^+e^-$ LC's \cite{LCW2,loew,pesmur,clic}.

\begin{table}[b]
\begin{center}
\begin{tabular}{l c c c c c c c}
\hline
\hline
& & & & & & &\\
 & TESLA  & JLC & NLC & CLIC & TESLA & NLC & CLIC \\ 
\hline\hline
$E_{CM}(GeV)$& 500 & 500 & 500 & 500 & 800 &1000 & 1000\\ 
${\cal L}(10^{33})$& 6 & 5.2 & 7.1 & 4.8&5.7& 14.5&10\\ 
$grad(MV/m)$& 25& 73&50&80&40&85&80\\
$LC length(km)$&33&10.4&15.6&11.2&33&18.7&20.0\\
\hline
\hline
\end{tabular}
\end{center}

\begin{description}
\item {\bf Table 1}: Parameters of proposed linear colliders and of
their upgrading.
${\cal L}$ is the final luminosity. 
\label{tab1}
\end{description}
\end{table}
Two distinct approaches have been followed; in the first (JLC and NLC)
one has developed a design, based on 
improving the efficiency in using standard
copper accelerating cavities, in the second (TESLA) one uses
superconducting cavities. CLIC project is based on beam acceleration
by traveling
wave structures powered by a superconducting
drive linac.

In the discussion of the physics at high energy LC's,
we start considering  the fermion annihilation  channels.
We would like to  analyze the effect
of new neutral vector bosons from a strong interacting
sector in cross-sections and asymmetries for the channels 
$e^+e^-\rightarrow l^+l^-$ and $e^+e^-\rightarrow q\bar q$.

In principle the neutral  resonances  could be produced as real resonances
if their mass is known to be
below the collider energy by just  tuning
 the beam energies. In this case the machine could
operate like a $Z^\prime$ factory and the properties of these
new resonances can be studied with precision by measuring the
line shape.

If instead the masses are higher than the maximum c.m. energy, they would 
give rise to indirect effects in the $e^+e^-\rightarrow f^+f^-$
that  we discuss below.

The clean environment of $e^+e^-$ colliders allows to study these
virtual
effects.
One could take into account radiative corrections and in particular
initial state QED corrections. However one can remove the effect of
these
corrections by a suitable cut on the photon
energy $E_\gamma /E_{beam} <1-M_Z^2/s$ \cite{rieman}.
 We perform first a general study of
the observables, giving the analytical formulas for cross sections
and asymmetries in the  case  of the BESS model or for a new triplet of vector
bosons.
The case of two  triplets can be easily derived from these equations
and can be found in \cite{debess}.

An analysis of the effect of new vector particles from a strong
electroweak symmetry breaking sector on the lepton and
hadron channels was also done in \cite{verz}. The
authors  use a formalism
where the corresponding oblique corrections to cross-sections and
asymmetries are expressed as once subtracted dispersion integrals.
The subtraction constants are provided by using LEP
results.

\subsection{Observables}
\label{obsf}

In models with additional gauge vector bosons one usually
performs the  analysis using the  following observables:
\bea
&\sigma^{\mu},~~R=\sigma^h/\sigma^{\mu}\nn\\
&A_{FB}^{e^+e^- \rightarrow \mu^+ \mu^-}
,~~ A_{FB}^{e^+e^- \rightarrow {\bar b} b}\nn\\
&A_{LR}^{e^+e^- \rightarrow \mu^+  \mu^-},~~A_{LR}^{e^+e^- \rightarrow
{\bar b} b},~~
  A_{LR}^{e^+e^- \rightarrow {had}}
\label{obsff}
\eea
with  $\sigma^{h(\mu)}$  the total hadronic ($\mu^+\mu^-$) cross section. 
 $A_{FB}$ is the forward-backward asymmetry given by
\be
A_{FB}=\frac{\sigma_F-\sigma_B}{\sigma_F+\sigma_B}
\ee
where $\sigma_{F,B}$ are  respectively the cross sections in the forward and
backward hemispheres of the detector, and
$A_{LR}$  the left-right asymmetry
\be
A_{LR}=\frac{\sigma_L-\sigma_R}{\sigma_L+\sigma_R}
\ee
where  $\sigma_{L,R}$ are the cross sections for
left and right longitudinal polarization states of the incoming electron.

The total cross section for the process $e^+e^-\rightarrow f^+f^-$ 
is given by (at tree level)
\be
\sigma = {\pi\alpha_{em}^2 s\over 3}\sum_{h_f,h_e}|F(h_f,h_e)|^2
\ee
with $\alpha_{em}=e^2/(4\pi)$ and
\be
F(h_f,h_e)=-{1\over s}e_f+{(v^f_Z+h_f a^f_Z)(v_Z+h_e a_Z)
\over {s-M_Z^2+iM_Z\Gamma_Z}}+{(v^f_V+h_f a^f_V)(v_V+h_e a_V)
\over {s-M_V^2+iM_V\Gamma_V}}
\ee
where $h_f,~h_e=\pm 1$ are the helicities of $f$ and $e$ respectively, $e_f$
is the electric charge of $f$ (in units of $e_{\rm proton}=1$), 
$v_{Z,V}=v^e_{Z,V}$ and $a_{Z,V}=a^e_{Z,V}$,
 and
$\Gamma_{Z,V}$ are the widths of the neutral gauge bosons.
For instance for the BESS model the couplings 
are given in eq. (\ref{cneubess}).

The forward-backward asymmetry in the present case is given by
\be
A_{FB}^{e^+e^-\rightarrow f^+ f^-}={x\over {1+{1\over 3}x^2}}
{{(1-P)\sum_{h_f,h_e}h_fh_e|F(h_f,h_e)|^2+2P\sum_{h_f}h_f|F(h_f,1)|^2}\over
{(1-P)\sum_{h_f,h_e}|F(h_f,h_e)|^2+2P\sum_{h_f}|F(h_f,1)|^2}}
\ee
where $x$ is the detector acceptance ($x\le 1$), and
$P$ is the degree of longitudinal polarization of the
electron beam. In the  analysis presented in Section \ref{bess}
and Section  \ref{debess}    $x=1$ is assumed.

The left-right asymmetry is given by
\be
A_{LR}^{e^+e^-\rightarrow f^+ f^-}=P{{\sum_{h_f,h_e}h_e|F(h_f,h_e)|^2}\over
{\sum_{h_f,h_e}|F(h_f,h_e)|^2}}
\ee
The notations are the same as for the forward-backward asymmetry.

\section{$e^+e^- \rightarrow WW$ channel}
\label{wwch}

In this Section we will consider the $WW$ channel, which is expected to
be more sensitive, at high energy, than the $f\bar f$ channel to effects
coming from a strongly interacting electroweak symmetry breaking sector.
In the case of a vector resonance this is simply due to the strong coupling
between  the resonance and the longitudinal $W$ bosons. Furthermore this
interaction, in general, destroys the fine cancellation between
 the $\gamma-Z$
exchange and the neutrino contribution occurring in the SM. This effect gives
rise, for instance in the case of the BESS model,
 to a differential cross-section
increasing linearly with $s$.
However this is no more true in the Degenerate BESS 
model due to the cancellation 
between the vector and axial-vector resonances.

Usually one  considers
 the $WW$ channel, with one $W$  decaying leptonically 
and the other hadronically.  The main reason for choosing this decay channel
is to get a clear signal to reconstruct the polarization of the $W$'s
(see for example \cite{fujii}). 

In fact by reconstructing the decay of $W$ pairs one can get information on 
their helicities, as shown by the decay angular distributions
\be
 {d \Gamma /d\cos\theta} \sim \cases{ (1+\cos\theta)^2  & $h_W = -1$ \cr
                                       2 \sin^2\theta   & $h_W = 0$ \cr
                                     (1-\cos\theta)^2  & $h_W = +1$\cr}\
\ee
where we have denoted by $\theta$   the polar
$W^-$ decay angle and by
$h_W$ its   helicity. Simulation studies and reconstruction of
decay angles of $W$ can be found in \cite{miya}.

Furthermore longitudinally polarized beams will be considered;
in fact by selecting initial 
polarized beams one can enhance the  final polarization.
For instance  an initial $e^-_R$ beam produces
 mostly longitudinal $W$ pairs.
 
\begin{figure}
\epsfysize=8truecm
\centerline{\epsffile{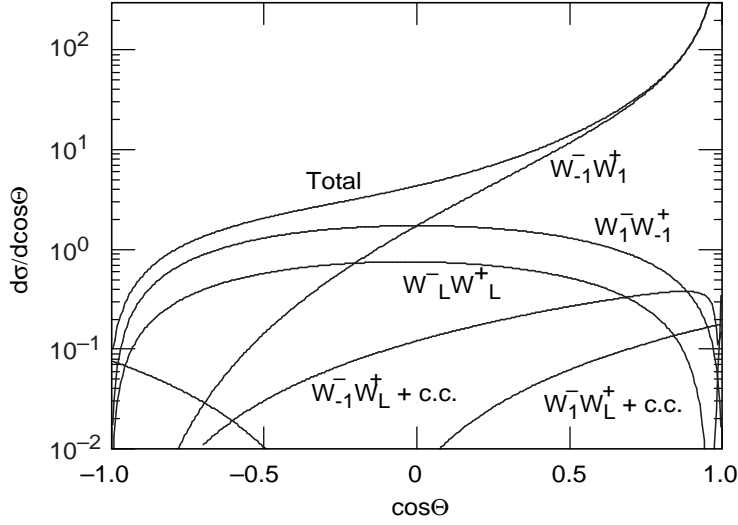}}
\caption[ww3]{Angular distributions for $W$ bosons of various 
 helicity in $e_L^- e_R^+ \rightarrow W^+W^-$.  The differential cross sections
are given in units of $R=86.8~fb/s(TeV^2)$ at $\sqrt{s}= 1~ TeV$,
from \cite{81}.}
\label{ww3}
\end{figure}

The reaction $e^-_Le^+_R \rightarrow W^+
W^-$ is shown in Fig. \ref{ww3}.  The 
subscripts  $1,-1$,
$L$,  denote the transverse and longitudinal helicity
 polarization
states of a massive vector boson.
For initial $e_R$ the cross section is dominantly $W_L W_L$ with a
rate 1/5 with respect to the $W_L W_L$ 
cross section shown in Fig. \ref{ww3}  \cite{81}. 

\subsection{Observables}
\label{obsw}

Let us consider first the following observables: the differential cross
section and the left right asymmetry
\bea
&&{d\sigma \over {d\cos\Theta}}(e^+ e^-\rightarrow W^+ W^-)\nn\\
& &A_{LR}^{{ e^+ e^- \rightarrow W^+ W^-}}=(
{d\sigma \over {d\cos\Theta}}(P_{e}=+P)-
{d\sigma \over {d\cos\Theta}}(P_{e}=-P))/
{d\sigma \over {d\cos\Theta}}
\eea
where $\Theta$ is the $e^+e^-$ center of mass scattering angle.
In the $e^+e^-$ center of mass frame the angular distribution
$d\sigma/d\cos\Theta$ and the left-right asymmetry read \cite{chife}
\bea
{{d\sigma}\over {d\cos\Theta}}&=& {{2\pi\alpha_{em}^2 p}\over {\sqrt{s}}}
\Big{\{} 2a_W^4\left [{4\over M^2_W}+p^2\sin^2\Theta
\left ( {1\over M_W^4}+{4\over t^2}\right )\right ]\nn\\
&&+G_1p^2\left [ {{4s}\over M^2_W}+
\left ( 3+{{sp^2}\over M^4_W}\right )\sin^2\Theta\right ]\nn\\
&&+G_1^{\prime}
\left [ 8\left ( 1+{{M^2_W}\over t}\right )+
16 {{p^2}\over M^2_W}+{{p^2}\over s}\sin^2\Theta
\left ( {{s^2}\over M_W^4}-2{s\over M_W^2}-4{s\over t}\right )
\right ] \Big{\}}
\label{dsig}
\eea
and
\bea
A_{LR}(\cos\Theta)&=&-P
{{2\pi\alpha_{em}^2 p}\over {\sqrt{s}}}
\Big{\{} 2a_W^4\left [{4\over M^2_W}+p^2\sin^2\Theta
\left ( {1\over M_W^4}+{4\over t^2}\right )\right ]\nn\\
&&+G_2p^2\left [ {{4s}\over M^2_W}+
\left ( 3+{{sp^2}\over M^4_W}\right )\sin^2\Theta\right ]\nn\\
&&+G_1^{\prime}
\Big [ 8\left ( 1+{{M^2_W}\over t}\right )+
16 {{p^2}\over M^2_W}\nn\\
&&+{{p^2}\over s}\sin^2\Theta
\left ( {{s^2}\over M_W^4}-2{s\over M_W^2}-4{s\over t}\right )
\Big ] \Big{\}} {\Big /}{{d\sigma}\over {d\cos\Theta}}
\label{ALR}
\eea
where
\bea
&p=&{1\over 2}\sqrt{s}(1-4M^2_W/s)^{1/2}\nn\\
&t=&M_W^2-{1\over 2}s[1-\cos\Theta (1-4M_W^2/s)^{1/2}]
\eea
The quantity $a_W$ is the $\nu e W$ coupling 
and is given in eq. (\ref{cchbess}) for the BESS model.
Furthermore 
\bea
G_1&=&\left ({{e_e}\over s}\right )^2+
(v_Z^2+a_Z^2)g^2_{ZWW}{1\over {(s-M_Z^2)^2+M_Z^2\Gamma_Z^2}}\nn\\
&&+2{e_e\over s}v_Zg_{ZWW} {{s-M_Z^2}\over {(s-M_Z^2)^2+M_Z^2\Gamma_Z^2}}\nn\\
&&+(v_V^2+a_V^2)g^2_{VWW}{1\over {(s-M_V^2)^2+M_V^2\Gamma_V^2}}\nn\\
&&+2{e_e\over s}v_Vg_{VWW} {{s-M_V^2}\over {(s-M_V^2)^2+M_V^2\Gamma_V^2}}\nn\\
&&+2 (v_Zv_V+a_Za_V)g_{ZWW}g_{VWW}\nn\\ 
&&~~~\times {{(s-M_Z^2)(s-M_V^2)+M_Z\Gamma_Z M_V\Gamma_V}
\over {[(s-M_Z^2)^2+M_Z^2\Gamma_Z^2][(s-M_V^2)^2+M_V^2\Gamma_V^2]}}
\eea
\bea
G_1^{\prime}&=&a^2_W [{{e_e}\over s}\nn\\
&&+g_{ZWW}(v_Z+a_Z)
{{s-M_Z^2}\over {(s-M_Z^2)^2+M_Z^2\Gamma_Z^2}}\nn\\
&&+g_{VWW}(v_V+a_V){{s-M_V^2}\over {(s-M_V^2)^2+M_V^2\Gamma_V^2}}]
\eea
\bea
G_2&=&2\Big ({e_e\over s}a_Z g_{ZWW}{{s-M_Z^2}\over {(s-M_Z^2)^2+M_Z^2
\Gamma_Z^2}}+
{e_e\over s}a_V g_{VWW}{{s-M_V^2}\over {(s-M_V^2)^2+M_V^2\Gamma_V^2}}\nn\\
&&+a_Zv_Zg_{ZWW}^2{1\over {(s-M_Z^2)^2+M_Z^2\Gamma_Z^2}}
+a_Vv_Vg_{VWW}^2{1\over {(s-M_V^2)^2+M_V^2\Gamma_V^2}}\nn\\
&&+(a_Zv_V+v_Za_V)g_{ZWW}g_{VWW}\nn\\
&&~~~\times {{(s-M_Z^2)(s-M_V^2)+M_Z\Gamma_Z M_V\Gamma_V}
\over {[(s-M_Z^2)^2+M_Z^2\Gamma_Z^2][(s-M_V^2)^2+M_V^2\Gamma_V^2)]}}\Big)
\eea
where $g_{VWW}$  
and $g_{ZWW}$, for the BESS model, are given in eq. (\ref{gvww}) 
and eq. (\ref{gzww}).

Assuming that the final $W$ polarization can be reconstructed, using the
$W$ decay distributions,
it is convenient to examine the cross sections for $W_LW_L$, $W_TW_L$, and
$W_TW_T$. One has \cite{lays}
\bea
{{d\sigma_{LL}}\over {d\cos\Theta}}&=& {{2\pi\alpha_{em}^2 p}\over {\sqrt{s}}}
\Big{\{} {{a_W^4}\over 8}{1\over M_W^4}{1\over t^2}
[ s^3 (1+\cos^2\Theta)-4M_W^4 (3s+4 M_W^2)\nn\\
&&-4(s+2M_W^2)p\sqrt {s} s \cos\Theta  ]\sin^2\Theta\nn\\
&&+{1\over {16}}G_1{1\over M_W^4}\sin^2\Theta
(s^3-12sM_W^4-16M_W^6)\nn\\
&&+G_1^{\prime}\sin^2\Theta {1\over {2t}} [
ps\sqrt{s}\cos\Theta{1\over {2M_W^4}}(s+2M_W^2)\nn\\
&&-{1\over {4 M_W^4}}(s^3-12sM_W^4-16M_W^6) ]
\Big{\}}
\label{sll}
\eea
\bea
{{d\sigma_{TL}}\over {d\cos\Theta}}&= &{{2\pi\alpha_{em}^2 p}\over {\sqrt{s}}}
\Big{\{} {{a_W^4}}{1\over {t^2 M_W^2}}
[ s^2 (1+\cos^4\Theta)+4M_W^4 (1+\cos^2\Theta)\nn\\
&&-4(4p^2+s\cos^2\Theta )p\sqrt {s}  \cos\Theta  
+2s(s-6M_W^2)\cos^2\Theta-4sM_W^2 ]\nn\\
&&+2G_1s{p^2\over M_W^2}(1+\cos^2\Theta)\nn\\
&&+2G_1^{\prime}
{{p\sqrt{s}}\over {tM_W^2}}[
\cos\Theta (4p^2+s\cos^2\Theta)-
2p\sqrt{s}(1+\cos^2\Theta)]
\Big{\}}
\label{stl}
\eea
\bea
{{d\sigma_{TT}}\over {d\cos\Theta}}&= &{{2\pi\alpha_{em}^2 p}\over {\sqrt{s}}}
\Big{\{} {2{a_W^4}}{1\over {t^2 }}
[ s (1+\cos^2\Theta)-2M_W^2-2p\sqrt {s}\cos\Theta ]\sin^2\Theta\nn\\
&&+2G_1{p^2}\sin^2\Theta+G_1^{\prime}
{{\sin^2\Theta}\over {2t}}[4p\sqrt{s}\cos\Theta -8p^2]
\Big{\}}
\label{stt}
\eea
The left-right asymmetries for longitudinal and/or transverse polarized 
$W$ can be easily obtained as in eq. (\ref{ALR}).

At LEPII we can add to the previous observables the $W$ mass measurement, 
coming from the $e^+e^- \rightarrow WW$ channel.

\section{Results for the BESS model}
\label{bess}

 Let us first discuss the fermion channel.

The results we will present are obtained by performing a $\chi^2$
comparison between the BESS predictions and the SM ones for the 
observables ${\cal O}_i$
discussed in Section \ref{obsf}:
\be
\chi^2=\sum_{i=1}^n \left [\frac {{\cal O}_i-{\cal O}_i^{SM}}
 {\delta  {{\cal O}_i}}\right ]^2
\ee

Following the existing studies of $500~
GeV$ $e^+e^-$ linear
colliders \cite{desy92,dleik}, we present the results assuming 
 a relative systematic error in luminosity of
${{\delta {\cal L}}/ {\cal L}}=1\%$, ${{\delta\epsilon_{\rm hadr}}/
\epsilon_{\rm hadr}}=1\%$, and
$\delta\epsilon_{\mu}/\epsilon_{\mu}=0.5\%$,
where $\epsilon_{hadr,\mu}$ denote the selection efficiencies.
The same systematic errors for the 1 and 2 $TeV$ machines are also
 assumed.
 
Finally  integrated luminosities  $L=20~fb^{-1}$ for
$\sqrt{s}=500~GeV$, $L=80~fb^{-1}$ for $\sqrt{s}=1~TeV$, and
$L=20~fb^{-1}$ for $\sqrt{s}=2~TeV$ have been considered.

 Unfortunately the most sensitive observables
are the left-right asymmetries, which means that, if the beams are not
polarized, one does practically get no advantage over LEP from this channel.

\begin{figure}
\epsfysize=8truecm
\centerline{\epsffile{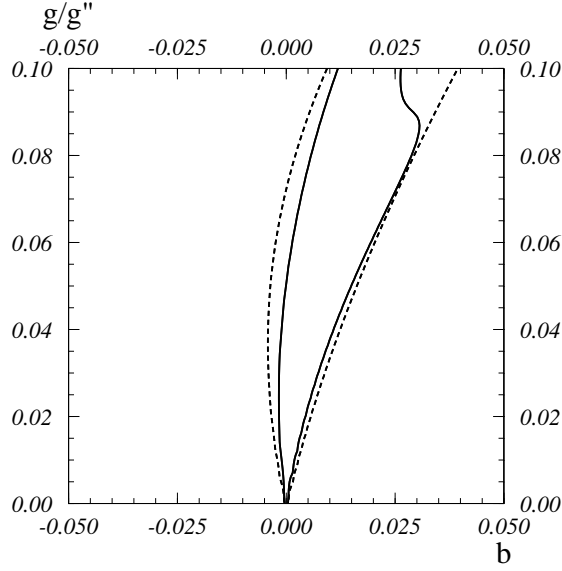}}
\noindent
\caption[ee2]{BESS $90\%$ C.L. contours in the plane 
$(b,g/\gs)$ for $\sqrt s=500~GeV$ and $M_V=600~GeV$
from the fermion channel. The solid line corresponds to
polarization $P=0.5$ while the dashed line is for unpolarized
electron beams.
The allowed regions are  the internal ones.}
\label{ee2}
\end{figure}

The contours shown in Fig. \ref{ee2} correspond
 to the regions which are allowed
at 90\% C.L. in the plane $(b,g/\gs)$, assuming that the BESS deviations
for the observables of eq. (\ref{obsff}) from the SM predictions are 
within the
experimental errors. The results are obtained assuming a longitudinal
polarization of the electron $P=0.5$ (solid line) and $P=0$ (dashed line).
In Fig. \ref{ee2} the energy of the collider is
 $\sqrt{s}=500~GeV$ and the mass of the vector
resonance $M_V=600~GeV$.
It is clear that there is no big improvement
with respect to the already existing bounds from LEP. Increasing the
energy of the machine does not drastically change the results.

All these conclusions become much more negative if one assumes a higher
systematic error for the hadron selection efficiency.
Therefore in conclusions the fermion channel is not much efficient
in this case.

Let us then consider the $W$ channel.

\begin{figure}
\epsfysize=8truecm
\centerline{\epsffile{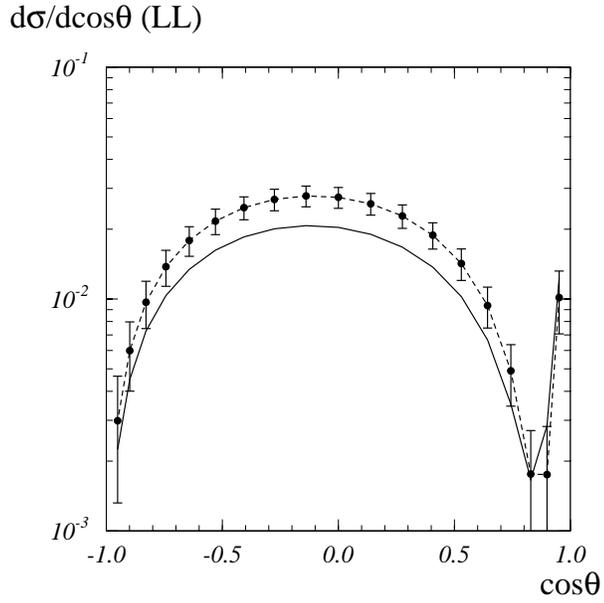}}
\noindent
\caption[wlwl]{Differential cross section $d\sigma/d\cos\Theta
(W_LW_L)$
for the BESS model for
$b=0.01$, $g/\gs=0.04$ and $M_V=600~GeV$ (continuous line) and for
the SM (dashed line). The error bars  are also shown.}
\label{wlwl}
\end{figure}

Fig. \ref{wlwl} shows the deviations of the BESS 
model with respect to the SM
for  $b=0.01$ and $g/\gs=0.04$,
 $M_V=600~GeV$,
for the   longitudinally polarized $W$ differential cross-section
at $\rs=500~GeV$.
The branching ratio
for decay of one $W$ leptonically and the other hadronically is included.
The continuous line represents 
the BESS prediction and the dashed line
the SM one.

The following systematics have been assumed:
 ${{\delta B}/ B}=0.005$ 
 \cite{frank}, where
$B$ denotes the product of the branching ratio for  $W\rightarrow hadrons$
and that for $W\rightarrow leptons$,  $1\%$ for the 
reconstruction efficiency,  $1\%$ for luminosity relative errors.
An angular cut has been imposed on the $W$ scattering angle 
($\vert \cos\Theta\vert \leq 0.95$) and 18 angular bins
have been considered.

An overall detection efficiency of 10\% including the branching ratio
$B=0.29$ and the loss of luminosity from
beamstrahlung \cite{fujii} has been assumed.

 As already
noticed \cite{iddir}, the bigger deviations are away from the forward
region. In the longitudinal channel the deviations are much bigger and
concentrated in the central region.

\begin{figure}
\epsfysize=8truecm
\centerline{\epsffile{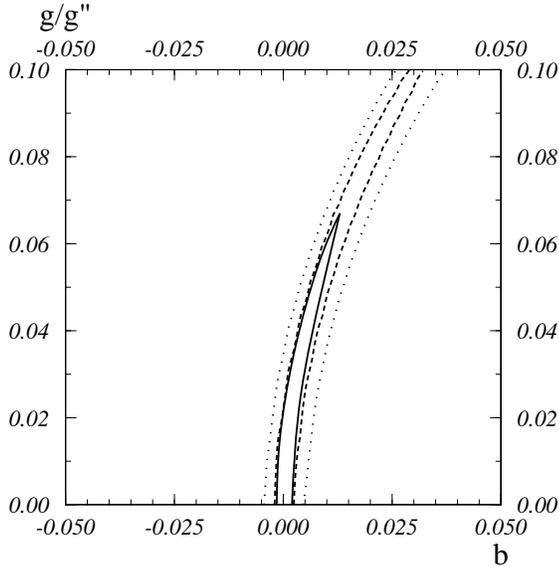}}
\noindent
\caption[ee3]{BESS $90\%$ C.L. contours 
in the plane $(b,g/\gs)$ for  $\sqrt s=500~GeV$ and $M_V=600~GeV$.
The dotted  line corresponds to the bound from
the $WW$ differential cross section, the dashed  line to the bound  
from $W_{L}W_{L}$ differential cross section, and
the continuous line to the bound combining the differential
$W_{L,T}W_{L,T}$ cross sections and $WW$ left-right asymmetries.}
\label{ee3}
\end{figure}

For a collider at
$\rs=500~GeV$ the results are illustrated in Fig. \ref{ee3}. 
 This figure illustrates the 90\% C.L. allowed regions
for $M_V=600~GeV$
obtained by considering the unpolarized $WW$ differential cross-section
(dotted line), the $W_LW_L$ cross section (dashed line),
and the combination of the left-right asymmetry with all the
differential cross-sections for the different final $W$ polarizations
(solid line). We see that
 measurements of cross-sections for polarized $W$ allow
to  get important restrictions
with respect to LEP. 

For colliders with $\rs=1,~2~TeV$ and for $M_V=1.2~{\rm and}~2.5~TeV$
respectively, the allowed region, combining all the observables,
reduces in practice to a line and the analysis is better discussed in the
plane $(M_V,g/\gs)$, as we shall see later on. Therefore, even the
unpolarized $WW$ differential cross section measurements can improve the
bounds.

\begin{figure}
\epsfysize=8truecm
\centerline{\epsffile{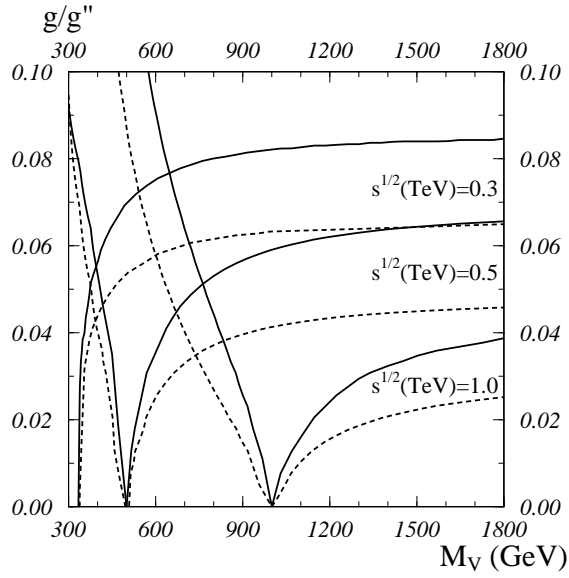}}
\noindent
\caption[ee4]{BESS $90\%$ C.L. contours in the plane $(M_V,g/\gs)$ for 
         $\sqrt s=0.3,~0.5,~1~TeV$, $L=20~fb^{-1}$ and $b=0$.
       The solid line corresponds to the bound from
       the unpolarized $WW$ differential cross section, 
       the dashed line to the bound  
       from  all the polarized
         differential cross sections $W_{L}W_{L}$, $W_{T}W_{L}$,
      $W_{T}W_{T}$   combined with 
       the $WW$ left-right asymmetries.
       The lines give the upper bounds on $g/\gs$.}
\label{ee4}
\end{figure}

In Fig. \ref{ee4} we show the restrictions in the plane $(M_V,g/\gs)$ for
three different choices of the collider energy, assuming the
same integrated luminosity of 20 $fb^{-1}$. It is  a 90\% C.L. contour
 for $\sqrt{s}=0.3,~0.5,~1~TeV$ and $b=0$.
The solid line represents the upper bound on $g/\gs$ from the 
 unpolarized $WW$ differential cross-section, the dashed line
from  the combination of the left-right asymmetry with all the
differential cross-sections for the different final $W$ polarizations. 

\begin{figure}
\epsfysize=8truecm
\centerline{\epsffile{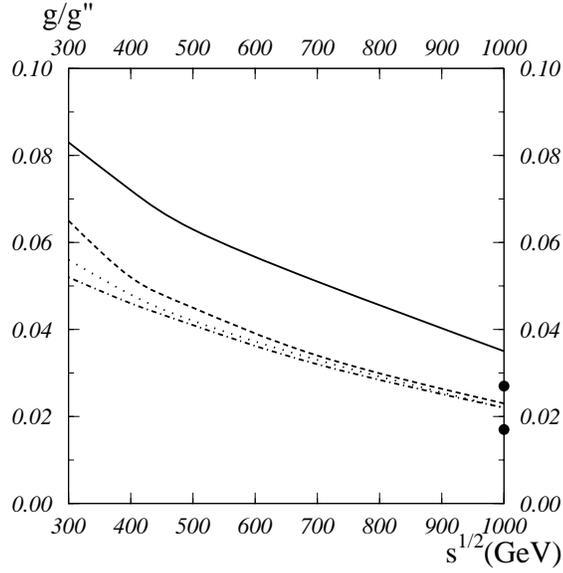}}
\noindent
\caption[ee5]{BESS $90\%$ C.L. contours in the plane $(\sqrt{s},g/\gs)$ for 
         $M_V=1.5~TeV$, $b=0$ and
       $L=20~fb^{-1}$ from the unpolarized
       $WW$ differential cross section (solid line), from  the 
       $W_{L}W_{L}$ differential cross section
       (dashed line), from all the 
       differential cross sections for $W_{L}W_{L}$, $W_{T}W_{L}$, 
       $W_{T}W_{T}$ combined with the $WW$ left-right 
       asymmetries (dotted line) and from all the WW and fermion
       observables with $P=0.5$ (dash-dotted line). 
       The black dots are the bounds for 
       the unpolarized $WW$ differential cross section and from
       all the WW and fermion observables by considering 
       $\sqrt{s}=1~TeV$ and $L=80~fb^{-1}$.
       The lines give the upper bounds on $g/\gs$.}
\label{ee5}
\end{figure}

In Fig. \ref{ee5} the 90\% C.L. upper bound on $g/\gs$ are
 shown for $M_V=1.5~TeV$ and $b=0$
as a function of the center of mass of the LC for
20$fb^{-1}$. The lines correspond to unpolarized $WW$
 differential cross-section (solid), $W_LW_L$ 
differential cross-section (dashed),
the combination of the left-right asymmetry with all the
differential cross-sections for the different final $W$ polarizations
(dotted line), all the $W$ and fermion observables (dash-dotted). 
The blacks dots correspond to 80$fb^{-1}$ of integrated luminosity.

\begin{figure}
\epsfysize=8truecm
\centerline{\epsffile{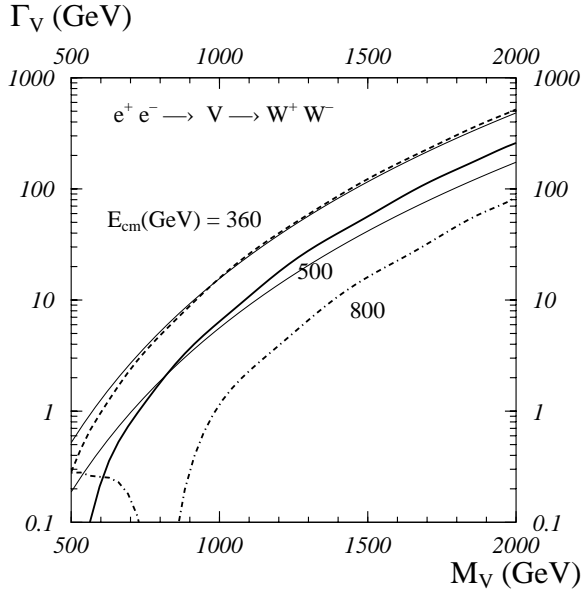}}
\noindent
\caption[gamma] {$90\%$ C.L. contour in the $(M_V,\Gamma_V)$ plane
from measurements of differential cross sections and left-right
asymmetries; $W$ polarizations are reconstructed from the decay in lepton
and quark jets. Energy and luminosity are the following:
$\sqrt{s}= 360~GeV,~L=10~fb^{-1}$ (dashed)
 $\sqrt{s}= 500~GeV,~L=20~fb^{-1}$ (solid)
and  $\sqrt{s}= 800~GeV,~L=80~fb^{-1}$ (dash-dotted).
 The lower narrow solid line
is the limit from LEP measurements.
 The upper  narrow solid line is obtained by 
considering the deviation with respect to the SM prediction and
assuming the LEP errors for the corresponding observables.
}
\label{gamma}
\end{figure}

In  Fig. \ref{gamma} the regions in the parameter space
$(M_V,\Gamma_V)$ which can be probed at 
 $\sqrt{s}=360~GeV$, $L=10~fb^{-1}$ (dashed),
 $\sqrt{s}=500~GeV$, $L=20~fb^{-1}$ (continuous) and
 $\sqrt{s}=800~GeV$, $L=50~fb^{-1}$ (dot-dashed) 
by measuring $W_{L,T} W_{L,T}$ differential cross
sections and left-right asymmetries are shown. 
 The lower narrow solid line
is the limit from LEP  measurements.
 The upper  narrow solid line is obtained by 
considering the deviation with respect to the SM prediction and
assuming the LEP errors for the corresponding observables.

For instance for  $\sqrt{s}=500~GeV$, one is sensitive 
for $M_V=2~TeV$ to $\Gamma_V\geq 250~GeV$, 
for $M_V=1.5 ~TeV$ to $\Gamma_V\geq 60~GeV$, in agreement
with Barklow results \cite{bark}.

In conclusion measurements of  cross sections with  
 different final $W$ polarizations at a LC with $\sqrt{s}=500~GeV$ improve
LEP bounds up to $M_V\sim 800~GeV$. At
$\sqrt{s}=
800~GeV$ the sensitivity exceeds the LEP bound for all values of
$M_V$.

 This is  a more conservative result
with respect to \cite{bark} and \cite{werner} mainly because the BESS
model takes into account the mixing of the new vector bosons to the
usual $\gamma$, $W$ and $Z$ gauge bosons and as a consequence
 LEP bounds have to be  considered.

\section{Results for the Degenerate BESS}
\label{debess}

In this Section we present an analysis done for the model
with  two new resonances,
vector and axial-vector degenerate in mass (Degenerate BESS model).
As already observed the model evades the LEP bounds and therefore
low masses are still allowed for these new resonances.

With respect to the previous analysis where the $WW$ channel
is the dominant one, here 
the fermion channel is much more relevant.

The comparison with LEPI bounds shows that LEPII do not
improve considerably the existing limits \cite{lep2}.

To further test the model the following  
 options for a high energy $e^+e^-$ collider have been studied: 
$\sqrt{s}=360~ GeV$ with an integrated luminosity of $L=10 fb^{-1}$,
$\sqrt{s}=500~ GeV$  $L=20 fb^{-1}$,
$\sqrt{s}=800~GeV$  $L=50 fb^{-1}$ and 
$\sqrt{s}=1~ TeV$  $L=80 fb^{-1}$.

\begin{figure}
\epsfysize=8truecm
\centerline{\epsffile{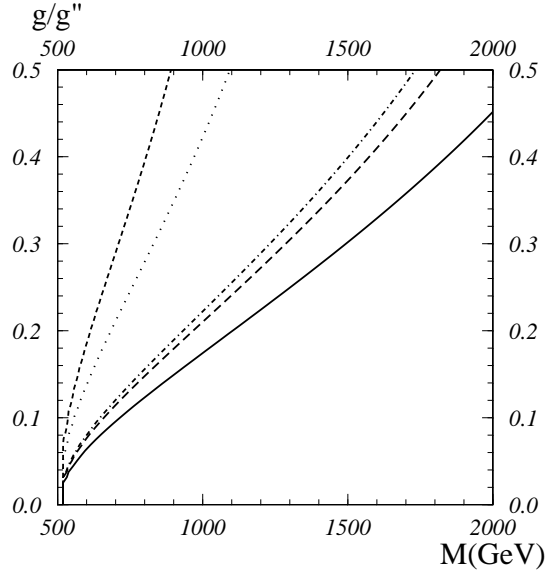}}
\noindent
\caption[fig5p]{Degenerate BESS
 90\% C.L. contours on the plane ($M$, $g/g''$) from 
$e^+e^-$
at $\sqrt{s}=500~GeV$ with an integrated luminosity of $20 fb^{-1}$ for 
various 
observables. The dash-dotted line 
represents the limit from $\sigma^h$ with an assumed error of 2\%; the dashed 
line near  the latter  is $\sigma^\mu$ (error 1.3\%); the dotted 
line 
is $A_{FB}^\mu$ (error 0.5\%); the uppermost dashed line is $A_{FB}^b$ (error 
0.9\%). The continuous line represents the combined limits.}
\label{fig5p}
\end{figure}

\begin{figure}
\epsfysize=8truecm
\centerline{\epsffile{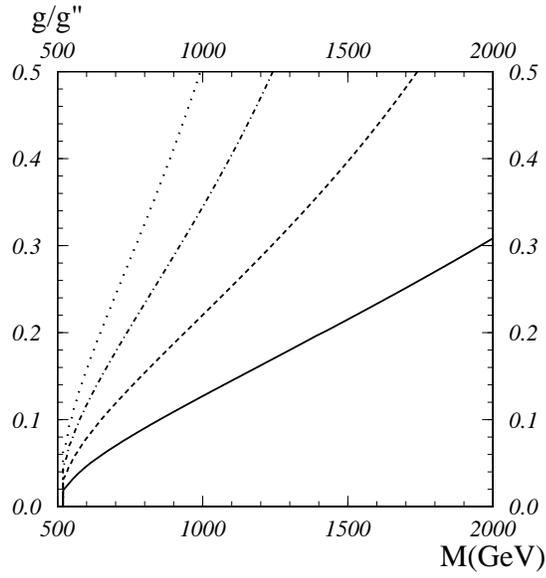}}
\noindent
\caption[fig6p]{Degenerate BESS 90\% C.L. contour on the plane 
($M$, $g/g''$) from 
$e^+e^-$
at $\sqrt{s}=500~GeV$ with an integrated luminosity of $20 fb^{-1}$ and a 
polarization $P= 0.5$ for various observables. The dash-dotted line 
represents 
the limit from $A_{LR}^\mu$ with an 
assumed error of 0.6\%; the dashed line is $A_{LR}^h$ (error 0.4\%); the 
dotted 
line is $A_{LR}^b$ (error 1.1\%). The continuous line 
is obtained by combining
the polarized and the unpolarized observables:
 $\sigma^h$, $\sigma^\mu$, 
$A_{FB}^\mu$, $A_{FB}^b$, $A_{LR}^\mu$, $A_{LR}^h$, $A_{LR}^b$.}
\label{fig6p}
\end{figure}

In Fig. \ref{fig5p}  the $90\% $ C.L. contour on the plane 
($M, g/\gs$) from $e^+e^-$ at $\sqrt{s}=500~ GeV$ with 
an integrated luminosity of $20 fb^{-1}$ for various 
observables are compared. The dash-dotted line 
represents the limit from $\sigma^h$; the dashed 
line near the  latter is $\sigma^\mu$, the dotted line 
is $A_{FB}^\mu$ and the uppermost dashed line is $A_{FB}^b$.
 It is evident that more stringent bounds come from the cross
section measurements. Asymmetries give less restrictive bounds because
of a compensation between  $L_3$ and $R_3$ exchange.
By combining all the deviations in the previously considered
observables we get the limit shown in Fig. \ref{fig5p} by
the continuous line.

Polarized electron beams
allow to get further limit on the parameter space as shown in Fig.
 \ref{fig6p}.
The error on the measurement of the polarization has been neglected
and the  
polarization $P= 0.5$.
The dash-dotted line represents the limit from $A_{LR}^\mu$, 
the dashed line from $A_{LR}^h$, the dotted line from $A_{LR}^b$.
The 
bound shown in Fig.  \ref{fig6p} by the continuous
line  is  obtained by
combining all the polarized and unpolarized beam observables. 
In conclusion even without polarized beams
 a substantial improvement with respect to the LEP bound is obtained.

As expected, increasing the energy of the collider and rescaling the
integrated luminosity result in stronger bounds on the 
parameter space (the comparison of different LC's
results is made in Section \ref{lclhc}).

Let us finally consider   the $WW$ final state.

An angular cut has been imposed on the $W$ scattering angle 
($\vert \cos\Theta\vert \leq 0.95$) and 18 angular bins
have been considered. An overall detection efficiency of 10\%,
including the branching ratio $B=0.29$ and the loss of luminosity
by beamstrahlung have been assumed.

All
the new bounds coming from the $WW$ channel do not alter the strong limits
obtained from the fermion final state. This is  because, as
 already noticed, the degenerate model has no strong 
enhancement of the $W_LW_L$ channel, present in the usual
strong electroweak models.

\section{Final state rescattering}
\label{fsin}

A different approach \cite{peskin,iddir} has also been  presented.
It consists in
considering the effects of a strong interacting
electroweak sector through final state rescattering.
 The final state  $W$ rescattering
is described making use of  the Omn\'es function $F_T$:
\be
M(e^+e^-\rightarrow W^+_LW^-_L)=M_0(e^+e^-\rightarrow W^+_LW^-_L)
F_T
\ee
where 
\be
F_T= \exp[\frac 1 \pi \int_0^\infty ds^\prime \delta(s^\prime) 
\{\frac {1}{s^\prime -s -i\epsilon}- \frac {1}  
{s^\prime} \}]
\ee
with $\delta$ the phase shift depending on the dynamics.
For instance Barklow \cite{bark} has considered
\be
\delta(s)= \frac {1}{96 \pi}\frac {s}{v^2}+\frac {3\pi}{8}
\big [ \tanh \left ( \frac {s-M_V^2}{M_V\Gamma_V}\right )+1\big ]
\ee
where $M_V$ and $\Gamma_V$ are mass and width of the vector resonance.

Note that for infinite $M_V$ one recovers the Low Energy
Theorem (LET) results.

A full final state helicity analysis on $e^+e^-\rightarrow W^+W^-$ 
was done \cite{bark} by considering  one $W$ decaying
hadronically and the second one leptonically.
The following multidifferential cross section has been considered
\be \frac {d\sigma (\cos\Theta, \cos\theta,\phi, \cos\bar\theta,\bar\phi)}
{d\cos\Theta d\cos\theta  d\phi d\cos\bar\theta d\bar\phi}
\label{multi}
\ee
where the angle $\Theta$ is defined to be the angle between the
initial state $e^-$  and the $W^-$ in the
$e^+e^-$ rest frame, $\theta$ and $\phi$ the polar and azimuthal
angles
of the fermion $f_1$ in the $W^-$ rest frame ($W^-\rightarrow
f_1\bar f_2$), $\bar \theta$ and $\bar \phi$ the polar and azimuthal
angles
of the antifermion $\bar f_4$ in the $W^+$ rest frame ($W^-\rightarrow
f_3\bar f_4$). These five angles are reconstructed by measuring
energies and angles of decay leptons and quarks. The masses of
the two $W$ are reconstructed by a kinematic fit of the momentum
of the lepton $\vec P_l$ and the four momentum of the
hadronically decaying $W$, $(E_H,\vec P_H)$, with
\be
E_H+E_l+\sqrt{\vec P^2_\nu}=\sqrt{s}
\ee
and
\be
\vec P_\nu=-(\vec P_l+\vec P_H)
\ee
where $E_l$ is the lepton energy.

Two cuts are imposed: $\vert \cos\Theta\vert< 0.9$ in order to ensure
the event to be within the detector, and the
$W^+W^-$ invariant mass is required to be within few $GeV$
of the $e^+e^-$ center of mass energy.

\begin{figure}
\epsfysize=8truecm
\centerline{\epsffile{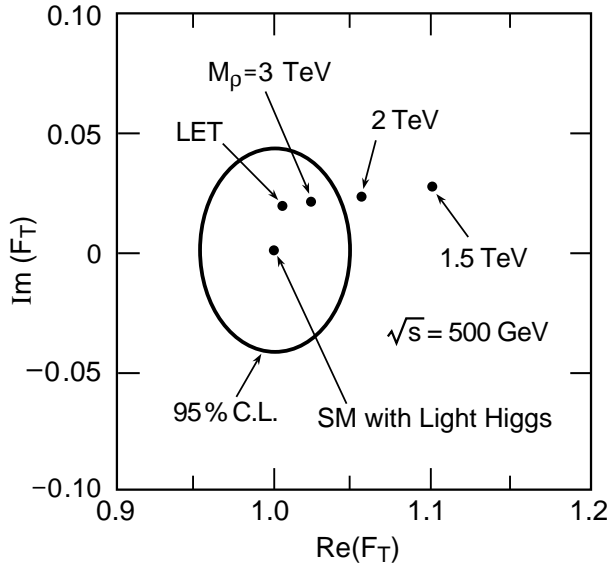}}
\noindent
\caption[bark]{$95\%$ C.L. contour for $ReF_T$ and $ImF_T$
 for $\sqrt{s}=500~GeV$ and predictions  for
various techni-$\rho$ masses, from \cite{bark}.}
\label{bark}
\end{figure}

\begin{figure}
\epsfysize=8truecm
\centerline{\epsffile{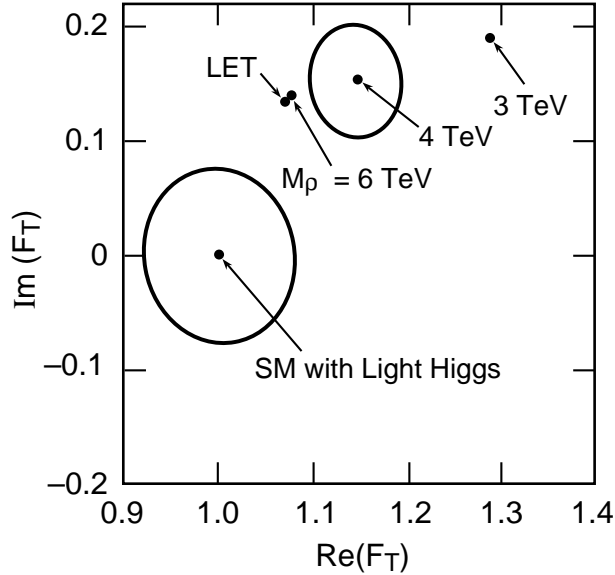}}
\noindent
\caption[bark2]{$95\%$ C.L. contour for $ReF_T$ and $ImF_T$ 
 for $\sqrt{s}=1.5~TeV$
 and predictions  for
various techni-$\rho$ masses, from \cite{bark}.}
\label{bark2}
\end{figure}

A maximum log likelihood analysis  is performed and the 
contour
for the real and imaginary parts of $F_T$ are shown in Figs. \ref{bark},
\ref{bark2}. Beamstrahlung, bremstrahlung and finite $W$ width
effects are
included.

Figure \ref{bark} shows the 95\% C.L. contour for the
real and imaginary parts of $F_T$ at $\sqrt{s}=500~GeV$.
  Also
indicated are values of $F_T$ for various techni-$\rho$ masses.
 We see that the LC at
$\sqrt{s}=500~GeV$ can exclude techni-$\rho$ masses up to about $2.5~TeV$ and
can discover techni-$\rho$ resonances with masses of more than $1.5~TeV$. The
significance of the $1.5~TeV$ techni-$\rho$ signal would be $6.7\sigma$. A
$1.0~TeV$ techni-$\rho$ would produce a $17.7\sigma$ signal. 
Notice that these values for the masses of the resonances are already
excluded by the LEP results if one assumes a QCD rescaled technicolor.

In Fig. \ref{bark2} 
the chosen parameters are $\sqrt{s} = 1.5~ TeV$
     with $190~ fb^{-1}$ .  The contour about 
  the light Higgs value is a $95\%$ C.L. contour; the contour about 
the point, corresponding to the technicolor with $M_\rho = 4~ TeV$,
 is a $68\%$ C.L. contour.

 The non-resonant
LET point is well outside the light Higgs $95\%$ C.L. region
and corresponds to a 4.5$\sigma$ signal.

 The results are obtained assuming $\Gamma_V/M_V=0.22$. However they
are not
much dependent on this assumption.
 In this model, when the vector resonance
mass is taken to infinity, its effect on the form factor decreases and
what is left behind is the residual scattering predicted by the LET. 
The values for high-mass techni-$\rho$ indicate this decoupling.
  A $2~TeV$ techni-$\rho$ would produce a $37\sigma$ signal. 

A similar analysis was done also in \cite{werner}. These authors
 consider
again a phase shift corresponding to a vector resonance and
investigate the cross section with
suitable cuts to enhance
 the ratio $W_LW_L$  to $W_TW_T$. Using
the cut $\cos\Theta\leq 0.95$ they study the ratio
\be
R=\frac{N(-0.95\leq \cos\Theta\leq \cos\Theta_0)}
{N(-0.95\leq \cos\Theta\leq 0.95)}
\ee
choosing $\Theta_0=0.6\pi$. This last cut enhances the 
$W_LW_L$ signal as it is clear from the differential distribution of
Fig. \ref{ww3}.
 Therefore with an integrated luminosity of $10fb^{-1}$ they find
that a vector resonance of $M_V=1~TeV$ and $\Gamma_V=54~GeV$
gives a $6\sigma$ effect
already at $\sqrt{s}=500~GeV$ and that even without the final state
helicity analysis it can be discovered. 

 A similar analysis was done in \cite{MHI} using $\sqrt{s}=500~GeV$
and $L=50~fb^{-1}$; with this choice the sensitivity to $M_V$
exceeds $2~TeV$.

\section{Anomalous trilinear gauge boson couplings}
\label{anom}
Another interesting sector of gauge boson interaction is given by
the triple
gauge boson vertices. Observing deviations of these vertices with
respect to their SM values can give some hints of new physics.
 
We review in this section the effective parametrization for anomalous
gauge boson couplings and their measurements  at future LC's.

 The  most general  CP invariant
effective Lagrangian describing such terms can be written
as \cite{zepp,appwu}
\bea
\f 1 {g_{WWV}}\LL_{WWV}&=&i g_1^V (W_{\mu\nu}^+ W^{-\mu} V^{\nu}-
W_{\mu\nu}^- W^{+\mu} V^{\nu})+ik_V  W^+_{\mu} W^-_{\nu}
V^{\mu\nu}\nn\\
&+&i \f {\lambda_V} {\Lambda^2}W_{\mu\nu}^+
W^{-\nu}_{~~~\rho}V^{\rho\mu}
+g_5^V\eps^{\mu\nu\rho\sigma} [W^+_\mu (\de_\rho W^-_\nu)
-(\de_\rho W^+_\mu)W^-_\nu]V_\sigma
\label{effl}
\eea
where $V\equiv \gamma$ or $Z$, $W_{\mu\nu}^\pm =\de_\mu W_\nu^\pm -
\de_\nu W_\mu^\pm$, $V_{\mu\nu}^\pm =\de_\mu V_\nu^\pm -
\de_\nu V_\mu^\pm$.  $\Lambda$ denotes the scale of the new physics
responsible for the symmetry breaking.

The couplings $g_{WW\gamma}$ and $ g_{WWZ}$ are given by
\be
g_{WW\gamma}=-e~~~g_{WWZ}= -e \f {c_\theta} {s_\theta}
\ee
with $\s$ defined in eq. (\ref{sint}).
It should be noted that in general these new couplings, that we assume
as constants,
are momentum dependent form-factors. In the SM one has
$g_1^V=k_V=1$ and $g_5^V=\lambda_V=0$. 
 
 The magnetic dipole moment $\mu_W$ and the
quadrupole moment $Q_W$ of the $W$ are
given by
\be
\mu_W=\f e{2M_W} (1+k_\gamma+\lambda_\gamma),
~~~
Q_W=- \f {e}{M_W^2} (k_\gamma-\lambda_\gamma)
\ee
Using eqs. (\ref{lp2}) and (\ref{lp4}) one finds
\bea
g_1^Z-1&=&\f {g^2}{c_{2\theta}}\beta_1+\f{g^2 s_\theta^2}{c_\theta^2
c_{2\theta}}\alpha_1+\f {g^2}{c_\theta^2}\alpha_3\nn\\
g_1^\gamma-1&=&0\nn\\
k_Z-1&=&\f {g^2}{c_{2\theta}} \beta_1 +\f{g^2 s_\theta^2}{c_\theta^2
 c_{2\theta}}\alpha_1+g^2 \f {s_\theta^2}{c_\theta^2}(\alpha_1-
 \alpha_2)+
g^2(\alpha_3-\alpha_8+\alpha_9)\nn\\
k_{\gamma}-1&=&g^2(-\alpha_1+\alpha_2+\alpha_3
-\alpha_8+\alpha_9)\nn\\
g_5^Z&=&\f {g^2 } {c_\theta^2} \alpha_{11}\nn\\
g_5^\gamma&=&\lambda_Z=\lambda_\gamma=0
\eea
In the previous expressions there are also contributions coming
from gauge boson two point functions.

In the literature there is also a second parametrization \cite{bmt}.
 Assuming
CP  invariance and for the electromagnetic interaction
also C invariance, one has  \cite{bmt}
\bea
\LL&=&-ie \{[(W_{\mu\nu}^- W^{+\nu} -
W_{\mu\nu}^+ W^{-\nu}) A^{\mu}+
(1+x_\gamma)F^{\mu\nu} W^+_\mu W^-_\nu]\nn\\
&+&(\f {c_\theta}{s_\theta} +\delta_Z)[(W_{\mu\nu}^- W^{+\nu} -
W_{\mu\nu}^+ W^{-\nu}) Z^{\mu}
+(1+\f {x_Z}{\f {\dd c_\theta}{\dd s_\theta} +\delta_Z})Z^{\mu\nu} W^+_\mu
W^-_\nu]
\nn\\
&+&\f 1 {M^2_W} (y_\gamma F^{\nu\lambda} +y_Z Z^{\nu\lambda})
W^+_{\lambda\mu}W^{-\mu}_{~~~\nu}\}\nn\\
&+&\f {e z_Z}{M^2_W} \de_\alpha \tilde Z_{\rho\sigma}
(\de^\rho W^{-\sigma} W^{+\alpha} -
\de^\rho W^{-\alpha} W^{+\sigma}+(\alpha\leftrightarrow\sigma))
\label{effl2}
\eea
where
\be
 \tilde Z_{\rho\sigma}=\f  1 2 \eps_{\rho\sigma\alpha\beta}
Z^{\alpha\beta}
\ee

The conversion is given by (assuming in eq. (\ref{effl}) $\Lambda\equiv M^2_W$)
\be
x_\gamma= \Delta k_\gamma~~
\delta_Z= \f {c_\theta}{s_\theta} \Delta g_Z^1~~
x_Z= \f {c_\theta}{s_\theta} ( \Delta k_Z-\Delta g_Z^1)~~
y_\gamma=\lambda_\gamma~~
y_Z= \f {c_\theta}{s_\theta} \lambda_Z
\ee
with
\be
\Delta k_\gamma=k_\gamma -1~~
\Delta k_Z=k_Z-1~~
\Delta g_1^Z= g_1^Z-1
\ee
For what concerns the last term of eq. (\ref{effl}) and
 eq. (\ref{effl2}) the comparison is less simple, because
the two operators are of dimension four and six respectively 
and therefore a $q^2$ dependence appears \cite{bouj}.

The best present limits on anomalous couplings come from hadron
collider
experiments. Preliminary CDF and D0 results using $W\gamma$, $WW$
and $WZ$ production give $95\%$ C.L.
bounds on the parameters which are of order unity \cite{cdfan,d0an}.
 LEPII can
improve these bounds to $O(10^{-1})$ \cite{lep2an} by considering
the process $e^+e^-\rightarrow W^+W^-$ with three main topologies:
those in which both $W$ decay hadronically, in which one decays
hadronically, the other leptonically, and in which both decay
leptonically. Anomalous couplings can be measured from the
multidifferential
cross section (\ref{multi}) by using  a maximum likelihood method.
This analysis has been also repeated for future LC's, by
considering events with $\vert\cos\Theta\vert <0.8$ and the
topology in which one  $W$  decays
hadronically, the other leptonically \cite{bmt,bouj,barkww,bile,bela,paver}.
The common result is that one can reach precisions of  $O(10^{-3})$,
at a LC of $\sqrt{s}=500~GeV$ especially if beam polarization
is available.
 
In a recent work the analysis of the anomalous trilinear gauge couplings
\cite{dafne} has been performed by considering at the same time also
fermion anomalous couplings, because in general there are strong
correlations.
The corrections to the fermion couplings, assuming
universality, are parametrized in terms
of the $\eps_3$ parameter (in many models the $SU(2)$ symmetry
guarantees $\eps_1=\eps_2=0$) and therefore existing bounds
on $\eps_3$ give also limits on anomalous fermion couplings.

A recent review on anomalous gauge couplings is given in \cite{barkan}.

\section{Fusion subprocesses and vector resonances }
\label{fusi}

Another mechanism to produce $W^+ W^- $ pairs is the fusion of a pair
of ordinary gauge bosons, each being initially emitted from an electron or a 
positron. This process allows to study $WW$ scattering and therefore
possible strong interacting electroweak breaking. This was first
investigated in \cite{tofi,mura,kuri,hawai}.

In the so called effective $W$ approximation the initial $W,Z,\gamma$
are assumed to be real and the cross section for producing a  $W^+ W^- $ pair
is obtained by a convolution of the fusion subprocess with the luminosities of
the initial $W,Z,\gamma$ inside the electrons and positrons.
There are two fusion subprocesses which contribute to produce $W^+ W^- $ pairs.
The first one is 
\be
e^+e^-
\rightarrow W^+_{L,T} W^-_{L,T} e^+ e^-
\ee
 It is mediated by $W^{\pm}$ 
and $V^{\pm}$ exchanges in the $t$ and $u$ channels. 
The second fusion subprocess   is 
\be
e^+e^-
\rightarrow W^+_{L,T} W^-_{L,T}{\overline \nu} \nu
\ee 
It is mediated by $\gamma, Z$ and $V^{0}$ exchanges in the 
$s$ and $t$ channels. Both  processes get a contribution from the gauge 
boson fourlinear couplings.

In principle, the fusion processes are interesting because they allow 
to study a wide range of mass spectrum for the $V$ resonance for one
given $e^+e^-$ c.m. energy.

In the $e^+e^-$ center-of-mass frame the invariant mass distribution
$d\sigma/dM_{WW}$ reads
\bea
{\frac {d\sigma} {dM_{WW}}}&=\frac {\dd 1}{ \dd 4\pi s} \frac {\dd 1}
{\dd  M^2_{WW}}
{\dd \sum_{i,j}}{\dd \sum_{l1,l2}} 
&\int^{M_{WW}^2/4}_{(p_T^2)_{min}} 
d{p_T^2}\int^{-\log{\sqrt{\tau}}}_{\log{\sqrt{\tau}}}dy ~f^{l1}_i(\sqrt{\tau}
\e^{y})f^{l2}_j(\sqrt{\tau}\e^{-y})\nn\\
&&\cdot {p'\over p} {1\over\sqrt{M^2_{WW}-4 p_T^2}}
|M({V^{l1}_i V^{l2}_j} \rightarrow W^+_{l3}
W^-_{l4})|^2
\eea
where  $p_T$ is the transverse momentum of the outgoing $W$,
$\tau=M_{WW}^2/s$, 
$p$ and $p'$ are the absolute values of the three momenta 
for incoming and outgoing pairs
of vector bosons: $p=(E_1^2-M_1^2)^{1/2}=(E_2^2-M_2^2)^{1/2}$ and 
$p'=(\sqrt{M_{WW}}/2) (1-4 M_W^2/M_{WW})^{1/2}$ 
with $E_i$ the 
energy of the vector boson $V_i$ with mass $M_i$ and helicity $l_i$.
The structure functions $f$ appearing in the previous formula are
\bea
f^+(x)&=&{{\alpha_{em}}\over{4\pi}} {{[(v+a)^2+(1-x)^2(v-a)^2]}\over{x}} 
\log{s\over{M^2}}\nn\\
f^-(x)&=&{{\alpha_{em}}\over{4\pi}} {{[(v-a)^2+(1-x)^2(v+a)^2]}\over{x}} 
\log{s\over{M^2}}\nn\\
f^{0}(x)&=&{{\alpha_{em}}\over{\pi}} (v^2+a^2) {{1-x}\over{x}}
\label{169}
\eea
and represent the probability of having inside the electron a vector boson of 
mass $M$ with fraction $x$ of the electron energy.
 In eq. (\ref{169})
 $v$ and $a$ 
are the vector and axial-vector couplings of the gauge bosons to fermions. 
We present here the results for the BESS model.
For the photon, $v=-1$ and $a=0$; for $Z$, $v=v^f_Z$ and 
$a=a^f_Z$ 
as given in eq. (\ref{cneubess}); and, for $W$, $v=a=a_W$ as
given in eq. (\ref{cchbess}). The quadrilinear 
couplings are
\bea
&g_{WWWW}=& {{e^2}\over{\sin^2\theta}} {{\cos^4\phi}\over{\cos^2\psi}} 
+ {{\gs^2}\over{4}} \sin^4\phi\nn\\
&g_{WWZZ}=&{{e^2}\over{\cos^2 \phi}} g^2_{ZWW}+
{{\gs^2}\over{4}} \sin^2\phi \sin^2\xi \cos^2\psi \nn\\
&g_{WW\gamma\gamma}= &e^2 \cos^2\phi + {{\gs^2}\over{4}} \sin^2\phi \sin^2\psi
=e^2
\nn\\
&g_{WWZ\gamma}=& e^2 g_{ZWW} - {{\gs^2}\over{4}} \sin^2\phi \sin\xi 
\sin\psi \cos\psi
\eea
with $g_{ZWW}$ as given in eq. (\ref{gzww})
 and the mixing angles $\csi$, $\psi$
and $\phi$ in eqs. (\ref{lcsi}), (\ref{lpsi}) and (\ref{lphi}).

The amplitudes of the vector bosons scattering processes within the BESS 
model are:
\bea
M(W^+_1W^-_2 \rightarrow W^+_3W^-_4)&=&-i e^2 {f_s\over s} -i e^2 
g^2_{ZWW} {f_s\over {s-M^2_Z}}\nn\\
& &-i e^2 g^2_{VWW} {f_s\over {s-M^2_V+i \Gamma_V M_V}} 
+(s\rightarrow t)\nn\\
&& + ig_{WWWW} [2(\eps_1 \cdot \eps_4^*)(\eps_2 \cdot \eps_3^*)
-(\eps_1 \cdot \eps_2) (\eps_3^* \cdot \eps_4^*)\nn\\
&&-(\eps_1 \cdot \eps_3^*)(\eps_2 \cdot \eps_4^*)]
\label{MWW}
\eea
\bea
M(\gamma_1\gamma_2 \rightarrow W^+_3W^-_4)&=&-i e^2 {h_t\over {t-M^2_W}} 
+(t\rightarrow u)\nn\\
&& - ig_{WW\gamma\gamma} [2(\eps_1 \cdot \eps_2)(\eps_3^* \cdot \eps_4^*)
-(\eps_1 \cdot \eps_3^*) (\eps_2 \cdot \eps_4^*)\nn\\
&&-(\eps_1 \cdot \eps_4^*)(\eps_2 \cdot \eps_3^*)]
\label{MGG}
\eea

\bea
M(\gamma_1 Z_2 \rightarrow W^+_3W^-_4)&=&-i e^2 g_{ZWW} {h_t\over {t-M^2_W}} 
+(t\rightarrow u)\nn\\
& &- ig_{WWZ\gamma} [2(\eps_1 \cdot \eps_2)(\eps_3^* \cdot \eps_4^*)
-(\eps_1 \cdot \eps_3^*) (\eps_2 \cdot \eps_4^*)\nn\\
&&-(\eps_1 \cdot \eps_4^*)
(\eps_2 \cdot \eps_3^*)]
\label{MGZ}
\eea

\bea
M(Z_1 Z_2 \rightarrow W^+_3W^-_4)&= &-i e^2 
g^2_{ZWW} {h_t\over {t-M^2_W}}\nn\\
&& -i e^2 g^2_{VWW} {h_t\over {t-M^2_V+i 
\Gamma_V M_V}} 
+(t\rightarrow u)\nn\\
& &- ig_{WWZZ} [2(\eps_1 \cdot \eps_2)(\eps_3^* \cdot \eps_4^*)
-(\eps_1 \cdot \eps_3^*) (\eps_2 \cdot \eps_4^*)\nn\\
&&-(\eps_1 \cdot \eps_4^*)
(\eps_2 \cdot \eps_3^*)]
\label{MZZ}
\eea
where
\bea
f_s&=&[(\eps_1 \cdot \eps_2) (p_2-p_1)_{\lambda} -2(\eps_1 \cdot p_2)
\eps_{2 \lambda} + 2 (\eps_2 \cdot p_1) \eps_{1 \lambda} ] \nn\\
&&\cdot [(\eps_3^* \cdot \eps_4^*) (p_3-p_4)^{\lambda} -2(\eps_4^* \cdot p_3) 
\eps_{3}^{* \lambda} + 2 (\eps_3^* \cdot p_4) \eps_4^{* \lambda} ]
\eea
and $f_t$ can be deduced from $f_s$ with the substitution $p_2 \leftrightarrow
-p_3$ and $\eps_2 \leftrightarrow \eps^*_3$, while
\bea
h_t&=&[2(\eps_1 \cdot p_3) \eps^*_{3 \lambda} - (\eps_1 - \eps^*_3)
(p_3 +p_1)_{\lambda}+2(p_1 \cdot \eps^*_3) \eps_{1 \lambda}] \nn\\
&&\cdot [2(\eps_2 \cdot p_4) \eps^{* \lambda}_4-(\eps_2 \cdot \eps_4^*)
(p_4+p_2)^{\lambda}+2(p_2 \cdot \eps^*_4) \eps^{\lambda}_2] \nn\\
&&+ (\eps_1 \cdot \eps^*_3) (p_1^2 - p_3^2) (\eps_2 \cdot \eps^*_4) 
(p^2_2 - p_4^2)/M_W^2
\eea
and $h_u$ can be deduced from $h_t$ with the substitution $p_3 \leftrightarrow
p_4$ and $\eps_3^* \leftrightarrow \eps^*_4$.

In eqs. (\ref{MWW}-\ref{MZZ}) $g_{ZWW}$ and $g_{VWW}$ are given in eq.
 (\ref{gzww}) and (\ref{gvww})
respectively and $\Gamma_V$ is the width of the $V$ resonance.

For comparison, the amplitudes within the SM
 can be obtained from eqs. (\ref{MWW}-\ref{MZZ})
by taking all the trilinear and quadrilinear vector boson couplings in the
limit $\gs\to\infty$ and $b\to 0$ and adding the contribution due to the
Higgs boson exchange.
 All the non-annihilation graphs 
contributing to the processes
$e^+e^-\rightarrow W^+W^-e^+e^-$ and $e^+e^-\rightarrow W^+W^-\nu\bar \nu$ 
(at the order $\alpha_{em}^2$) in which the final $W$'s are emitted from the
electron (positron) legs are not considered. 
This is because  their contribution is expected mostly
in a
kinematical region different from the one in which $p_T\sim
M_W$. This is the typical  momentum of a longitudinal $W$
radiated by an initial electron or positron.

The result of the analysis  for
the BESS model is that there are not significative differences 
with respect to the SM differential cross section
in the case of the process $e^+e^-
\rightarrow W^+W^-e^+e^-$. This is due, first of all, to the
absence of the $s$ channel 
exchange of the $V$ resonance, secondly, 
to the dominance of the $\gamma\gamma$ fusion  contribution, and to the
 fact that in the BESS model 
the couplings of the 
photon to the fermions and to $W^+W^-$ are the  same as in the SM.

Concerning the process
$e^+e^-\rightarrow W^+W^-\nu {\overline \nu}$, the 
differential cross sections $d\sigma/d M_{WW}$ both for the SM
with $M_H=100~GeV$ and for the
BESS model have been evaluated \cite{hawai,desy93}. 
The only channel which turns out to be useful is the one 
corresponding
to longitudinally polarized final $W$'s.
The results of the analysis
are illustrated in \cite{hawai,desy93}
 for two different choices of
the BESS parameters.

In both cases cuts  are  not applied except for $(p_T)_{min}=10~GeV$.

Although theoretically there is a clear difference between the curves 
of the two models, the experimental situation is quite different.

Let us first consider the case of $\sqrt{s}=1.5~TeV$.
By integrating the differential cross section for $500<M_{WW}(GeV)<1500$
and considering an integrated luminosity of 80 $fb^{-1}$ one  obtains
127 $W_L$ pairs for the SM and 158 for the BESS model (with $M_V=1~TeV$, 
$\gs=13$ and $b=0.01$)
corresponding to 
a statistical significance $S/\sqrt{B}$ of 2.7.
 By considering, for comparison with other calculations, 
 the higher luminosity
of 200 $fb^{-1}$ this ratio becomes 4.4.
In this analysis   the branching
ratio $BR(W\to jj)$ in final dijets is not taken into account.

The case of a resonance with
 $M_V=1.5~TeV$, 
$\gs=13$ and $b=0.01$ has been considered for an
energy of  $2 ~TeV$ and by considering an integrated luminosity
of  $20~fb^{-1}$. 
By integrating the differential cross section for $1000<M_{WW}(GeV)<2000$
one gets 7 $W_L$ pairs for the SM and 13 for the BESS model 
corresponding to 
a statistical significance of 2.3 (which increases to 7.2 by considering
200 $fb^{-1}$ ).

The channel corresponding to transverse-longitudinal final $W$'s leads to
a very small bump in the region of the resonance above the SM backgrounds
which is not observable.

In conclusion this  
analysis shows that it is possible to discover new
vector
resonances from the fusion process at high energy LC's, but
a high luminosity of the order $200~fb^{-1}$ is required, in 
agreement with the results of \cite{bargerepem}.

\section{Fusion: comparison of models}
\label{fusi2}

A more detailed analysis, taking  into account not only the $W^+W^-\rightarrow
W^+W^-$ but also $W^+W^-\rightarrow ZZ$, and considering a more
accurate 
choice
of cuts and also the dijet mass resolution in
$W,Z\rightarrow jj$
 was done
in  \cite{bargerepem}.
The authors study not only the vector model, but also the chirally
coupled
scalar (CCS)
 model of Section \ref{ccm}, the LET model, given by the
simplest chiral Lagrangian
and using $K$ matrix unitarization, and the SM Heavy Higgs model.
The calculations for the last two models were done from
the complete SM  and defining  the signal 
as the excess with respect to
the case of $M_H=0$. Calculations for the CCS and the vector models
were done using the effective $W$ approximation.

While in studies at LHC it is necessary
for identification of $W$ and
$Z$  to consider their
leptonic decays which have small branching fractions, at LC's
$W$ and $Z$ bosons are detected via the dijet mode and identified using
the invariant masses $M(W\rightarrow jj)$, $M(Z\rightarrow jj)$ with
an 
identification
probability of the order of $75\%$ for $WW$ and $60\%$ for $ZZ$.

Concerning the background coming from
\be
e^+e^-\rightarrow ZW^+W^-\rightarrow \bar \nu\nu W^+W^-
\ee
it can be reduced using the so called recoil mass or the invariant mass
of all the final particles excluding $WW$. The recoil mass spectrum 
has a peak at $M_Z$,
and therefore a cut as $M_{rec}>200~GeV$ suppresses
this background.

To isolate $WW$ scattering at high energy, the authors of
\cite{bargerepem} require also
high invariant mass $M_{WW}$, high transverse momentum $p_T(W)$, and
a large angle with respect to the beam axis:
\be
M_{WW}>500~GeV,~~p_{T}(W)>150~GeV,~~\vert \cos\theta (W)\vert<0.8
\ee

The remaining major background comes from
\be
e^+e^-\rightarrow e^+e^-W^+W^-,e^+e^-ZZ,e^\pm\nu W^\mp Z
\ee
where the final state electron disappears along the beam pipe.

This is reduced through the following cuts (at $\sqrt{s}=1.5~TeV$)
\be50~GeV<p_{T}(WW)<300~GeV,~~~20~GeV<p_T(ZZ)<300~GeV
\ee and a veto on hard electrons
\be
no~e^\pm ~with~ E>50~GeV~and~ \vert\cos\theta_e\vert<\cos(0.15~rad)
\ee

\begin{figure}
\epsfysize=8truecm
\centerline{\epsffile{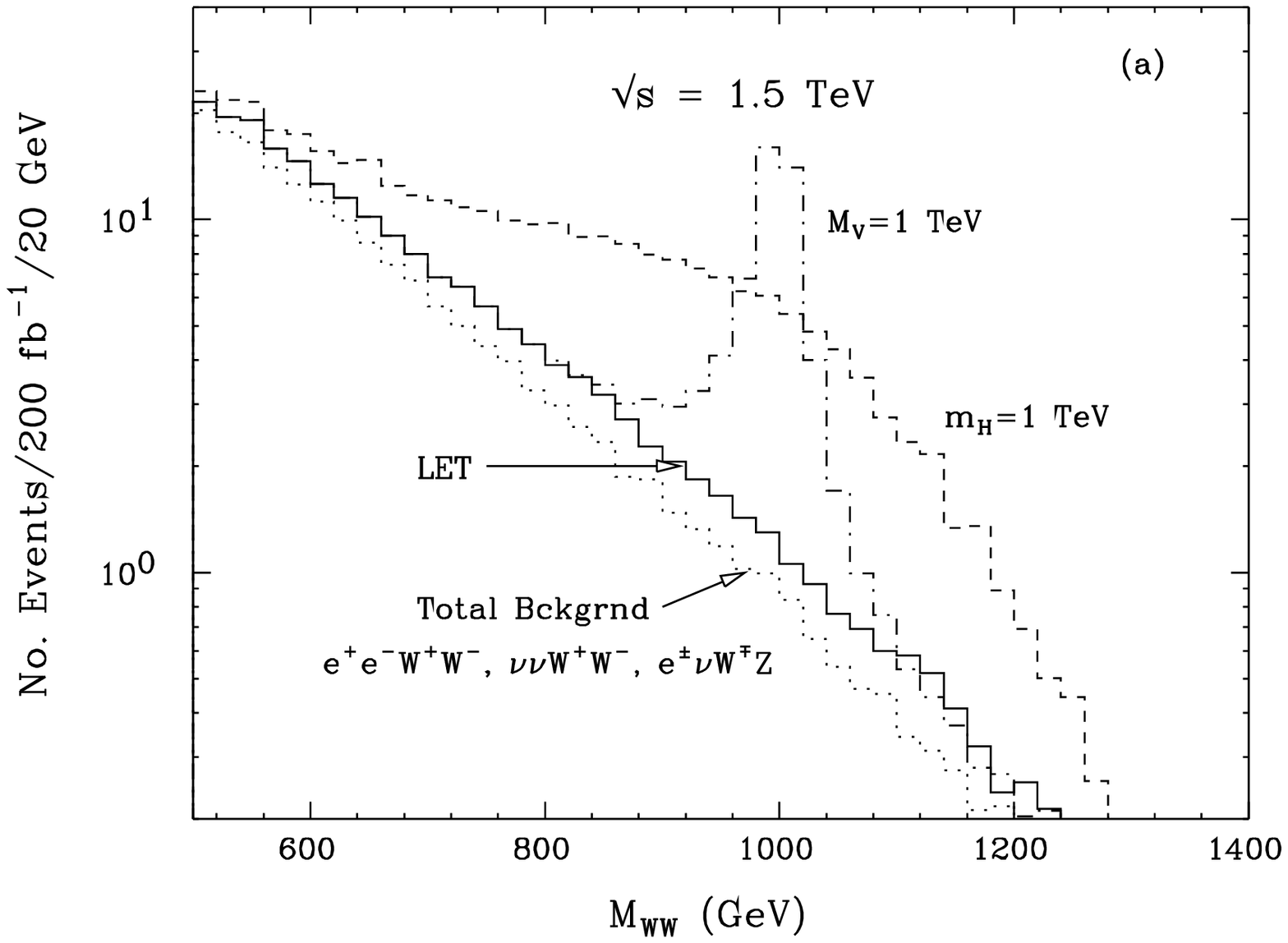}}
\noindent
\caption[barga]{Signal and background events versus the diboson invariant
mass
in the channel $e^+e^-\rightarrow \bar \nu\nu W^+W^-$ for 
$200~fb^{-1}$ and $\sqrt{s}=1.5~TeV$. The dashed curve
denotes the SM with $M_H=1~TeV$, the solid one the SM
 with $M_H\to\infty$ (LET);
 the  dot-dashed curve
denotes the vector model with $M_V=1~TeV$ and $\Gamma_V=30~GeV$,
 the dotted curve the total
background. The CCS model is close to the SM with $M_H=1~TeV$,
 from \cite{bargerepem}.}
\label{barga}
\end{figure}

\begin{figure}
\epsfysize=8truecm
\centerline{\epsffile{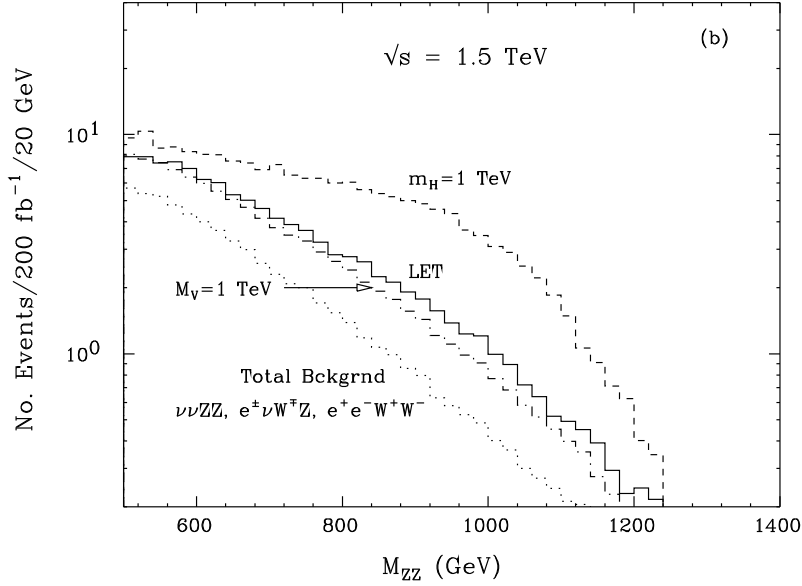}}
\noindent
\caption[bargb]{Signal and background events 
versus the diboson invariant mass
in the channel $e^+e^-\rightarrow \bar \nu\nu ZZ$ for 
$200~fb^{-1}$ and $\sqrt{s}=1.5~TeV$. The dashed curve
denotes the SM with $M_H=1~TeV$, the solid one the SM
 with $M_H\to\infty$ (LET);
the   dot-dashed curve
denotes the vector model with $M_V=1~TeV$ and $\Gamma_V=30~GeV$,
the dotted curve the total
background. The CCS model is close to the SM 
 with $M_H=1~TeV$, from \cite{bargerepem}.}
\label{bargb}
\end{figure}

The results for the collisions at $\sqrt{s}=1.5~TeV$ are
presented in Fig. \ref{barga}, \ref{bargb}.
 The branching fractions $BR(W\rightarrow jj)=67.8\%$
and $BR(Z\rightarrow jj)=69.9\%$ and the $W,Z$ identification 
/miseidentifications
factors are included.
The results  summarized in Table 2
 show that in order to have a statistical significant signal
one needs a high luminosity $(200~fb^{-1})$.
\begin{table}
\begin{center}
\begin{tabular}{l c c c c}
\hline
\hline
& & & & \\
{\rm Channels} &SM $(M_H=1~TeV)$ & CCS $(M_S=1~TeV)$& 
V $(M_V=1~TeV)$& LET \\ 
\hline\hline
S($e^+e^-\rightarrow \bar\nu\nu WW)$& 160 & 160 & 46 & 
31 \\ 
B&170&170&4.5&170\\
S/$\sqrt{B}$&12&12&22&2.4\\
& & & & \\
S($e^+e^-\rightarrow \bar\nu\nu ZZ$)&120 &130
&36  &45  \\ 
 B& 63 & 63&63&63 \\
S/$\sqrt{B}$& 15&17&4.5&5.7\\
\hline
\hline
\end{tabular}
\end{center}

\begin{description}
\item {\bf Table 2}: Total number of $WW$, $ZZ$ signal and background at
$\sqrt{s}=1.5~TeV$ and integrated luminosity $200~fb^{-1}$. Events are
summed over
the mass range $0.5<M_{WW}(TeV)<1.5$ except for the vector case  
$V$ ($0.9<M_{WW}(TeV)<1.1$).
\end{description}
\label{table2}
\end{table}

 The authors find a 5.7 $\sigma$ signal for LET 
amplitudes in the $ZZ$
final state, and 22 $\sigma$  in the $WW$
channel for the vector model with $M_V=1~TeV$ and
$\Gamma_V=30~GeV$ (This would correspond to a BESS parameter
$g/\gs=0.08$).

At the LC
 it is interesting to consider also the effect of polarization.
The results, assuming initial $e_L^+$ and $e_R^-$, are 
 listed in Table 3.

\begin{table}
\begin{center}
\begin{tabular}{l c c c c}
\hline
\hline
& & & & \\
{\rm Channels} &SM $(M_H=1~TeV)$ & CCS $(M_S=1~TeV)$& 
V $(M_V=1~TeV)$& LET \\ 
\hline\hline
S($e^+e^-\rightarrow \bar\nu\nu WW)$& 330 & 320 & 92& 
62 \\ 
B&280&280&7.1&280\\
S/$\sqrt{B}$&20&20&35&3.7\\
& & & & \\
S($e^+e^-\rightarrow \bar\nu\nu ZZ$)&240 &260
&72  &90 \\ 
 B& 110 & 110&110&110 \\
S/$\sqrt{B}$& 23&25&6.8&8.5\\
\hline
\hline
\end{tabular}
\end{center}

\begin{description}
\item {\bf Table 3}: Total number of $WW$, $ZZ$ signal and background at
$\sqrt{s}=1.5~TeV$ and integrated luminosity $200~fb^{-1}$ with $100\%$
polarized beams. Events are summed over
the mass range $0.5<M_{WW}(TeV)<1.5$ except for the vector case  
$V$ ($0.9<M_{WW}(TeV)<1.1$).
\end{description}
\label{table3}
\end{table}

In conclusion $W^+W^-\rightarrow W^+W^-$ and $W^+W^-\rightarrow ZZ$ 
channels are sensitive to different dynamics of the electroweak
breaking. Signals are enhanced at $\sqrt{s}=2~TeV$ colliders and at
$\mu^+\mu^-$ colliders \cite{bargermu}.

\section{Measuring chiral parameters }
\label{chpar}	

At energies below the new resonances, vector boson scattering amplitudes
can   generally be described, as we have
already seen, by  electroweak chiral Lagrangians.

For a LC with center of mass  energy of 
$\sqrt{s}=1.6~ TeV$ and an integrated luminosity of $200~
fb^{-1}$  the sensitivity on the chiral parameters
$\alpha_4$ and $\alpha_5$, defined in 
eq. (\ref{a4a5}), has been recently investigated \cite{kilian}.
The processes $e^+e^-\rightarrow
W^+W^-\bar\nu\nu$ and $e^+e^- \rightarrow ZZ\bar\nu\nu$
 have been considered, by performing a 
complete calculation which includes all relevant Feynman diagrams at
tree level, without relying on the Equivalence Theorem or the
effective $W$ approximation.
 The following  set of optimized cuts as in \cite{bargerepem}
 have been applied to the $WW$ channel
\begin{center}
  $|\cos\theta(W)|<0.8$ \nonumber\\
  $150~GeV<p_T(W)$ \nonumber\\
  $50~ GeV < p_T(WW) < 300~ GeV$ \nonumber\\
  $200~GeV < M_{ inv}(\bar\nu\nu)$ \nonumber\\
  $700 ~ GeV < M_{ inv}(WW) < 1200~ GeV$
\end{center}
The lower bound on $p_T(WW)$ is necessary because of the large $W^+W^-
e^+e^-$ background which is concentrated at low $p_T$ if both
electrons disappear into the beampipe.
The window $ 700~ GeV < M_{ inv}(WW) < 1200~ GeV$ is used
to avoid regions of possible violation of unitarity.
For the $ZZ$ final state the same cuts have been applied  except
for $p_T^{min}(ZZ)=30~GeV$.
Hadronic decays of the $W^+W^-$ pairs and
hadronic as well as $e^+e^-$ and $\mu^+\mu^-$ decays 
of the $ZZ$ pairs have been
used.

 From this analysis in the case of a LC  with center of mass  energy  
$\sqrt{s}=1.6~ TeV$ and an integrated luminosity of $200~
fb^{-1}$
with polarized beams ($90\%$ for electrons
and $60\%$ for positron)   one gets
$\alpha_4\leq 0.003$ and $\alpha_5\leq 0.002$ at $91.7\%$ C.L.

\section{Comparison LC - LHC}
\label{lclhc}

Complementarity of LC and LHC has been already advocated referring
to different aspects of the electroweak symmetry breaking problem
\cite{peskcomp,pesmur,snowrep,LCW}.
 We perform here a critical discussion of the results
that can be obtained at future LC's and LHC on strong electroweak
sector.

Let us first consider the case of models for strong electroweak
sector with no resonance or with a new very
heavy scalar resonance. Then one can measure
the chiral lagrangian parameters given in eq. (\ref{a4a5}).

By rescaling the  result \cite{kilian} from
the channels $e^+e^-\rightarrow\nu\bar\nu ZZ$ and $e^+e^-\rightarrow 
e^+e^-W^+W^-$
to $100~fb^{-1}$ and
longitudinal polarization $P=0.8$ one gets \cite{snow}
$\alpha_5\leq 1.8\times 10^{-3}$ at $95\%$ C.L. to be compared with
$\alpha_5\leq 3.5\times 10^{-3}$  at $95\%$ C.L \cite{pela}
from the reaction $qq\rightarrow qqZZ$ at LHC, using the scheme of cuts of
CMS \cite{CMS}.

This can be translated, using eq. (\ref{a4}), in a lower  limit  
on the scalar resonance
\be
M_S~(TeV)\geq 1.8 ~@~LC (E=1.6~TeV,~~L=100~fb^{-1})
~~
M_S~(TeV)\geq 1.6 ~@~LHC
\ee
Therefore to get a result which is comparable or better
than  LHC, one needs
a high energy and luminosity LC.

In presence of new vector resonances the annihilation channels are much
more efficient than the fusion ones because all the center
 of mass energy is used to produce the new particles.

\begin{figure}
\epsfysize=8truecm
\centerline{\epsffile{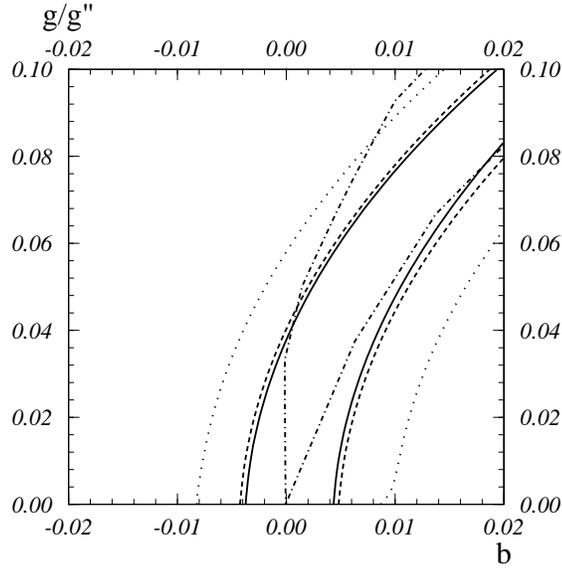}}
\noindent
\caption[cb1000]{BESS 90\% C.L. contours in the
 plane $(b,g/\gs)$ for $M_V=1~TeV$.
The dotted  line corresponds to the bound from
the $WW$ differential cross section, the dashed  line from   
 $W_{L}W_{L}$ differential cross section and
the continuous line from  the differential
$W_{L,T}W_{L,T}$ cross sections and $WW$ left-right asymmetries at
a $500~GeV$ LC.
The dot-dashed line corresponds to the bound from the total cross
section 
$pp\to W^\pm,V^\pm\to  W^\pm Z$ at LHC.
The allowed regions are  the internal ones.}
\label{cb1000}
\end{figure}

 The result for the BESS model has  already been shown
in
Fig. \ref{gamma} on the plane $(M_V,\Gamma_V)$. The comparison between
LHC and LC is shown in Fig. \ref{cb1000} on the plane 
$(b,g/\gs)$ 
for a LC of $\sqrt{s}=500~GeV$ and $L=20~fb^{-1}$, assuming
$M_V=1~TeV$ \cite{ecfa}.
The dotted  line corresponds to the $90\%$ C.L.
bound from
the $WW$ differential cross section, the dashed  line to the bound  
from $W_{L}W_{L}$ differential cross section, and
the continuous line to the bound combining the differential
$W_{L,T}W_{L,T}$ cross sections and $WW$ left-right asymmetries.
The dot-dashed line represents the bound from LHC. The 
allowed regions are those between the lines.
The bound from
LHC is obtained by considering the total cross section
$pp\rightarrow W^\pm,V^\pm\rightarrow  W^\pm Z\rightarrow \mu\nu \mu^+\mu^-$ 
and assuming that no deviation is
observed
with respect to the SM within the experimental  errors:
a systematic error of $5\%$ and the  statistical one have been
considered.

\begin{figure}
\epsfysize=8truecm
\centerline{\epsffile{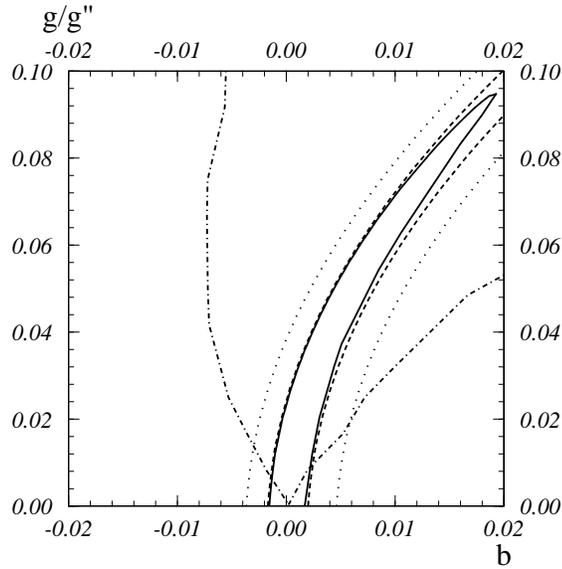}}
\noindent
\caption[cb2000]{Same as in Fig. \ref{cb1000} for $M_V=2~TeV$ and
a $800~GeV$ LC with $L=80~fb^{-1}$.}
\label{cb2000}
\end{figure}

\begin{figure}
\epsfysize=8truecm
\centerline{\epsffile{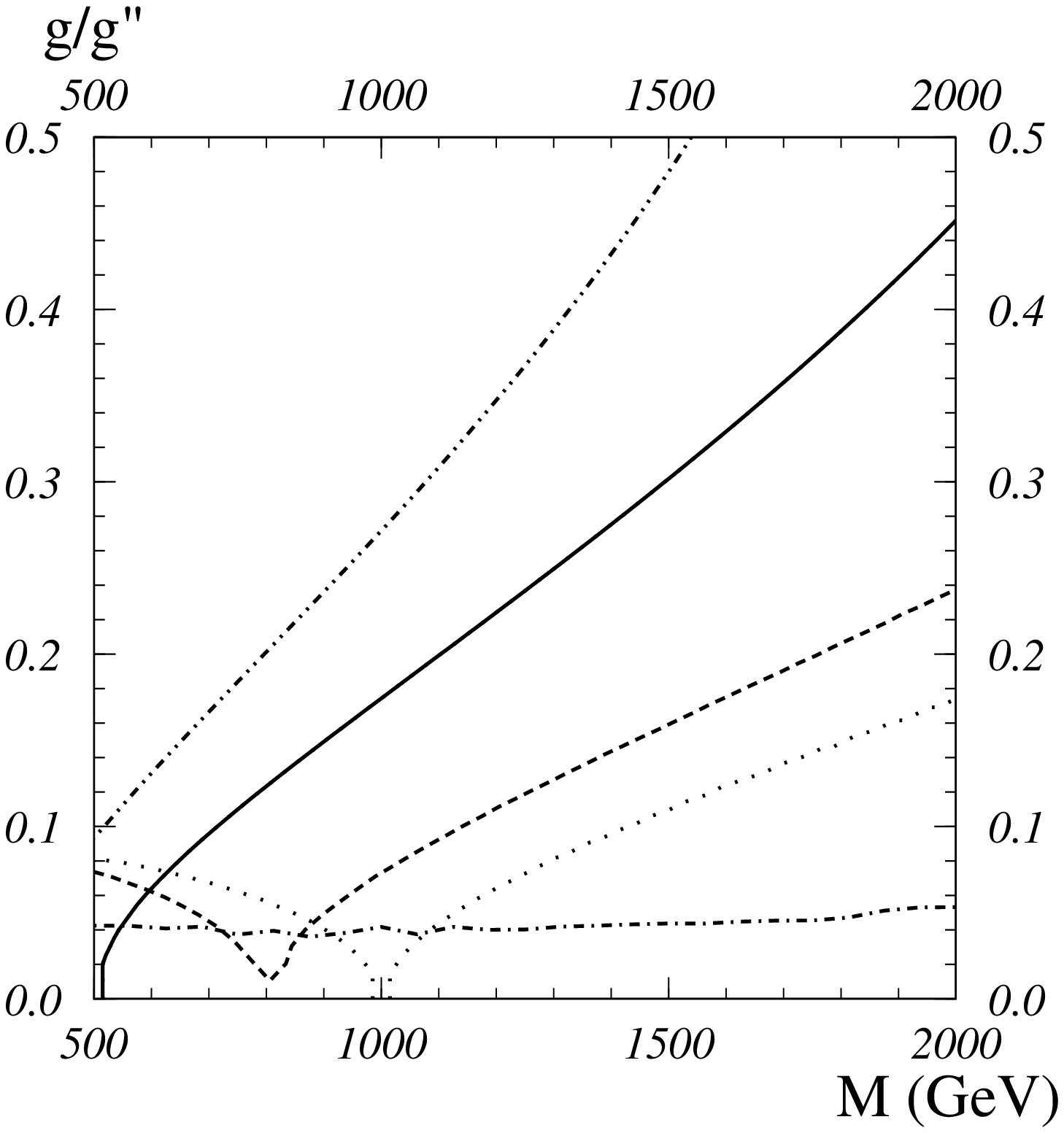}}
\noindent
\caption[deall]{Degenerate BESS
 90\%  C.L. contour on the plane ($M$, $g/g''$) from 
$e^+e^-$ at different $\sqrt{s}$ values:  the dash-double 
dotted line represents 
the limit from the 
combined unpolarized observables at $\sqrt{s}=360~GeV$ with an 
integrated luminosity of $L=10 fb^{-1}$,
the continuous line  at $\sqrt{s}=500~GeV$ and $L=20 fb^{-1}$,
the dashed  line  at $\sqrt{s}=800~GeV$ and $L=50 fb^{-1}$,
 the dotted line at $\sqrt{s}=1000~GeV$ 
and  $L=80 fb^{-1}$. The dot-dashed line represents
the limit from LHC.}
\label{deall}
\end{figure}
The above channel is for LHC the more efficient 
one in the case of a vector 
resonance strongly coupled to longitudinal $W$.
LHC can discover new vector resonances in a large region of
the parameter space up to masses $M_V=1.5-2~TeV$ in the channel
$pp\rightarrow W^\pm,V^\pm\rightarrow  W^\pm Z$ 
\cite{lhc1,CMS,ATLAS,bargerlhc}.
If such a $2~TeV$ resonance is not found at LHC then an upgrade of the
LC to $\sqrt{s}=800~GeV$ and $L=50 fb^{-1}$ can be very effective in
getting bounds
on the parameter space of the model, as shown in Fig. \ref{cb2000},
where  $90\%$ C.L. contours in the
 plane $(b,g/\gs)$ for $M_V=2~TeV$ are presented \cite{ecfa}.
The bounds come from the same observables as in Fig. \ref{cb1000} at
a LC of $\sqrt{s}=800~GeV$ and $L=50~fb^{-1}$.
Again the dot-dashed line corresponds to the bound from the total cross
section 
$pp\rightarrow W^\pm,V^\pm\rightarrow  W^\pm Z$ at LHC.

 The neutral channel at LHC 
$pp\rightarrow \gamma ,Z,V\rightarrow  W^+ W^-$ suffers 
of  severe background from $t\bar t$  production.

Nevertheless the new neutral vector bosons can be studied at LHC by
considering their lepton decay up to masses of the
order $1~TeV$ \cite{lhc2}.

Notice that while LHC is more 
sensitive to the charged new vector bosons,
the LC is sensitive to the neutral ones (at least in the 
annihilation channels).

As a general conclusion it seems that for models of new vector
resonances, if one is able to reconstruct the final $W$ polarization,
the measurement of polarized $e^+e^-\rightarrow W_LW_L$ gives strong bounds on
the parameter space of the model.

In the case of the model with vector and axial-vector
resonances degenerate in mass (degenerate BESS),  LHC is sensitive to
 the new particles in the channels
$pp\rightarrow W^\pm,L^\pm\rightarrow\mu\nu$ and 
$pp\rightarrow\gamma,Z,L_3,R_3\rightarrow
\mu^+\mu^-$ up to masses of the order $2~TeV$ as shown in
Fig. \ref{deall}. The bounds are obtained by assuming no 
deviations with respect to the SM in the total
cross section $pp\rightarrow W^\pm,L^\pm\rightarrow\mu\nu$
within the experimental errors:
a systematic error of $5\%$ and the  statistical one have been
considered.
 The limits  are compared with those coming
from LC's in  Fig. \ref{deall} \cite{ecfa}.  The double dot-dashed 
line represents the limit from the combined unpolarized observables at 
$\sqrt{s}=360 ~GeV$ with an integrated luminosity of $L=10 fb^{-1}$;
the continuous line from  $\sqrt{s}=500 ~GeV$ and  $L=20 fb^{-1}$;
the dashed line from  $\sqrt{s}=800 ~GeV$ and  $L=50 fb^{-1}$
and the dotted 
 line from  
 $\sqrt{s}=1000$ GeV $L=80 fb^{-1}$. The dot-dashed line represents
the limit from LHC. Therefore to be competitive with LHC one
has to consider a LC of still  higher energy than those
considered.

The statistical significances of strong symmetry breaking signals at
the LC and LHC are summarized in Table~
4
from \cite{snowrep}.

\begin{table}[htb]
\bigskip\begin{center}
\begin{tabular}{|l|l|c|c||c|c|c|} \hline \hline
Collider & Process & $\sqrt{s}$ & ${\cal L}$&
$M_V$& $M_H$& LET \\
 &  & (TeV) & $(\rm{fb}^{-1})$ &
1.5 TeV & 1 TeV & \\
\hline
LC & $e^+e^-\rightarrow W^+W^-$ & .5 & 80 & $7\sigma$ & -- & -- \\
LC & $e^+e^-\rightarrow W^+W^-$ & 1.0 & 200 & $35\sigma$ & -- & -- \\
LC & $e^+e^-\rightarrow W^+W^-$ & 1.5 & 190 & $366\sigma$ & -- & $5\sigma$ \\
LC & $W^+W^-\rightarrow  ZZ$ & 1.5 & 190 & -- & $22\sigma$ & $8\sigma$ \\
LC & $W^-W^-\rightarrow  W^-W^-$ & 1.5 & 190 & -- & $4\sigma$ & $6\sigma$
\\[2ex]
LHC & $W^+W^-\rightarrow W^+W^-$ & 14 & 100 & -- & $14\sigma$ & -- \\
LHC & $W^+W^+\rightarrow  W^+W^+$ & 14 & 100 & -- & $3\sigma$ & $6\sigma$ \\
LHC & $W^+Z\rightarrow  W^+Z$ & 14 & 100 & $7\sigma$ & -- & -- \\
\hline \hline
\end{tabular}
\end{center}
\begin{description}
\item {\bf Table 4}: Statistical significance of strong electroweak
sector
at LC and LHC, from \cite{snowrep}. .
\label{tab4}
\end{description}
\end{table}
The LHC
results are taken from the ATLAS design report~\cite{ATLAS}. If an
entry is blank it means that the process is insensitive to the
corresponding model or that the analysis has not been done. An electron beam
with $ 90\%$ left--handed polarization has been assumed.
 For both the LC and LHC
results it was assumed that $M_V= 1.5~TeV$  and $\Gamma_V=
0.33~TeV$.

For vector resonances a LC with $\sqrt{s}=500~GeV$ and $L=80~fb^{-1}$
has the same sensitivity as LHC.

At a  LC with $\sqrt{s}=1500~GeV$ the effects of a strong
interacting sector becomes more relevant  even if the $WW$
channel is not resonant.

Notice however that systematic
errors have largely been ignored  both for the LHC
and for the LC.

\section{Conclusions}
\label{conc}
In this paper we have reviewed some  theoretical aspects for a strong
interacting electroweak sector and its phenomenological consequences.
Different models of strong breaking, all based on the common
assumption
of a chiral symmetry breaking $SU(2)_L\otimes SU(2)_R\to SU(2)_{L+R}$,
but with a different content of particles (Goldstones, spin one vector
and axial-vector resonances, scalars) have been surveyed. 
These models are built using the effective lagrangian approach
which makes use of the chiral symmetry and of an expansion in the
energy. Bounds from already existing measurements have been 
taken into account by computing the effects beyond the SM on
the corresponding observables.
Since the Goldstone scattering amplitudes are at high energy 
equivalent to the corresponding scattering of longitudinal components 
of $W$ and $Z$ gauge bosons, the scattering of these gauge bosons
at future colliders in principle can give access to the mechanism
of electroweak symmetry breaking.
In this review we have considered how future $e^+e^-$ linear colliders
can investigate  this phenomenon. Different channels have been
studied:
the
annihilation
processes $e^+e^-\rightarrow f^+f^-$ and $e^+e^-\rightarrow W^+W^-$ are in
particular
relevant if a new vector resonance mixed with $Z$ is present.
These channels are already important 
at a LC of a center of mass energy of $500~GeV$ for a  new
resonance up to masses 
of the order of  $1~TeV$.
In principle if LHC has already discovered such a new vector boson, one can
tune the LC energy to study its properties and decays. Otherwise the
LC can give bounds on its couplings and masses.
 The process $e^+e^-\rightarrow W^+W^-$ is particularly relevant
because, using the topology in which one $W$ decays hadronically
and the other leptonically, the angular distributions  of the
$W$ can be measured and therefore
the corresponding $W$ polarization reconstructed. In this way one has
access to $W_LW_L$ scattering.

The fusion processes 
$e^+e^-\rightarrow \bar \nu\nu W^+W^-$ and
$e^+e^-\rightarrow \bar \nu\nu ZZ$ 
can be used to study $WW$ scattering
also in absence of new resonances but, as we have seen, for
this   much
higher
energy (of the order $1.5~TeV$) and luminosity ($200~fb^{-1}$)
would be required.
One can reach a 5.7 $\sigma$ signal for LET amplitudes,
12 $\sigma$ for the CCS with $M_S=1~TeV$ and $\Gamma_S=0.35~TeV$,
 and 22 $\sigma$ for the vector
model with
$M_V=1~TeV$, $\Gamma_V=30~GeV$.

All these processes are suitable for neutral vector bosons. 
LHC is complementary because it is more efficient for
the channel 
$pp\rightarrow W^\pm,V^\pm\rightarrow  W^\pm Z\rightarrow \mu\nu
\mu^+\mu^-$
when the 
 new vector 
resonance is strongly coupled to longitudinal $W$.
LHC can discover new charged vector resonances in a large region of
the parameter space up to masses $M_V=1.5-2~TeV$. 
The neutral channel
$pp\rightarrow \gamma ,Z,V\rightarrow  W^+ W^-$ suffers 
of  background from $t\bar t$  production,
nevertheless the new neutral vector bosons can be studied at LHC by
considering their lepton decay  up to masses of the
order $1~TeV$. 

In the case of the model with vector and axial-vector
resonances degenerate in mass,  LHC is sensitive to
 the new particles in the channels
$pp\rightarrow W^\pm,L^\pm\rightarrow\mu\nu$ and 
$pp\rightarrow\gamma,Z,L_3,R_3\rightarrow
\mu^+\mu^-$ up to masses of the order $2~TeV$. To be
competitive with LHC one needs  to consider a LC with
a c.m.  energy of the order $1.5~TeV$.

Finally let us mention that  the symmetry group can be larger than 
$SU(2)_L\times SU(2)_R$ like in the one family technicolor model
based on the chiral symmetry $SU(8)_L\times SU(8)_R$ \cite{farhi}.
In this review we  have not considered 
 such 
 models. These models have a rich particle spectrum with new
pseudo-Goldstone bosons which in principle could be produced at future
LC's. Their phenomenology has been for instance considered in
 \cite{pseu,lubi,leptq,ruc}. 
Furthermore we have not considered the $\gamma\gamma$ and the
$e\gamma$
options of LC's (see for example
\cite{han,jikia} for recent reviews).

\section{Appendix  }
\label{appA}

The most simple description of the symmetry breaking in
the SM is obtained by considering a  $2\times 2$ unitary
matrix field $U$, satisfying $U^\dagger U=1$. This field transforms
as $U\to g_L U g_R^\dagger$ under the group  $SU(2)_L\otimes SU(2)_R $
and describes the spontaneous breaking $SU(2)_L\otimes SU(2)_R\to
SU(2)_{L+R}$.
This corresponds to the
infinite
Higgs mass limit in the SM. In this non linear realization of the
symmetry breaking no Higgs is left in the spectrum. 
 Due to the condition  $U^\dagger U=1$ the symmetry
is non linearly realized.

We will therefore briefly review this technique of non linear 
group realization of a group $G$ which breaks spontaneously to a
subgroup
$H$ \cite{ccwz}.

Let $G$ be a compact, connected and semisimple Lie group of dimension
$n$ and $H$ a
subgroup.
Let us denote by $V_i$ $(i=1,\dots ,n-d)$ the generators of $H$ and by
$A_l$
$(l=1,\dots ,d)$
the remaining generators.
Every group element $g\in G$ can be decomposed as
\be
g=e^{\csi\cdot A} e^{ u\cdot V}
\ee
where $\csi\cdot A =\csi_l A_l$ and $u\cdot V =u_i V_i$ and
$u$ and $v$ are real parameters.
For every element $g_0\in G$ one has
\be
g_0 e^{\csi\cdot A}=e^{\csi^\prime \cdot A} e^{u^\prime\cdot V}
\label{nonlin}
\ee
where $\csi^\prime =\csi^\prime (\csi,g_0) $ and
 $u^\prime =u^\prime (\csi,g_0) $.

Let  $D$ denote a linear representation  of the subgroup $H$
\be
h:~~~\psi\rightarrow D(h)\psi
\ee
then the transformation 
\be
g_0:~~\csi\rightarrow\csi^\prime ,~~~~
\psi\rightarrow D(e^{u^\prime\cdot V})\psi
\ee
gives a nonlinear realization of $G$. In fact if
\be
g_1 e^{\csi^\prime\cdot A}=e^{\csi^{\prime\prime} \cdot A}
 e^{u^{\prime\prime}\cdot V}
\ee
then
\be
g_1g_0 e^{\csi\cdot A}=e^{\csi^{\prime\prime} \cdot A}
 e^{u^{\prime\prime\prime} \cdot V}
\ee
with
\be
 e^{u^{\prime\prime\prime} \cdot V}=  
e^{u^{\prime\prime} \cdot V}e^{u^\prime \cdot V}
\ee
Since $D$ is a representation
\be
 D(e^{u^{\prime\prime\prime} \cdot V})=  
D(e^{u^{\prime\prime} \cdot V})D(e^{u^\prime \cdot V})
\ee
If $g_0=h\in H$ then
\be
e^{\csi\cdot A}\rightarrow h e^{\csi\cdot A} h^{-1},
~~~~\psi\rightarrow D(h)\psi
\ee
and the symmetry is linearly realized.

There is a special case in which the transformation on $\csi$ can be
simplified. This is the case in which the group has a parity like
transformation $P:g\to P(g)$ such that
\be
V\to V, ~~~A\to -A
\ee
Applying this operation to  eq. (\ref{nonlin}) one gets
\be
P(g_0) e^{-\csi\cdot A}=e^{-\csi^\prime \cdot A} e^{u^\prime\cdot V}
\label{Rnonlin}
\ee
and combining eqs. (\ref{nonlin}) and (\ref{Rnonlin}), 
\be
g_0  e^{2\csi\cdot A} P(g_0^{-1})=e^{2\csi^\prime \cdot A}
\label{186}
\ee
One can easily verify in this form that the transformation on
$\csi$ is a realization of the group which becomes
linear when restricted to the subgroup.

In the  case we are interested, $G=SU(2)_L\otimes SU(2)_R$
\be
g=\pmatrix{L&0\cr 0&R}
\ee
with $L(R)\in SU(2)_{L(R)}$. 
Furthermore
\be
V^i=T_{L}^i+T_{R}^i=i\pmatrix{\f {\dd \tau^i}{\dd 2}&0\cr
0&\f {\dd\tau^i}{\dd 2}},~~~~~
A^i=T_{L}^i-T_{R}^i=i\pmatrix{\f {\dd \tau^i}{\dd 2}&0\cr
0&-\f {\dd\tau^i}{\dd 2}}
\ee
A parity transformation $P$ exists, $P\colon L\to R$, 
and using eq. (\ref{186}) one has the following
transformation
properties
\be
\pmatrix{\exp(i \csi^i \tau^i)&0\cr
0&\exp(-i \csi^i \tau^i)}\rightarrow
\pmatrix{L&0\cr 0&R} \pmatrix{\exp(i \csi^i \tau^i)&0\cr
0&\exp(-i \csi^i \tau^i)} \pmatrix{R^\dagger&0\cr 0&L^\dagger}
\ee
Therefore one  can  construct the Lagrangian using the field $\exp(i\csi \cdot
\tau)$
or introducing the usual normalization 
$U(x)=\exp(i\pi^a(x)\tau^a/v)$,
 transforming  under $G$
as $(2,2)$ or $U\to g_L Ug_R^\dagger$,  $g_{L(R)}\in SU(2)_{L(R)}$. 

\vskip1.5truecm
{\bf Acknowledgments}
\vskip0.5truecm
 I would like to thank R.Casalbuoni, P.Chiappetta,
A.Deandrea, S.De Curtis, 
F.Feruglio,
R.Gatto and M.Grazzini for the fruitful and enjoyable collaboration 
on the topics covered here.
I am grateful to  R.Casalbuoni,
S.De Curtis and R.Gatto for a critical reading of the
manuscript.

This work is part of the EEC project ``Test of electroweak symmetry
breaking and future european colliders'', CHRXCT94/0579.

\end{document}